\documentclass[pra,twocolumn,superscriptaddress,amssymb]{revtex4-2}
\usepackage[colorlinks=true,linkcolor=magenta,citecolor=magenta,urlcolor=magenta,linktocpage=true]{hyperref}
\usepackage{tabularx}
\newcolumntype{Y}{>{\centering\arraybackslash}X}
\usepackage{array, makecell}
\usepackage{amsmath}
 \usepackage{longtable} 
 \usepackage{xcolor}
 \usepackage{breakcites}

\definecolor{codegreen}{rgb}{0,0.6,0}
\definecolor{red}{rgb}{1,0,0}
\def\tallvdots{
   \mathrel{\vcenter{\offinterlineskip
      \kern-1.4ex\hbox{$\vdots$}\kern-.7ex\hbox{$\vdots$}
   }}
}
\newcommand{\binomial}[2]{\left(\begin{array}{c}#1 \\#2 \end{array}\right)}
\newcommand{\ket}[1]{\left |#1 \right\rangle}

\newcommand{\ketbra}[2]{\left|#1\right\rangle\left\langle#2\right|}

\usepackage{graphicx}
\begin{document}
\title{Input-Output Hierarchical Equations Of Motion}
\author{Mauro Cirio}
\email{cirio.mauro@gmail.com}
\affiliation{Graduate School of China Academy of Engineering Physics, Haidian District, Beijing, 100193, China}
 \author{Pengfei Liang}
  \email{pfliang@imu.edu.cn}
 \affiliation{Center for Quantum Physics and Technologies,  School of Physical Science and Technology, Inner Mongolia University, Hohhot 010021, China}
 \affiliation{Graduate School of China Academy of Engineering Physics, Haidian District, Beijing, 100193, China}
 \author{Neill Lambert}
 \email{nwlambert@gmail.com}
\affiliation{Theoretical Quantum Physics Laboratory, RIKEN Cluster for Pioneering Research, Wako-shi, Saitama 351-0198, Japan}
\affiliation{RIKEN Center for Quantum Computing (RQC), Wakoshi, Saitama 351-0198, Japan}
\date{\today}
\setcounter{section}{0}

\begin{abstract}
We derive an extended version of the hierarchical equations of motion (HEOM) to compute output physical properties of a bosonic environment, which is allowed to be initially prepared at an earlier time in a non-Gaussian input state and then non-perturbatively interact with a quantum system with a linear environmental operator. 
While spectral assumptions analogous to the ones used in the regular HEOM are imposed to compute  dynamical output bath observables, they are not required to model input states  or output observables at a fixed time,  in this case leading to time-dependent contributions to the equations.  In the Markovian limit, we use this formalism to derive an input-output Lindblad equation which can be used to extend the applicability of the regular version.
For a given desired input state and output observable, the range of the indexes extending the regular HEOM is, by construction, bounded. Overall, the aim of this formalism is to take advantage of the efficiency   and generality of the HEOM framework to  model non-Gaussian input states and the dynamics of environmental observables in bosonic, non-Markovian open quantum systems.
\end{abstract}
\pacs{}
\maketitle
A complete description of an open quantum system \cite{Petruccione,Gardiner} requires, on one side, to model the reduced dynamics of the system and, on the other side, to compute useful observables related to the external continuum. In the Markovian limit, this can be done solving perturbative master equations for the reduced system dynamics \cite{Redfield,Lindblad,Gorini,Fruchtman}, and analyzing input-output relations for the environment \cite{PhysRevA.31.3761}  which can be used to model input-output networks \cite{Hudson,5286277,PhysRevA.87.032117,Gough,doi:10.1080/23746149.2017.1343097} such as those characterized by cascaded quantum systems \cite{PhysRevLett.70.2269,PhysRevLett.70.2273}, time-delay \cite{PhysRevLett.112.110401,PhysRevLett.115.060402,Tabak2016,PhysRevA.94.023806,PhysRevLett.116.093601}, and feedback  \cite{PhysRevA.49.4110,PhysRevA.78.032323,PhysRevA.81.023804,Mabuchi,ZHANG20171}. 

A more refined analysis to characterize system-bath hybridization often requires a description of the coherent interaction between the system and collective environmental degrees of freedom. To model this interaction, it is possible to proceed using non-perturbative techniques such as  the reaction coordinate method \cite{Garg,Martinazzo,iles2014environmental,Woods,PhysRevB.97.205405,Melina}, chain mapping \cite{Chin_2010,PhysRevLett.105.050404,Woods,PhysRevLett.123.090402,PhysRevB.101.155134}, the polaron transformation \cite{Holstein1,Holstein2,Jackson,Silbey,Silbey2,PhysRevB.57.347,PhysRevB.65.235311,Weiss,Jang,Jang2,PhysRevLett.103.146404,McCutcheon,Jang3,PhysRevB.83.165101,Kolli,PollockNazir,PollockThesis,Xu},   by directly calculating the Feynman-Vernon influence functional using tensor networks \cite{PhysRevB.93.075105,PhysRevA.94.053637,PhysRevA.97.012127,tempo,PhysRevLett.123.240602,Strathearn_book,PRXQuantum.3.010321,ace,PhysRevLett.128.167403}, or by discretizing the bath in frequency domain \cite{PhysRevB.92.155126,10.1063/1.5135363,Shuai_review}. 

In parallel, it is possible to achieve compatible results by using purely effective non-perturbative methods which focus on describing the effects of the environment by introducing ancillary degrees of freedom which do not have a direct correspondence to the original environment.  A typical example are the auxiliary density matrices in the Hierarchical Equations of Motion (HEOM) method \cite{Tanimura_3,Tanimura_1,YAN2004216,Ishizaki_1,Zhou_2005,Tanimura_2,Ishizaki_2,PhysRevLett.104.250401,doi:10.1143/JPSJ.81.063301,PhysRevLett.109.266403,PhysRevA.85.062323,Moix,Tanimura_2014,Yan_Shao,Song_Shi,Hsieh_1,Hsieh_2,FreePoles,Dan_Shi,xu2023universalframeworkquantumdissipationminimally,Tanimura_2020,Tanimura_2021}, dissipatons \cite{Yan_1,10.1063/1.4905494,Yan_2,Yan_3,Yan_4,10.1063/5.0123999,10.1063/5.0155585,10.1063/5.0151239,Chen_Yan,li2024quantumsimulationnonmarkovianopen}, the stochastic
Hierarchy of Pure States (HOPS) method \cite{PhysRevLett.113.150403,Hartmann2017,Hartmann2021-zx,10.1063/5.0192075},  deterministic \cite{Nakajima,Zwanzig} and stochastic  \cite{Diosi_2,Diosi_1,PhysRevA.58.1699,PhysRevLett.82.1801} memory kernels, minimally extended state spaces \cite{xu2023,vilkoviskiy2023}, Liouville space operators \cite{Stockburger_JCP1999,Stockburger_CP2001,PhysRevLett.88.170407,Shao,STOCKBURGER2004159,PhysRevLett.100.230402,Stockburger_2016,PhysRevLett.123.050601,PhysRevA.100.042112,Chernyak1,Chernyak2,PhysRevE.102.062134,Yun-AnShao}, or pseudomodes \cite{PhysRevA.50.3650,PhysRevA.55.2290,PhysRevLett.110.086403,PhysRevB.89.165105,PhysRevB.92.245125,Schwarz,Dorda,Mascherpa,Lemmer_2018,Tamascelli,Lambert,Chen_2019,PhysRevLett.122.186803,PhysRevResearch.2.043058,PhysRevA.101.052108,8820133,PhysRevResearch.5.033011,LuoSi,PhysRevResearch.6.033237,PhysRevA.110.022221,albarelli2024}. 
 
While the ``effective'' nature of these methods is designed to allow compact and optimized descriptions of the reduced system dynamics, it comes at the price of a less obvious way to  compute general environmental properties. In fact, while it might be argued that their ancillary degrees of freedom should constitute a representation of the original environment, an explicit mapping to observables in the original quantum system is usually limited to collective quantities related to the interaction operator, such as energy flows \cite{10.1063/1.4890441,10.1063/1.4971370,Wiedmann_2020,10.1063/5.0093666,PhysRevApplied.20.024038,10.1063/5.0192075,PhysRevResearch.6.033237}. For these observables, this leads to results which are consistent with perturbative expansions of ``physically-oriented'' methods such as the reaction coordinate \cite{iles2014environmental,10.1063/1.4940218,Melina}. 

In general, the computation of bath observables can be done by introducing non-Markovian extension to the input-output formalism \cite{PhysRevA.85.034101,PhysRevA.87.032117,Propp:22,PRXQuantum.3.020348} or, more generally, by computing system correlations \cite{PhysRevLett.126.093601,Gribben2022usingenvironmentto} and by numerically propagating the environmental physical degrees of freedom using matrix product states and tensor networks \cite{PhysRevLett.113.263604,tempo,PhysRevA.97.012127,Strathearn_book,PhysRevLett.123.240602,PhysRevA.102.062414,10.1063/5.0047260,ace,PhysRevLett.128.167403,PRXQuantum.3.010321,PhysRevB.107.195306,PhysRevB.107.L201115,PhysRevB.107.125103,PhysRevResearch.5.033078,PhysRevX.14.011010,PhysRevLett.132.053602,PhysRevLett.132.200403,cygorek2024}, possibly in conjunction with collision  models \cite{PhysRevA.100.052113,PhysRevA.101.023807,PhysRevResearch.3.023030,PhysRevA.106.013714} or other analytical tools \cite{PhysRevA.96.023831,PhysRevA.102.013726,PhysRevA.102.013709,PhysRevResearch.2.043014,PhysRevLett.131.073602,rodallordes2023,vega2024,dibenedetto2024}, such as the polaron transformation \cite{PhysRevB.93.155423,PhysRevB.98.045309,PhysRevA.99.013807,PhysRevA.102.023702,PhysRevA.104.053701,PhysRevA.106.063717}. Other methods have been also used to model baths characterized by finite bandwidths \cite{HJCarmichael_1973,Gardiner:87,PhysRevLett.61.1097,PhysRevA.38.4657,PhysRevA.42.4352,PhysRevA.46.4354,PhysRevA.50.1792,PhysRevA.50.1700,doi:10.1080/09500349608232870,doi:10.1080/09500349708230713,PhysRevA.59.R2579,PhysRevA.61.033811,doi:10.1080/09500340108232462,PhysRevA.105.023721,PhysRevResearch.6.013150},  Poisson statistics \cite{PhysRevLett.132.170402}, time-delays \cite{PhysRevB.107.205301}, and even  the non-perturbative incorporation of the system in terms of a general non-linear response theory in the context of superconducting quantum circuits \cite{Hyyppa2022,vadimov2023}. 
Remarkably, the power of non-perturbative methods can also manifest in the possibility to study response functions \cite{10.1063/5.0193530,10.1063/1.5134745} and the effects of initial correlations between the system and the environment \cite{doi:10.1143/JPSJ.81.063301,10.1098/rsta.2011.0203,10.1063/1.4890441} thanks their capability to evolve separable states into entangled ones \cite{PhysRevLett.104.250401}.

Here, we aim to achieve these goals within the context of the Hierarchical Equations of Motion. In fact, alongside their formal elegance \cite{Tanimura_3,Tanimura_1,YAN2004216,Ishizaki_1,Zhou_2005,Tanimura_2,Ishizaki_2,PhysRevLett.104.250401,doi:10.1143/JPSJ.81.063301,PhysRevLett.109.266403,PhysRevA.85.062323,Moix,Tanimura_2014,Yan_Shao,Song_Shi,Hsieh_1,Hsieh_2,FreePoles,Dan_Shi,xu2023universalframeworkquantumdissipationminimally,Tanimura_2020,Tanimura_2021}, these equations have been optimized numerically and applied to a range of physical systems~\cite{doi:10.1021/ct3003833,Qutip1,Qutip2,https://doi.org/10.1002/wcms.1269,https://doi.org/10.1002/jcc.25354,10.1063/5.0007327,Lambert_Bofin,Huang2023,quick_heom2}, but, so far, are  are usually limited in capability to computing system or system-environment energy-flow properties. Because of their broad range of application, introducing the capability to obtain arbitrary bath observables and prepare arbitrary non-Gaussian initial environment states is highly desirable. In fact, this would allow the use of the HEOM to analyze input-output and scattering between propagating photons and quantum emitters along the lines of \cite{Shen:05,PhysRevA.76.062709,PhysRevB.79.205111,PhysRevA.82.063821,Pletyukhov,PhysRevLett.113.183601,Caneva,PhysRevA.92.053834,PhysRevLett.119.153601,PhysRevLett.120.153602,PhysRevA.97.043850,PhysRevLett.126.023603,Le_Jeannic_2022,González-Tudela2024}.
 
To achieve this goal, we analyze correlation functions (for linear  environmental fields which encode input-output information) which depend on the full system-bath interaction (which itself is assumed to be linear in the environmental coupling operators). We derive an exact representation for these correlations in terms of a finite series which involves free-bath correlations (in the fields) and free-bath  cross correlations (between the fields and the environmental coupling operators).
We show that the evaluation of this series is equivalent to solving a set of hierarchical equations of motion which extend the regular ones for the system dynamics.  The range of the indexes extending the regular HEOM is, by construction, bounded by the parameters characterizing the corresponding input-output task. 

 The possibility to solve these extended HEOM  allows, for example, to model baths initially prepared in non-Gaussian states. While this might appear surprising, it is a direct consequence of the fact that the input and output properties are ultimately encoded in correlation functions defined in terms of fields which are linear in the environmental modes and in terms of an underlying Gaussian equilibrium state. The mapping of these  correlations into a series is  designed to reshape the full combinatorial power of Wick's theorem into a form which can be readily used to derive the proposed input-output hierarchical equations of motion (io-HEOM), thereby offering an intuitive justification for these results. In the Markovian limit, this formalism allows the derivation of an input-output version of the Lindblad master equation which can be used to extend the application of the regular version to include the effects of wave-packets propagating in the environment and the estimation of bath observables.

This article is organized as follows. We start by introducing the main formalism for open quantum systems in section \ref{sec:Introduction} and derive the regular HEOM in section \ref{sec:HEOM_main_rsd}. In section \ref{sec:environmentalcorrelations}, we present a formal representation for environmental correlation which we then map to an extended, input-output version of the HEOM in section \ref{sec:input_output}. There, we proceed by gradually increasing the level of complexity by writing the extended HEOM for pure output (section \ref{subsec:o}), pure input (section \ref{subsec:i}), and input-output (section \ref{subsec:io}). This leads to the presentation of a more abstract version of the extended HEOM in section \ref{sec:GeneralCase}. In section \ref{sec:io-Lindblad}, we show that, in the Markovian limit, this formalism  leads to an input-output version of the Lindblad equation.  The concluding section \ref{sec:Conclusions} gives an overall summary of these results, including an outlook on possible future directions.

\section{Introduction}
\begin{figure}[t!]
\includegraphics[clip,width=\columnwidth]{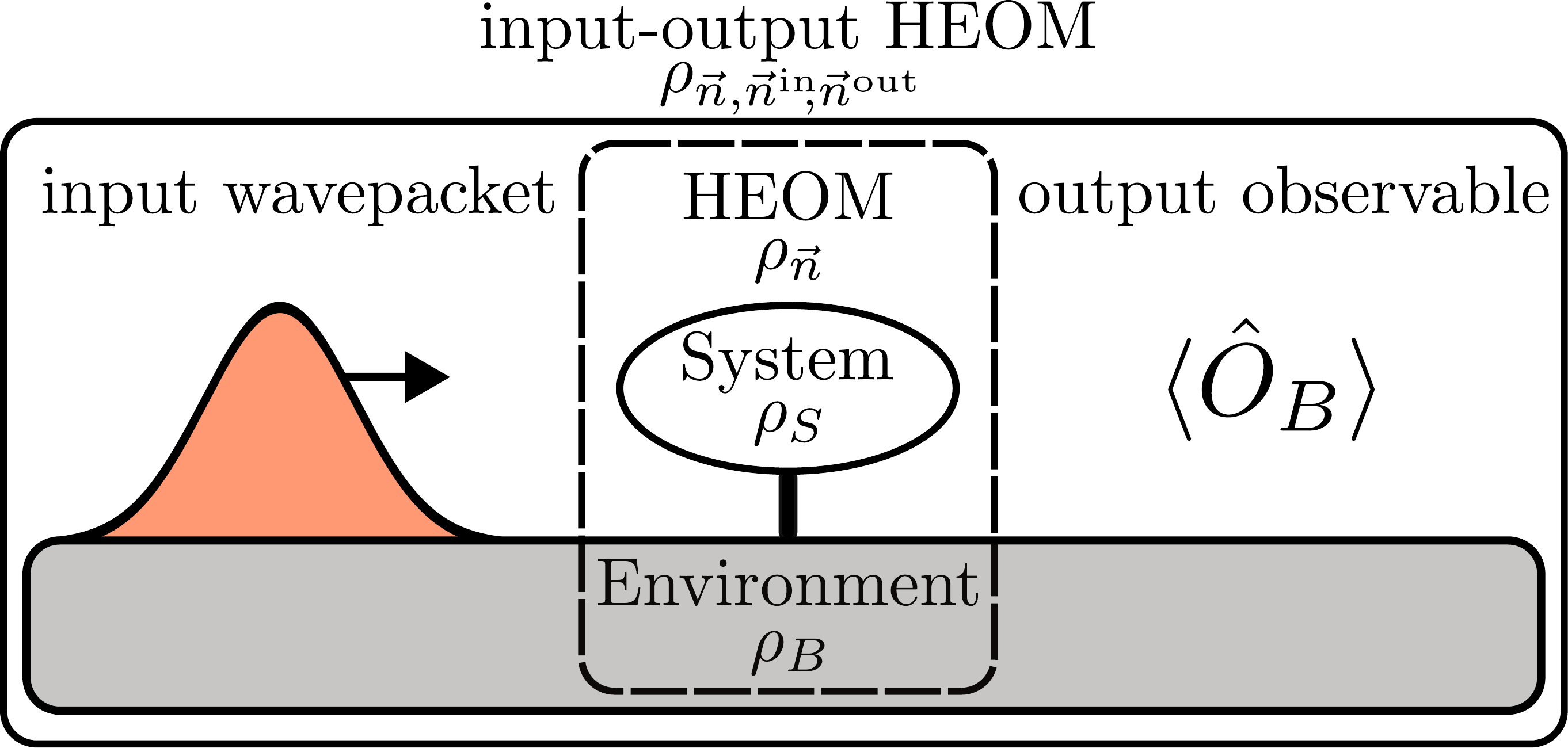}
\caption{Schematic representation of the input-output HEOM (io-HEOM). The main setting consists of a quantum system interacting with an environment. The regular-HEOM allow to compute the reduced system dynamics $\rho_S$ for a bath initially prepared in a Gaussian state $\rho_B$, by solving a differential equation involving different auxiliary system density matrices $\rho_{\vec{n}}$. The range of the label $\vec{n}$ depends on the spectral decomposition of the required exponential ansatz for the correlation between the environmental coupling operators and the system.  The regular-HEOM can also be used to compute environmental properties in case they can be directly related to such coupling operators. Here, we derive an input-output HEOM to further model both the preparation of the environment in a non-Gaussian state, such as an input wavepacket, and the computation of environmental output observables. This is done by considering an extension of the auxiliary space to $\rho_{\vec{n},\vec{n}^\text{in}\!\!,\vec{n}^\text{out}}$ to encode cross correlations between the operators defining the input, the output, and the environmental coupling operators. Interestingly, input indexes are characterized by binary entries, they do not require any spectral ansatz, and they correspond to time dependent terms in the io-HEOM. Output properties can be either computed imposing a spectral ansatz for the corresponding cross-correlations or, without such an assumption, by fixing the output observation time.}
\label{fig:schematics}
\end{figure}
\label{sec:Introduction}
We start by defining the main setting for the open quantum systems to be analyzed in this article throughout which, we assume units such that $\hbar=1$.

We consider a system $S$ with Hamiltonian $H_S$, interacting with a bosonic environment $B$ made of non-interacting harmonic modes $b_k$ having energy $\omega_k$ with free Hamiltonian $H_B=\sum_k\omega_k b_k^\dagger b_k$.
In the following we are going to work in the interaction picture so that operators evolve following $H_S$ and $H_B$. In this frame, the dynamics of the states depends on the interaction Hamiltonian describing the coupling between the system and the bath as
\begin{equation}
\label{eq:HI}
    H^{\text{I}}_t=\sum_{q=1}^{N^H_{\text{I}}}{s}^q_t X^q_t\;,
\end{equation}
which is written in terms of $N^H_{\text{I}}$ pairs of generic system operators ${s}^q_t$ and bath operators $X^q_t$, where $X^q_t$ is linear in the operators $b_k$ and $b_k^\dagger$. For concreteness, here we assume that the environmental operators are the ones carrying the dimension of energy. The reduced system dynamics can be computed as
\begin{equation}
\label{eq:rhoS}
\rho_S(t)=\text{Tr}_B\left[\rho(t)\right]\;,
\end{equation}
i.e., by tracing out the environment on the full system-bath density matrix $\rho(t)$ which satisfies the Shr\"{o}dinger equation
\begin{equation}
\label{eq:SE}
\dot{\rho}(t)=-i[H^{\text{I}}_t,\rho(t)]\;.
\end{equation}
Equivalently, the same physical setting can also be described using superoperators, which will constitute the natural formalism for the following analysis. In fact,  since the evolution only depends on the iterative application of the commutator with the interaction Hamiltonian, it is possible to write Eq.~(\ref{eq:HI}) and Eq.~(\ref{eq:SE}) as
\begin{equation}
\label{eq:SE_SO}
\dot{\rho}(t)=-i\mathcal{H}^{\text{I}}_t\;\rho(t)\;,
\end{equation}
in terms of the superoperator version of the interaction Hamiltonian
\begin{equation}
\label{eq:sup_int_H}
\mathcal{H}^{\text{I}}_t=\sum_{\alpha=1}^{N_{\text{I}}}\mathcal{S}^\alpha_t \chi^\alpha_t \;,
\end{equation}
where $\mathcal{S}^\alpha_t$ and $\chi^\alpha_t$ are simply the superoperator versions of the operators  $ s^{q}_t$ and $X^{q}_t$, respectively. Specifically,  since the commutator in Eq.~(\ref{eq:SE}) generates two superoperators for each of the operators in Eq.~(\ref{eq:HI}), the index $\alpha$ labels
$N_{\text{I}}=2N_{\text{I}}^H$ terms, i.e., we can write $\alpha=(p,q)$ to encapsulate both indices, where $p=\text{l},\text{r}$ (left, right) with 
\begin{equation}
\label{eq:leftright}
\begin{array}{llllll}
\chi^{(\text{l},q)}_t &=&X^q_t[\cdot]\;, &\chi^{(\text{r},q)}_t &=&[\cdot]X^q_t\;,\\
\mathcal{S}^{(\text{l},q)}_t &=&s^q_t[\cdot]\;, &\mathcal{S}^{(\text{r},q)}_t &=&-[\cdot]s^q_t\;.
\end{array}
\end{equation}
We note that Eq.~(\ref{eq:sup_int_H}) can also be taken as a starting point for a more abstract dynamics, not necessarily related to an underlying Hamiltonian in Eq.~(\ref{eq:HI}), see, for example, 
\cite{Stockburger_JCP1999,Stockburger_CP2001,PhysRevLett.88.170407,Shao,STOCKBURGER2004159,PhysRevLett.100.230402,Stockburger_2016,PhysRevLett.123.050601,PhysRevA.100.042112,Chernyak1,Chernyak2,PhysRevE.102.062134,Yun-AnShao,PhysRevResearch.6.033237}.
For the simplest case where 
\begin{equation}
\label{eq:sX}
H^I_t=s_t X_t\;,
\end{equation}
 in terms of single Hermitian environmental and system interaction operators $X_t$ and $s_t$,  the index $\alpha$ takes only two values, corresponding to the left and right part of the commutator, i.e., omitting the index $j$ in Eq.~(\ref{eq:leftright}).

In order to make progress in the calculation of the partial trace in Eq.~(\ref{eq:rhoS}), we now assume the initial state of the open system to factorize as $\rho(0)=\rho_S(0)\rho_B$ in terms of a generic system state $\rho_S(0)$ and  a Gaussian environmental state $\rho_B$. Because of the linearity of the environmental coupling operators $X^{q}_t$ in terms of the bosonic degrees of freedom of the bath, this implies that all the effects of the bath on the system dynamics are determined by the two-point correlation matrix (recalling that $\alpha$ now labels both $q$ and the left or right action of the superoperator)
\begin{equation}
\label{eq:Corr}
C^{\alpha_2\alpha_1}_{t_2,t_1}=\text{Tr}_B\left[\chi^{\alpha_2}_{t_2}\chi^{\alpha_1}_{t_1}\rho_B\right]\;.
\end{equation}
Here, we assume that the environmental coupling operators have zero expectation value, which translates to imposing $\text{Tr}_B\left[\chi^{\alpha}_{t}\rho_B\right]=0$, which can always be accomplished by subtracting a possible non-zero expectation from the interaction term while adding a renormalization to the system Hamiltonian.

As a summary of the logic outlined so far, we can start from  Eq.~(\ref{eq:SE_SO}) and simply assume that $\rho_B$ and $\chi_t^\alpha$ are such that Wick's theorem holds.

In this Gaussian setting, we can explicitly trace out the degrees of freedom of the bath and  write the reduced density matrix of the system at time $t$ as
\begin{equation}
\label{eq:F}
    \rho_S(t)=\mathcal{T}_S e^{\mathcal{F}_t}\rho_S(0)\;,
\end{equation}
in terms of an influence superoperator $\mathcal{F}_t$ which can be represented as, see Appendix \ref{sec:app:openquantumsystems},
Here,
\begin{equation}
\label{eq:F_rep}
\begin{array}{lll}
\mathcal{F}_t&=&\displaystyle-\int_0^t d t_2 \sum_{\alpha_2}\mathcal{S}^{\alpha_2}_{t_2} \int_0^{t_2} d t_1 \sum_{\alpha_1}C^{\alpha_2\alpha_1}_{t_2,t_1} \mathcal{S}^{\alpha_1}_{t_1}\\
    &\equiv&\displaystyle\int_0^t d t_2 \sum_{\alpha}\mathcal{A}^{\alpha}_{t_2} \int_0^{t_2} d t_1 \sum_{\beta}D^{\alpha\beta}_{t_2,t_1} \mathcal{B}^{\alpha\beta}_{t_1}\;,
    \end{array}
\end{equation}
where, in order to allow the notation to include different, but equivalent, representations for  $\mathcal{F}_t$, we introduced the superoperators $\mathcal{A}^\alpha$ and  $\mathcal{B}^{\alpha\beta}$ alongside the functions $D^{\alpha\beta}_{t_2,t_1}$.
 Importantly, here the system time-ordering  $\mathcal{T}_S$ acts on the superoperators $\mathcal{S}^{\alpha}_t$ present in the powers of the influence superoperator obtained by expanding the exponential in Eq.~(\ref{eq:F}). In other words, Eq.~(\ref{eq:F}) is a purely formal result, i.e., it does allow a direct  computation of $\rho_S(t)$.  For this reason, different techniques such as the pseudomode model and the HEOM have been proposed to evaluate such an expression. In this context,  the
 different representations for the influence superoperator present in the second line of Eq.~(\ref{eq:F_rep}) can lead to different formulations for the HEOM while still being equivalent in terms of the system dynamics, as we are going to analyze in the next section.

\subsection{Regular HEOM}
\label{sec:HEOM_main_rsd}
In order to derive the HEOM, we follow \cite{Tanimura_3,Tanimura_1,YAN2004216,Ishizaki_1,Zhou_2005,Tanimura_2,Ishizaki_2,PhysRevLett.104.250401,doi:10.1143/JPSJ.81.063301,PhysRevLett.109.266403,PhysRevA.85.062323,Moix,Tanimura_2014,Yan_Shao,Song_Shi,Hsieh_1,Hsieh_2,FreePoles,Dan_Shi,xu2023universalframeworkquantumdissipationminimally,Tanimura_2020,Tanimura_2021,JianMa}, and assume the bath correlations to be written as a sum of complex exponentials which, in the notation introduced in the previous section, corresponds to 
\begin{equation}
\label{eq:D_ansatz}
    D^{\alpha\beta}_{t_2,t_1}\equiv \sum_{k=1}^{N_{\alpha\beta}}a^{\alpha\beta k} e^{-b^{\alpha\beta k}(t_2-t_1)}\;,
\end{equation}
where $a^{\alpha\beta k},b^{\alpha\beta k}\in\mathbb{C}$ and where $N_{\alpha\beta}\in\mathbb{N}$ bounds the number of exponentials for each of the correlations labeled by $\alpha$ and $\beta$. For simplicity, in the following we assume that $N_{\alpha\beta}\equiv N_{\text{exp}}$, a condition which can always be relaxed. We further introduce the multi-index $\sigma\equiv(\sigma_1,\sigma_2,\sigma_3)=(\alpha,\beta,k)$ which takes $\sum_{\alpha\beta} N_{\alpha\beta}\sim N^2_{\text{I}}N_{\text{exp}}$ values. 
To keep the notation compact, we will use the full multi-index $\sigma$ also to indicate partial dependencies, such as $\mathcal{A}^{\alpha}_t= \mathcal{A}^{\sigma_1}_t\rightarrow\mathcal{A}^\sigma_t$ and $\mathcal{B}^{\alpha\beta}_t\rightarrow\mathcal{B}^\sigma_t$.
This formalism allows us to write the influence superoperator in Eq.~(\ref{eq:F_rep}) as
\begin{equation}
\label{eq:F_rep2}
    \mathcal{F}_t=\int_0^t d t_2 \;\sum_{\sigma_1,\sigma_2,\sigma_3} \mathcal{A}^{\sigma_1}_{t_2}  \Theta^{(\sigma_1,\sigma_2,\sigma_3)}_{t_2}=\int_0^t d t_2 \;\sum_{\sigma} \mathcal{A}^{\sigma}_{t_2}  \Theta^{\sigma}_{t_2}\;,
\end{equation}
where
\begin{equation}
\label{eq:defTheta}
    \Theta^{\sigma}_t\equiv a^{\alpha\beta k} \int_0^t d\tau \; e^{-b^{\alpha\beta k}(t-\tau)} \mathcal{B}^{\alpha\beta}(\tau)\;.
\end{equation}
Using the ansatz in Eq.~(\ref{eq:defTheta}), we now look for a differential equation whose solution is the reduced system dynamics $\rho_S(t)$. To do this, we start by taking the time derivative of Eq.~(\ref{eq:F}) to find
\begin{equation}
\label{eq:rhodot}
\begin{array}{lll}
\dot{\rho}_S(t)&=&\displaystyle\mathcal{T}_S\dot{\mathcal{F}}_te^{\mathcal{F}_t}\rho_S(0)= \sum_{\sigma} \mathcal{A}^{\sigma}_t \mathcal{T}_S\Theta^{\sigma}_t e^{\mathcal{F}_t}\rho_S(0)\\
&=&\displaystyle\sum_{\sigma} \mathcal{A}^{\sigma}_t \mathcal{T}_S \Theta^{\sigma}_t \rho_S(t)\;.
\end{array}
\end{equation}
Despite the appearances, this differential equation is still just a formal expression. In fact, the quantities $\Theta_t^{\sigma}$ are a function of the superoperators $\mathcal{B}_\tau^{\alpha\beta}$ for all $\tau\in[0,t]$ which require to be time-ordered. However, the exponential ansatz above implies that
\begin{equation}
\label{eq:temptemp2}
    \dot{\Theta}^{\sigma}_t=a^{\sigma} \mathcal{B}^{\sigma}_t-b^{\sigma} \Theta^{\sigma}_t\;,
\end{equation}
which offers a strategy to compute the reduced system dynamics by iteratively postponing the explicit evaluation of the time-ordering. This can be done by defining the following auxiliary density matrices
\begin{equation}
\label{eq:rhoN}
\rho^{(N)}_{\vec{n}}(t)\equiv \alpha_0^N\mathcal{T}_S\prod_{\sigma}\left[\Theta^{\sigma}_t\right]^{n_\sigma}\rho_S(t)\;,
\end{equation}
where $\vec{n}=\{n_\sigma\}$ is a $N^2_{\text{I}}N_{\text{exp}}$-dimensional vector whose  components $n_\sigma$ track the number of times the superoperator $\Theta^{\sigma}_t$ is applied to $\rho_S(t)$. The total number of times that any of the superoperators $\Theta^{\sigma}_t$ appears in the expression is $N =\sum_{\sigma} n_\sigma$. Here, we also introduced an arbitrary, non-zero parameter $\alpha_0\in\mathbb{C}$.
With these definitions, the derivation of the HEOM in the canonical formalism is simply a matter of taking the derivative of Eq.~(\ref{eq:rhoN}) which, together with Eq.~(\ref{eq:rhodot}) and Eq.~(\ref{eq:temptemp2}), directly gives the regular version of the HEOM as
\begin{equation}
\label{eq:HEOM}
\begin{array}{lll}
\dot{\rho}^{(N)}_{\vec{n}}(t)&=&\displaystyle\text{HEOM}_{0}\left[\rho^{(N)}_{\vec{n}}(t)\right]\\
&\equiv&\displaystyle -\sum_{\sigma} n_\sigma b^{\sigma}\rho^{(N)}_{\vec{n}}+\alpha_0\sum_{\sigma}n_\sigma a^{\sigma}\mathcal{B}^{\sigma}\rho^{(N-1)}_{\vec{n}-\vec{e}{(\sigma)}}\\
&&\displaystyle +\alpha^{-1}_0\sum_{\sigma}\mathcal{A}^{\sigma}\rho^{(N+1)}_{\vec{n}+\vec{e}(\sigma)}\;,
\end{array}
\end{equation}
 where $\vec{e}(\sigma)$ denotes the $N^2_{\text{I}}N_{\text{exp}}$-dimensional vector whose single non-zero entry has value $1$ at position $\sigma$, i.e., $[\vec{e}(\sigma)]_{\sigma_0}=\delta_{\sigma,\sigma_0}$. The reduced density matrix can then be computed as
 \begin{equation}
 \rho_S(t)=\rho^{(N=0)}_{\vec{n}=\vec{0}}(t)\;.
 \end{equation}
The rather unique elegance of the HEOM  is manifested by the fact that the difficulty in computing the time-ordering in the original expression is iteratively postponed while, at the same time, taken advantage of, to relocate the superoperators emerging when taking derivatives of the auxiliary density matrices to ultimately write Eq.~(\ref{eq:HEOM}). In fact, while the time-ordering does not allow, in general, to evaluate the expressions for the auxiliary density matrices in Eq.~(\ref{eq:rhoN}), it allows to factor on the left the extra superoperators appearing when taking derivatives, leading to Eq.~(\ref{eq:HEOM}).

As already implicitly highlighted by the presence of the arbitrary complex parameter $\alpha_0\neq 0$, the HEOM can take different equivalent forms.  These representations are equivalent in the sense that they reproduce the same reduced system dynamics. They can be obtained by considering alternative definitions of the auxiliary density matrices in Eq.~(\ref{eq:rhoN}).
In  Appendix \ref{sec:HEOMrepresentations}, we analyze two such representations for a specific open system. 
Here, we note that it is possible to write a version of the HEOM which is uniform in terms of physical dimensions. In fact,  the auxiliary density matrices defined in  Eq.~(\ref{eq:rhoN}) have different units since  they rely on the action of the operator $\Theta^\sigma_t$ having dimension of frequency (which follows from the assumption that the environmental coupling operators have dimension of energy). It is then possible to follow \cite{Yan_HEOM} to define
\begin{equation}
\rho^{(N)}_{\vec{n}}(t)\rightarrow\left[\prod_{\sigma}(n_\sigma)!(a^\sigma)^{n_\sigma}\right]^{-1/2}\rho^{(N)}_{\vec{n}}(t)\;,
\end{equation}
in terms of an extra factor which can be multiplied on both sides of the HEOM to get
\begin{equation}
\label{eq:HEOM_Yan}
\begin{array}{lll}
\dot{\rho}^{(N)}_{\vec{n}}(t)&=&\displaystyle -\sum_{\sigma} n_\sigma b^{\sigma}\rho^{(N)}_{\vec{n}}+\alpha_0\sum_{\sigma}\sqrt{n^\sigma a^\sigma}\mathcal{B}^{\sigma}\rho^{(N-1)}_{\vec{n}-\vec{e}{(\sigma)}}\\
&&\displaystyle +\alpha^{-1}_0\sum_{\sigma}\sqrt{(n^\sigma+1)a^\sigma}\mathcal{A}^{\sigma}\rho^{(N+1)}_{\vec{n}+\vec{e}(\sigma)}\;.
\end{array}
\end{equation}
In the next section, we are going to consider a generalization of these HEOM to describe, alongside the system dynamics, also environmental properties.

\section{Environmental correlations}
\label{sec:environmentalcorrelations}
The aim of this work is to provide a formalism to allow the preparation of the bath in a non-Gaussian initial state and the computation of bath observables or correlations. Interestingly, these different concepts can all be analyzed through a single formal quantity which represents environmental correlations. Specifically, in this section we consider correlation in the form
\begin{equation}
\label{eq:CO}
\Phi(t_j;t)\equiv\text{Tr}\left[\mathcal{T}\prod_{j=1}^m{{\phi}}_j(t_j)\rho(t)\right]\;,
\end{equation}
where $\rho(t)$ is the full  density matrix defined in Eq.~(\ref{eq:SE}) in the interaction picture with initial condition at $t=0$, so that $t_j\geq 0$. The superoperators ${{\phi}}_j(t_j)$, $j\in M\equiv\{1,\cdots,m\}$ are assumed to be written in terms of $m$ fields whose support is non-trivial only on the environmental Hilbert space. In other words, these superoperators are linear combinations of the environmental bosonic modes acting on either the left or the right. For example,  an environment prepared in a state $b^\dagger_0 \rho_B b_0$ defined in terms of a single environmental mode $b_0$ on an environmental equilibrium state $\rho_B$ could be written using two superoperators $\phi_1=b^\dagger_0[\cdot]$ and $\phi_2=[\cdot]b_0$  acting  on the left and right of their argument, respectively.

As done in the section \ref{sec:Introduction}, we assume the fields have zero expectation value in the free-bath state, i.e., $\text{Tr}_B\left[\phi_j(t_j)\rho_B\right]=0$, for all $j$, which can always be achieved through a redefinition of the fields. Furthermore, we note that, assuming the state $\rho_B$ to be an equilibrium state for the free bath, these conditions can also be written as $\text{Tr}_B\left[\phi_j(0)\rho_B\right]=0$.
The time-dependence of the superoperators $\phi_j(t_j)$ originates from the fact that the underlying operators are written in the interaction picture. However, we note that it is also possible to proceed with  an independent definition to allow such time-dependency to specify an arbitrary order for the evaluation of the underlying operators. In fact, the operator $\mathcal{T}$ is intended to order both the superoperators implicit in the expression for $\rho(t)$ and the superoperators  ${{\phi}}_j(t_j)$. We note that it is also possible to relax the support restrictions for the fields to allow them to have non-trivial components also on the system space, see Appendix \ref{app:a_generalization}. We further present more analysis on this correlation in Appendix \ref{sec:app:environmentalcorrelationsanalysis}.

The trace over the degrees of freedom in the full system+environment space can be decomposed by defining the following ``reduced correlations''
\begin{equation}
\label{eq:redCO}
\rho_S^m(t_j;t) \equiv\text{Tr}_B\left[\mathcal{T}\prod_{j=1}^m{{\phi}}_j(t_j)\rho(t)\right]\;.
\end{equation}
As a consequence,  $\rho_S^m(t)$  contains more information than  Eq.~(\ref{eq:CO}). In fact, while the full correlation can be recovered by taking a trace over the system, i.e.,  $\Phi(t_j;t)=\text{Tr}_S\left[\rho_S^m(t_j;t) \right]$, such an expression also allows to further compute generalized expectation values of system observables. In this way, $\rho_S^m(t)$ can be interpreted as a special case of a process tensor \cite{tempo,PhysRevA.97.012127,PhysRevLett.123.240602,Strathearn_book,PhysRevA.102.062414,10.1063/5.0047260,ace,PhysRevLett.128.167403,PRXQuantum.3.010321,PhysRevB.107.195306,PhysRevB.107.L201115,PhysRevB.107.125103,PhysRevResearch.5.033078,PhysRevX.14.011010,PhysRevLett.132.053602,PhysRevLett.132.200403,cygorek2024}. 
In  Appendix \ref{sec:BathCorrelations},  we show that, by direct combinatorial arguments, these reduced correlation matrices can be expressed, omitting the explicit dependencies on $t_j$, as the following formal series
         \begin{equation}
         \label{eq:rhomSt_main}
   \begin{array}{lll}
\rho^{m}_S(t)  &=&\displaystyle \sum_{k=0}^{[m/2]}\sum_{\mathfrak{a}\in \mathfrak{m}(m,2k)}\left\langle\prod_{j\in \mathfrak{a}}{{\phi}}_j(t_j)\right\rangle(-i)^{(m-2k)}\rho_S^{\mathfrak{a}}(t),
      \end{array}
   \end{equation}
   in terms of the definition  $\langle[\cdot]\rangle\equiv\text{Tr}_B\left\{\mathcal{T}_B[\cdot]\rho_B\right\}$  as a free-bath expectation. While referring to Appendix \ref{sec:BathCorrelations} for more details, here we are now going to define the notation and intuitively justify the different terms present in this expression. In fact, Eq.~(\ref{eq:rhomSt_main}) simply originates from expanding the dynamics of the correlations in Eq.~(\ref{eq:redCO}) and applying Wick's theorem.  In other words, this allows to \emph{reduce} the interacting, multi-point correlation in Eq.~(\ref{eq:redCO}) in terms of a finite series only involving  free-bath  correlations. More specifically, these free correlations are either written in terms of the fields ${{\phi}}_j(t_j)$ only, as explicitly shown in Eq.~(\ref{eq:rhomSt_main}), or written in terms of both the fields and the environmental coupling superoperators $\chi^\alpha_\tau$. The dependency on the latter is implicit in the term $\rho_S^{\mathfrak{a}}(t)$. 
   Before analyzing this term more specifically, we note that  Eq.~(\ref{eq:rhomSt_main}) is a finite series whose terms are labeled by the number $k$ of field-pairs which contribute to pure field-correlations. Since the total number  of fields present in Eq.~(\ref{eq:redCO}) is $m$, the number of possible field-pairs $k$ can take $m/2$ values for even $m$ and $(m-1)/2$ values for odd $m$. To include both cases, we defined  $[m/2]$ as the closest integer to $m/2$ from below,  justifying the first sum in Eq.~(\ref{eq:rhomSt_main}).

To reiterate, the label $k$ is used to describe the choice of all the possible sets $\mathfrak{a}$ made out of $2k$ elements (i.e., $k$ pairs) from $M=\{1,\cdots,m\}$. These elements are used to label the fields involved in the correlation $\left\langle\prod_{j\in \mathfrak{a}}{{\phi}}_j(t_j)\right\rangle$ present in Eq.~(\ref{eq:rhomSt_main}). To finish, it is necessary to sum over all possible $\mathfrak{a}$, whose collection is denoted as $\mathfrak{m}(m,2k)$. In other words, we need to sum over all possible ways to choose  $k$ pairs from the set $\{1,\cdots,m\}$. The set of all the possible sets $\mathfrak{a}$  is the collection  $\mathfrak{m}(m,2k)$. This justifies the second sum in Eq.~(\ref{eq:rhomSt_main}). 
Importantly, all the fields which were not chosen with this procedure, i.e. all the fields whose labels are not in $\mathfrak{a}$, still contribute through cross-correlations with the interaction superoperators $\chi^\alpha_\tau$. This dependence is encoded in the quantities $\rho^{\mathfrak{a}}_S(t)$, which can be explicitly written as
      \begin{equation}
   \label{eq:Sat_main}
   \rho_S^{\mathfrak{a}}(t)=\mathcal{T}_S\prod_{j\in \mathfrak{a}_{\text{c}} } \sum_{\alpha}\int_0^t d\tau \; \langle{{\phi}}_{j}(t_j)\chi^\alpha_\tau\rangle\mathcal{S}^{\alpha}_\tau\rho_S(t)\;,
   \end{equation}
   Here, $\mathfrak{a}_{\text{c}}$ is used to specify all the fields which do not contribute through a field-field correlation. In more formal terms, $\mathfrak{a}_{\text{c}}$ defines the  complement of the set $\mathfrak{a}$ in $M$. It is important to note that, in the expression above, the order in which the superoperators are presented in the cross correlations is irrelevant because of the implicit presence of a time-ordering in the definition of $\langle[\cdot]\rangle$.
   The matrices in Eq.~(\ref{eq:Sat_main}) represent the most challenging part in the evaluation of the original correlation. Under the assumption described in this section, the series in Eq.~(\ref{eq:rhomSt_main}) is exact, i.e.,  equivalent to  Eq.~(\ref{eq:redCO}), and constitutes the basis for the derivations to follow. In fact, despite its formal nature, the  series in Eq.~(\ref{eq:rhomSt_main})  has a specific structure which originates from Wick's theorem and which can be used to explicitly derive an extended version of the HEOM to obtain its solution.

   {In summary, the reduced correlations in Eq.~(\ref{eq:redCO}) can be expanded using Eq.~(\ref{eq:rhomSt_main}) in terms of free correlations for the fields and the matrices given in Eq.~(\ref{eq:Sat_main}). These matrices encode the most challenging part of the calculation and are consequently going to be the main focus for the rest of the article.}

In the next section, where we consider different specific instances of Eq.~(\ref{eq:redCO}) and derive an extended version of the HEOM for the quantities in Eq.~(\ref{eq:Sat_main}) which constitute the main computational challenge.
   
  \section{Input-output HEOM}
  \label{sec:input_output}
   In this section, we  consider the case in which the superoperators $\phi_j(t_j)$ in Eq.~(\ref{eq:redCO}) are used to represent environmental linear operators in the interaction picture evaluated at time $t_j$ and which act either on the left or on the right of their argument. Our goal is to derive a generalized, input-output, version of the HEOM in order to compute the corresponding reduced correlations in Eq.~(\ref{eq:redCO}). 
To introduce more general cases in a gradual manner, here we further restrict the analysis to fields which are evaluated either at the final or the initial time for the dynamics. In other words, we consider either output fields, i.e.,  ${{\phi}}^\text{out}_j(t^\text{out})$, $j=1,\dots,m_\text{out}$, evaluated at the final time for the dynamics $t^{\text{out}}=t$,   and  input fields, i.e., ${{\phi}}^\text{in}_{j'}(t^{\text{in}})$, $j'=1,\dots,m_\text{in}$, evaluated at the initial time $t^{\text{in}}=0$, so that, in the notation used in the previous section, $m=m_\text{out}+m_\text{in}$. 
With these assumptions, the reduced correlations in Eq.~(\ref{eq:redCO}) can be expressed as
\begin{equation}
\label{eq:CO_in_out}
\rho_S^{m_\text{in},m_\text{out}}(t)\equiv\text{Tr}_B\left[\mathcal{T}\prod_{j=1}^{m_\text{out}}{{\phi}}^{\text{out}}_{j}(t)\prod_{j'=1}^{m_\text{in}}{{\phi}}^{\text{in}}_{j'}(0)\;\rho(t)\right]\;.
\end{equation}
In the next subsections, we explicitly derive a HEOM representation for these quantities in the simplest cases involving at most two input and two output field operators. The HEOM for the general case are presented in section \ref{sec:GeneralCase}.
\subsection{Output}
\label{subsec:o}
In this section, we consider the HEOM representation for  reduced correlations involving either a single or a pair of output fields.
\subsubsection{Linear output}
 We start by considering one of the simplest cases in which we are interested in considering a single output environmental field $\phi^{\text{out}}(t)$, so that $m_\text{in}=0$ and $m_\text{out}=1$, in Eq.~(\ref{eq:CO_in_out}). For example, this could correspond to computing the expectation value for the electric field in an electromagnetic environment coupled to a given system. Using $m=m_\text{in}+m_\text{out}=1$,  in Eq.~(\ref{eq:rhomSt_main}) and Eq.~(\ref{eq:CO_in_out}), only a single term in the sum (i.e., $k=0$) contributes and we can write
   \begin{equation}
   \label{eq:CO_1_out_field}
   \begin{array}{lll}
\rho_S^{0,1}(t)&=&\displaystyle\text{Tr}_B\left[\mathcal{T}{{\phi}}^{\text{out}}(t)\;\rho(t)\right]\\
&=&\displaystyle -i \sum_{\alpha}\int_0^t d\tau \; \langle{{\phi^{\text{out}}}}(t)\chi^\alpha_\tau\rangle\left[ \mathcal{T}_S\mathcal{S}^{\alpha}_\tau\rho_S(t)\right]\;.
\end{array}
   \end{equation}
   To derive a HEOM representation for  this quantity, we first assume the following spectral decomposition of the cross-correlations between the field and the environmental interaction operators:
       \begin{equation}
   \label{eq:ansatzOchi_1_out_field}
   \langle{{\phi}}^{\text{out}}(t)\chi^\alpha_\tau\rangle=\sum_{k}c^{\alpha k} e^{-\gamma^{\alpha k}(t-\tau)}\;.
   \end{equation}
In order to take advantage of this ansatz, we further write Eq.~(\ref{eq:CO_1_out_field}) as
   \begin{equation}
   \label{eq:rho01}
  \rho^{0,1}_S(t)=-i\mathcal{T}_S\sum_{\alpha k}\mathcal{Y}^{\text{out};\alpha k}_t\rho_S(t)\;,
  \end{equation}
  in terms of the superoperators
     \begin{equation}
     \label{eq:out_sup}
            \mathcal{Y}^{\text{out};\alpha k}_t=c^{\alpha k}\int_0^td\tau \;e^{-\gamma^{\alpha  k}(t-\tau)}\mathcal{S}^\alpha_\tau\;.
   \end{equation}
 These quantities satisfy the important relation
     \begin{equation}
   \label{eq:Yprop_1_out_field}
       \dot{\mathcal{Y}}^{\text{out};\alpha k}_t=    c^{\alpha  k} \mathcal{S}^{\alpha}_t-\gamma^{\alpha  k}{\mathcal{Y}}^{\text{out};\alpha k}_t\;.
   \end{equation}
 We now reached a key moment in the derivation. In fact, it is now time to generalize the definition in Eq.~(\ref{eq:rhoN}) for the auxiliary density matrices used in the regular HEOM. In fact, the formal definition of these auxiliary density matrices must now also include information about the output field. This can be done by upgrading Eq.~(\ref{eq:rhoN}) to
         \begin{equation}
   \label{eq:LrhoNnt_main_1_out_field}
   \begin{array}{lll}
\rho^{(N^{\text{out}}\!\!,N)}_{\vec{n}^{\text{out}}\!,\vec{n}}(t)&\equiv&   \mathcal{T}_S\prod_{\alpha,k} \left[\mathcal{Y}^{\text{out};\alpha k}_t\right]^{n^{\text{out}}_{\alpha k}}\prod_{\sigma}\left[\Theta^{\sigma}_t\right]^{n_\sigma}\rho_S(t)\;,
\end{array}
   \end{equation}
   where  ${n_\sigma}$ is the one defined in section \ref{sec:HEOM_main_rsd}, and used to derive the regular HEOM, while $n^{\text{out}}_{\alpha k}\in\mathbb{N}$ counts the number of times the superoperator $\mathcal{Y}^{\text{out};\alpha k}_t$ appears in the expression. The vectors $\vec{n}^{\text{out}}$ and $\vec{n}$ indicate a specific set of values for the variables  ${n_\sigma}$ (for all $\sigma$) and $n^{\text{out}}_{\alpha k}$ (for all $\alpha$, and $k$). To keep track of the total number of operators present in the expression, we also defined  $N^{{\text{out}}}=\sum_{\alpha,k}n^{\text{out}}_{\alpha k}$, similarly to the number $N$ in section \ref{sec:HEOM_main_rsd}.
   By taking the time-derivative of this expression, and using Eq.~(\ref{eq:Yprop_1_out_field}), we obtain the following extended HEOM
      \begin{equation}
   \label{eq:HEOM_extended_main_1_out_field}
\begin{array}{lll}\dot{\rho}^{(N^{\text{out}}\!\!,N)}_{\vec{n}^{\text{out}}\!,\vec{n}}&=&\displaystyle \text{HEOM}_{0}\left[{\rho}^{(N^{\text{out}}\!\!,N)}_{\vec{n}^{\text{out}}\!,\vec{n}}\right] -\sum_{\alpha,k} n^{{{\text{out}}}}_{\alpha k} \gamma^{\alpha k}\rho^{(N^{\text{out}}\!\!,N)}_{\vec{n}^{\text{out}}\!,\vec{n}}\\
&&\displaystyle+\sum_{\alpha, k}n^{{{\text{out}}}}_{\alpha k} c^{\alpha k}\mathcal{S}_t^{\alpha}\rho^{(N^{\text{out}}\!\!-1,N)}_{\vec{n}^{\text{out}}\!-\vec{e}{(\alpha,k)},\vec{n}}\;,
\end{array}
\end{equation}
where we omitted the explicit time-dependencies, and where $\vec{e}{(\alpha,k)}$ is a vector in the $(\alpha,k)$-space whose components are $[\vec{e}{(\alpha,k)}]_{\alpha_0,k_0}=\delta_{\alpha,\alpha_0}\delta_{k,k_0}$. The regular part of the HEOM is defined in Eq.~(\ref{eq:HEOM}), and it acts on the indexes $\vec{n}$. The initial condition for this first-order differential equation is
  \begin{equation}
  \label{eq:in_cond_1_out_field}
{\rho}^{(N^{\text{out}}\!\!,N)}_{\vec{n}^{\text{out}}\!,\vec{n}}(0)=\delta_{\vec{n}^{\text{out}}\!,\vec{0}}\;\delta_{\vec{n},\vec{0}}\;\rho_S(0)\;.
\end{equation}
Upon solving Eq.~(\ref{eq:HEOM_extended_main_1_out_field}), the matrix in Eq.~(\ref{eq:CO_1_out_field}) can be obtained using Eq.~(\ref{eq:LrhoNnt_main_1_out_field}) as
\begin{equation}
\rho^{0,1}_S(t)=-i \sum_{\vec{n}^{\text{out}}:N^{{\text{out}}}=1}{\rho}^{(N^{{\text{out}}}=1,N=0)}_{\vec{n}^{\text{out}}\!,\vec{n}=\vec{0}}\;,
\end{equation}
i.e., by summing over all the ancillary density matrices in Eq.~(\ref{eq:LrhoNnt_main_1_out_field}) which do not contain any superoperators $\Theta$, and only a single instance of the superoperators $\mathcal{Y}^\text{out}$ for each of its indexes, as required by Eq.~(\ref{eq:rho01}). This concludes the general mapping to the HEOM to evaluate a single, linear output environmental operator.

In the next section, we consider the case of two output fields. 
\subsubsection{Quadratic output}
\label{sec:quadratic_output}
Here, we analyze the presence of an additional output field with respect to the case in the previous section. Specifically, we consider two output fields $\phi^\text{out}_2(t)$ and $\phi^\text{out}_1(t)$, i.e.,  $m_\text{in}=0$ and $m_\text{out}=2$ in Eq.~(\ref{eq:CO_in_out}). This can be used to compute observables which are quadratic in the bosonic environmental operators, such as energy. In this case, $m=m_\text{in}+m_\text{out}=2$ in Eq.~(\ref{eq:rhomSt_main}), which leads to
     \begin{equation}
   \label{eq:CO_out_quadr_field_0}
\rho^{0,2}_S(t)=\langle\phi_1^{\text{out}}(t)\phi_2^{\text{out}}(t)\rangle\rho_S(t)-\sum_{\alpha,\alpha'}\rho^{\alpha\alpha'}_S(t)\;,
   \end{equation}
   where, omitting the explicit time-dependence,
   \begin{equation}
   \rho^{\alpha\alpha'}_S\equiv\int_0^t d\tau d\tau' \; \langle{{\phi_1^{\text{out}}}}(t)\chi^\alpha_\tau\rangle\langle{{\phi_2^{\text{out}}}}(t)\chi^{\alpha'}_{\tau'}\rangle\mathcal{T}_S\mathcal{S}^{\alpha}_\tau\mathcal{S}^{\alpha'}_{\tau'}\rho_S(t)\;.
   \end{equation}
   Following the arguments provided in the previous sections, we impose the ansatz in Eq.~(\ref{eq:ansatzOchi_1_out_field}) for both output fields, i.e., 
      \begin{equation}
      \label{eq:spectr_ansatz}
      \langle{{\phi}}_j^{\text{out}}(t)\chi^\alpha_\tau\rangle=\sum_{k}c^{j \alpha k} e^{-\gamma^{j \alpha k}(t-\tau)}\;,
  \end{equation}
  for $j=1,2$, and which leads to the following superoperators corresponding to  Eq.~(\ref{eq:out_sup}), written as
       \begin{equation}
     \label{eq:out_sup_quadr}
            \mathcal{Y}^{\text{out};j\alpha k}_t=c^{j\alpha k}\int_0^td\tau \;e^{-\gamma^{j\alpha k}(t-\tau)}\mathcal{S}^\alpha_\tau\;.
   \end{equation}
   and which satisfies
        \begin{equation}
   \label{eq:Yprop_1_out_field_quadr}
       \dot{\mathcal{Y}}^{\text{out};j\alpha k}_t=    c^{j\alpha  k} \mathcal{S}^{\alpha}_t-\gamma^{j\alpha  k}{\mathcal{Y}}^{\text{out};j\alpha k}_t\;.
   \end{equation}
 With this notation, we can write
   \begin{equation}
  \rho^{\alpha\alpha'}_S(t)=\mathcal{T}_S\sum_{k,k'}\mathcal{Y}^{\text{out};1\alpha k}_t\mathcal{Y}^{\text{out};2\alpha'k'}_t\rho_S(t)\;.
  \end{equation}
  As in the previous case, we now need to generalize the definition of the auxiliary density matrices in Eq.~(\ref{eq:rhoN})  in order to include the  information about the output fields present in the correlation. This can be accomplished by defining
         \begin{equation}
   \label{eq:LrhoNnt_main_in_out_field_quadr}
\rho^{(N^{\text{out}}\!\!,  N)}_{\vec{n}^{\text{out}}\!,\vec{n}}\equiv \mathcal{T}_S\prod_{j,\alpha,k} \left[\mathcal{Y}^{\text{out};j\alpha k}_t\right]^{n^{\text{out}}_{j\alpha k}}\prod_{\sigma}\left[\Theta^{\sigma}_t\right]^{n_\sigma}\rho_S.
   \end{equation}
   Analogously to the previous case, we introduced    $n^{\text{out}}_{j \alpha k}\in\mathbb{N}$, to count the appearances of the superoperator $\mathcal{Y}^{\text{out};j \alpha k}_t$. The vector $\vec{n}^{\text{out}}$ represents the values of $n^{\text{out}}_{j \alpha k}$ in vector form. Furthermore, we have $N^{{\text{out}}}=\sum_{j,\alpha,k}n^{\text{out}}_{j\alpha k}$.
   The time-derivative of the matrices in Eq.~(\ref{eq:LrhoNnt_main_in_out_field_quadr})
  can be analyzed using the properties in Eq.~(\ref{eq:Yprop_1_out_field_quadr}) to obtain
        \begin{equation}
   \label{eq:HEOM_extended_main_in_out_field_quadr}
\begin{array}{l}\dot{\rho}^{(N^{\text{out}}\!\!,  N)}_{\vec{n}^{\text{out}}\!,\vec{n}}=\displaystyle \text{HEOM}_{0}\!\!\left[\rho^{(N^{\text{out}}\!\!, N)}_{\vec{n}^{\text{out}}\!,\vec{n}}\right] \\
\displaystyle+\sum_{j,\alpha,k} n^{{{\text{out}}}}_{j\alpha k}\left[c^{j\alpha k}\mathcal{S}_t^{\alpha}\rho^{(N^{{\text{out}}}\!-1,N)}_{\vec{n}^{{{\text{out}}}}-\vec{e}{(j,\alpha,k)},\vec{n}}-\gamma^{j \alpha k}\rho^{(N^{\text{out}}\!\!,  N)}_{\vec{n}^{\text{out}}\!,\vec{n}}\right],\\
\end{array}
\end{equation}
where $[\vec{e}{(j,\alpha,k)}]_{j_0,\alpha_0,k_0}=\delta_{j,j_0}\delta_{\alpha,\alpha_0}\delta_{k,k_0}$, and whose initial condition is
  \begin{equation}
  \label{eq:in_cond_in_out_field}
{\rho}^{(N^{\text{out}}\!\!,  N)}_{\vec{n}^{\text{out}}\!,\vec{n}}(0)=\delta_{\vec{n}^{{{\text{out}}}},\vec{0}}\;\delta_{\vec{n},\vec{0}}\;\rho_S(0)\;.
\end{equation}
Upon solving Eq.~(\ref{eq:HEOM_extended_main_in_out_field_quadr}), the matrix in Eq.~(\ref{eq:CO_out_quadr_field_0}) can be written explicitly as
\begin{equation}
\renewcommand{\arraystretch}{1.9}{
\begin{array}{lll}
\rho^{0,2}_S(t)&=&\displaystyle\langle\phi_2^{\text{out}}(t)\phi_1^{\text{out}}(t)\rangle{\rho}^{(N^{\text{out}}\!=0, N=0)}_{\vec{n}^{\text{out}}=0,\vec{n}=\vec{0}}\\
&&\displaystyle - \sum_{\vec{n}^{\text{out}}:\{N^{\text{out}}_1=N^{\text{out}}_2=1\}}{\rho}^{(N^{\text{out}}\!=2, N=0)}_{\vec{n}^{\text{out}}\!,\vec{n}=\vec{0}}\;,
\end{array}}
\end{equation}
where $N^{\text{out}}_j=\sum_{\alpha,k} n^\text{out}_{j\alpha k}$, for $j=1,2$.

This concludes this section for the extended HEOM written in terms of output fields. In the next section, we are going to analyze input fields.

\subsection{Input}
\label{subsec:i}
In this section, we consider the case of an environment initially prepared in a state which can be written by the action of a single or a pair of input fields. Interestingly, the corresponding HEOM are going to have qualitatively different features with respect to the output ones considered in the previous sections. In fact, here no ansatz for the cross-correlations between the input  fields and the environmental interaction operators is required. On one hand, this is going to reduce the range of the extra-indexes in the corresponding HEOM. On the other hand, it will also lead to a time-dependent term in the differential equation.
\subsubsection{Linear Input}
For clarity of exposition, we start with the simplest input case  involving a single input environmental field $\phi^\text{in}(t^\text{in})$, i.e., $m_\text{out}=0$ and $m_\text{in}=1$ in Eq.~(\ref{eq:CO_in_out}).  Importantly, as assumed at the beginning of this section, the input fields are all evaluated at the initial time for the dynamics, i.e., $t^\text{in}=0$. 
  As for the output case, we can use $m=m_\text{in}+m_\text{out}=1$ in Eq.~(\ref{eq:rhomSt_main}) to write
   \begin{equation}
   \label{eq:CO_1_in_field}
\rho^{1,0}_S(t)= -i \sum_{\alpha}\int_0^t d\tau \; \langle{{\phi^{\text{in}}}}(0)\chi^\alpha_\tau\rangle\left[ \mathcal{T}_S\mathcal{S}^{\alpha}_\tau\rho_S(t)\right]\;.
   \end{equation}
   The key difference with respect to the corresponding equation for the output correlation, i.e., Eq.~(\ref{eq:CO_1_out_field}), is that the correlation inside the integral  in Eq.~(\ref{eq:CO_1_in_field}), does not depend on the final time $t$. In fact, we are here simply considering the effect of a field operator applied at the initial time. This is going to 
 affect the resulting HEOM in two ways. First, in this case, there is no need for the usual spectral ansatz for the cross correlation between the input field and the interaction operator since their direct time-derivative is never taken. Second, since the correlation does \emph{not} depend on $(t-\tau)$, evaluation at $\tau\rightarrow t$ will not lead to a time-independent term. In other words, the corresponding term in the HEOM will be time-dependent. To see these points explicitly, we can directly define
     \begin{equation}
     \label{eq:in_sup}
            \mathcal{Y}^{\text{in}}_t= \sum_\alpha\int_0^td\tau\;\langle{\phi}^{\text{in}}(0)\chi^{\alpha}_{\tau}\rangle\mathcal{S}^\alpha_\tau\;.
   \end{equation}
whose time-derivative satisfies
     \begin{equation}
   \label{eq:Yprop_1_in_field}
       \dot{\mathcal{Y}}^{\text{in}}_t=    \sum_\alpha\langle{\phi}^{\text{in}}(0)\chi^{\alpha}_{t}\rangle \mathcal{S}^{\alpha}_t\equiv \mathcal{G}_t\;.
   \end{equation}
As we did for the output fields, we can now define the auxiliary density matrix to include information about the input field as
         \begin{equation}
   \label{eq:LrhoNnt_main_1_in_field}
   \begin{array}{lll}
\rho^{(N^{\text{in}}\!\!,N)}_{n^{\text{in}}\!,\vec{n}}(t)&\equiv&   \mathcal{T}_S\left[\mathcal{Y}^{\text{in}}_t\right]^{n^{\text{in}}}\prod_{\sigma}\left[\Theta^{\sigma}_t\right]^{n_\sigma}\rho_S(t)\;,
\end{array}
   \end{equation}
   where  $N^{{\text{in}}}, n^{\text{in}}=0,1$.
   By taking the time derivative of this expression, and using Eq.~(\ref{eq:Yprop_1_in_field}), we obtain the following extended HEOM
      \begin{equation}
   \label{eq:HEOM_extended_main_1_in_field}
\begin{array}{l}\dot{\rho}^{(N^{\text{in}}\!\!,N)}_{n^{\text{in}}\!,\vec{n}}=\displaystyle \text{HEOM}_{0}\left[\rho^{(N^{\text{in}}\!\!,N)}_{n^{\text{in}}\!,\vec{n}}\right]+n^{\text{in}}\mathcal{G}_t\rho^{(N^{{\text{in}}}\!-1,N)}_{{n}^{{{\text{in}}}}\!-1,\vec{n}}\;,
\end{array}
\end{equation}
The initial condition for this first-order differential equation is
  \begin{equation}
  \label{eq:in_cond_1_in_field}
{\rho}^{(N^{{\text{in}}}\!\!,N)}_{n^{\text{in}},\vec{n}}(0)=\delta_{n^{\text{in}}\!,0}\;\delta_{\vec{n},\vec{0}}\;\rho_S(0)\;.
\end{equation}
Upon solving Eq.~(\ref{eq:HEOM_extended_main_1_in_field}), the matrix in Eq.~(\ref{eq:CO_1_in_field}) can be written as
\begin{equation}
\rho^{1,0}_S(t)= -i\;{\rho}^{(N^{{\text{in}}}=1,N=0)}_{n^{\text{in}}=1,\vec{n}=\vec{0}}\;.
\end{equation}
As mentioned, the extra-indexes introduced in the HEOM in Eq.~(\ref{eq:HEOM_extended_main_1_in_field}) have minimal range at the cost of an extra time-dependence in the term $\mathcal{G}_t$.

In the next section, we are going to analyze the case where a pair of input fields are present. 

\subsubsection{Quadratic Input}
\label{sec:quadraticInput}
  Here, we consider the case in which the initial state of the environment can be modeled in terms of  two input environmental fields $\phi_1^\text{in}(t^\text{in})$, $\phi_2^\text{in}(t^\text{in})$, i.e., $m_\text{out}=0$ and $m_\text{in}=2$ in Eq.~(\ref{eq:CO_in_out}). For example, this could correspond to preparing the bath with an additional single photon wave-packet (so that two fields are present, corresponding to the action on the left and right of the initial thermal state).  Importantly, as assumed at the beginning of this section, the input fields are all evaluated at the initial time for the dynamics, i.e., $t^\text{in}=0$. 
We can now use $m=m_\text{in}+m_\text{out}=2$ in Eq.~(\ref{eq:CO_in_out}) to write
       \begin{equation}
   \label{eq:CO_1_in_field_quadr}
\rho^{2,0}_S(t)=\langle\phi_1^{\text{in}}(0)\phi_2^{\text{in}}(0)\rangle\rho_S(t)-\sum_{\alpha,\alpha'}\rho^{\alpha\alpha'}_S(t)\;,
   \end{equation}
      where, omitting the explicit time-dependence,
   \begin{equation}
   \rho^{\alpha\alpha'}_S\equiv\int_0^t d\tau d\tau' \; \langle{{\phi_1^{\text{in}}}}(0)\chi^\alpha_\tau\rangle\langle{{\phi_2^{\text{in}}}}(0)\chi^{\alpha'}_{\tau'}\rangle\mathcal{T}_S\mathcal{S}^{\alpha}_\tau\mathcal{S}^{\alpha'}_{\tau'}\rho_S(t)\;.
   \end{equation}
   As for the single input-field  case, the only difference with the case in section \ref{sec:quadratic_output}, lies in the independence of the integrand on the final time.
To find the HEOM, we can directly define
     \begin{equation}
     \label{eq:in_sup_quadr}
            \mathcal{Y}^{\text{in};j}_t= \sum_\alpha\int_0^td\tau\;\langle{\phi}_j^{\text{in}}(0)\chi^{\alpha}_{\tau}\rangle\mathcal{S}^\alpha_\tau\;.
   \end{equation}
   for $j=1,2$, whose time-derivative satisfies
     \begin{equation}
   \label{eq:Yprop_1_in_field_quadr}
       \dot{\mathcal{Y}}^{\text{in};j}_t=    \sum_\alpha\langle{\phi}_j^{\text{in}}(0)\chi^{\alpha}_{t}\rangle \mathcal{S}^{\alpha}_t\equiv \mathcal{G}^j_t\;.
   \end{equation}
As commented before, it is worth noting that, in the expression above, the cross correlation contains a time-ordering acting at the superoperator level. In other words, the order in which ${\phi}_j^{\text{in}}(0)$ and $\chi^{\alpha}_{t}$ are written in the equation above is irrelevant as, in this input case, the field operator is always evaluated first because of the implicit time ordering.
   
   We can now define the auxiliary density matrices as
         \begin{equation}
   \label{eq:LrhoNnt_main_1_in_field_quadr}
   \begin{array}{lll}
\rho^{(N^{\text{in}}\!\!,N)}_{\vec{n}^{\text{in}}\!,\vec{n}}(t)&\equiv&   \mathcal{T}_S\prod_j\left[\mathcal{Y}^{\text{in};j}_t\right]^{n_j^{\text{in}}}\prod_{\sigma}\left[\Theta^{\sigma}_t\right]^{n_\sigma}\rho_S(t)\;,
\end{array}
   \end{equation}
   where  $n_j^{\text{in}}=0,1$,  $N^{{\text{in}}}=\sum_jn_j^{\text{in}}$ and where $\vec{n}^{\text{in}}=(n_1^{\text{in}},n_2^{\text{in}})$.
   By taking the time derivative of this expression, and using Eq.~(\ref{eq:Yprop_1_in_field_quadr}), we obtain the following extended HEOM
      \begin{equation}
   \label{eq:HEOM_extended_main_1_in_field_quadr}
\begin{array}{l}\dot{\rho}^{(N^{\text{in}}\!\!,N)}_{\vec{n}^{\text{in}}\!,\vec{n}}=\displaystyle \text{HEOM}_{0}\left[\rho^{(N^{\text{in}}\!\!,N)}_{\vec{n}^{\text{in}}\!,\vec{n}}\right]+\sum_j n_j^{\text{in}}\mathcal{G}^j_t\rho^{(N^{{\text{in}}}\!-1,N)}_{\vec{n}^{{{\text{in}}}}\!-\vec{e}(j),\vec{n}}\;,
\end{array}
\end{equation}
where $[\vec{e}(j)]_{j_0}=\delta_{j,j_0}$ and whose initial condition is
  \begin{equation}
  \label{eq:in_cond_1_in_field_quadr}
{\rho}^{(N^{{\text{in}}}\!\!,N)}_{\vec{n}^{\text{in}},\vec{n}}(0)=\delta_{\vec{n}^{\text{in}}\!,0}\delta_{\vec{n},\vec{0}}\;\rho_S(0)\;.
\end{equation}
Upon solving Eq.~(\ref{eq:HEOM_extended_main_1_in_field_quadr}), the matrix in Eq.~(\ref{eq:CO_1_in_field_quadr}) can be written as
\begin{equation}
\label{eq:rho20}
\rho^{2,0}_S(t)=\langle\phi_1^{\text{in}}(0)\phi_2^{\text{in}}(0)\rangle{\rho}^{(N^{{\text{in}}}=0,N=0)}_{\vec{n}^{\text{in}}=\vec{0},\vec{n}=\vec{0}}- {\rho}^{(N^{{\text{in}}}=2,N=0)}_{\vec{n}^{\text{in}}=(1,1),\vec{n}=\vec{0}}\;.
\end{equation}
It is interesting to further increase in specificity and write the HEOM considered in this section in a pure-dephasing limit where the system coupling operators commute with the free system dynamics.  We do this in Appendix \ref{sec:PureDephasing}. In parallel, we also refer to Appendix \ref{sec:app:ioHEOM_ex} for an exemplification in terms of a specific open quantum system characterized by a single pair of system and bath coupling operators.

 In the next section, we are going to consider  the presence of both input and output fields.
\subsection{Input-output}
\label{subsec:io}
 In this section, we join the results of the two previous sections together to consider two input-output cases, corresponding to input and output operators which are either linear or quadratic in the environmental modes.
 \subsubsection{Linear Case}
 We start with  the simplest input-output example, corresponding to a  single output  field and a single input field, i.e., $m_\text{in}=1$ and $m_\text{out}=1$ in Eq.~(\ref{eq:CO_in_out}) so that $m=m_\text{out}+m_\text{in}=2$. 
More specifically, we impose $m_\text{in}=1$ and $m_\text{out}=1$ in  Eq.~(\ref{eq:CO_in_out}) to write
       \begin{equation}
   \label{eq:CO_in_out_field_0}
\rho^{1,1}_S(t)=\langle\phi^{\text{out}}(t)\phi^{\text{in}}(0)\rangle\rho_S(t)-\sum_{\alpha,\alpha'}\rho^{\alpha\alpha'}_S(t)\;,
   \end{equation}
where
   \begin{equation}
   \label{eq:CO_in_out_field}
   \begin{array}{lll}
\rho^{\alpha\alpha'}_S&=&\displaystyle\int_0^t d\tau d\tau' \; \langle{{\phi^{\text{out}}}}(t)\chi^\alpha_\tau\rangle\langle{{\phi^{\text{in}}}}(0)\chi^{\alpha'}_{\tau'}\rangle \mathcal{T}_S\mathcal{S}^{\alpha}_\tau\mathcal{S}^{\alpha'}_{\tau'}\rho_S(t)\;.
\end{array}
   \end{equation}
   Following the arguments provided in the previous sections, we impose the ansatz in Eq.~(\ref{eq:ansatzOchi_1_out_field}) for the definitions in Eq.~(\ref{eq:out_sup}) and Eq.~(\ref{eq:in_sup}) to write
   \begin{equation}
  \rho^{1,1}_S(t)=\langle\phi^{\text{out}}(t)\phi^{\text{in}}(0)\rangle\rho_S(t)+\mathcal{T}_S\sum_{\alpha,k}\mathcal{Y}^{\text{out};\alpha k}_t\mathcal{Y}^{\text{in}}_t\rho_S(t)\;.
  \end{equation}
  To compute this quantity, we need to combine the information about both input and output fields in the definition of the auxiliary density matrices. To do this, we define
         \begin{equation}
   \label{eq:LrhoNnt_main_in_out_field}
\rho^{(N^{\text{out}}\!\!, N^{\text{in}}\!\!, N)}_{\vec{n}^{\text{out}}\!,n^\text{in}\!,\vec{n}}\equiv \mathcal{T}_S\prod_{\alpha, k} \left[\mathcal{Y}^{\text{out};\alpha k}_t\right]^{n^{\text{out}}_{\alpha k}}\left[\mathcal{Y}^{\text{in}}_t\right]^{n^{\text{in}}}\prod_{\sigma}\left[\Theta^{\sigma}_t\right]^{n_\sigma}\rho_S,
   \end{equation}
whose time derivative can be analyzed using the properties in Eq.~(\ref{eq:Yprop_1_out_field}) and Eq.~(\ref{eq:Yprop_1_in_field}) to obtain
        \begin{equation}
   \label{eq:HEOM_extended_main_in_out_field}
   \renewcommand{\arraystretch}{1.9}{
\begin{array}{l}\displaystyle\dot{\rho}^{(N^{\text{out}}\!\!, N^{\text{in}}\!\!, N)}_{\vec{n}^{\text{out}}\!,n^\text{in}\!,\vec{n}}\!\!=\!\displaystyle \text{HEOM}_{0}\!\!\left[\rho^{(N^{\text{out}}\!\!, N^{\text{in}}\!\!, N)}_{\vec{n}^{\text{out}}\!,n^\text{in}\!,\vec{n}}\right]\!\!+\!n^{\text{in}}\mathcal{G}_t\rho^{(N^{{\text{out}}}\!\!,N^{{\text{in}}}\!-1,N)}_{\vec{n}^{{{\text{out}}}}\!,{n}^{{{\text{in}}}}\!-1,\vec{n}} \\
\displaystyle+\sum_{\alpha,k} n^{{{\text{out}}}}_{\alpha k}\left[c^{\alpha k}\mathcal{S}_t^{\alpha}\rho^{(N^{{\text{out}}}\!-1,n^\text{in}\!\!,N)}_{\vec{n}^{{{\text{out}}}}-\vec{e}{(\alpha,k)},n^\text{in}\!,\vec{n}}- \gamma^{\alpha k}\rho^{(N^{\text{out}}\!\!, N^{\text{in}}\!\!, N)}_{\vec{n}^{\text{out}}\!,n^\text{in}\!,\vec{n}}\right].\\
\end{array}}
\end{equation}
with initial condition
  \begin{equation}
  \label{eq:in_cond_in_out_field}
{\rho}^{(N^{\text{out}}\!\!, N^{\text{in}}\!\!, N)}_{\vec{n}^{\text{out}}\!,n^\text{in}\!,\vec{n}}(0)=\delta_{{n^\text{in}},{0}}\delta_{\vec{n}^{{{\text{out}}}},\vec{0}}\delta_{\vec{n},\vec{0}}\;\rho_S(0)\;.
\end{equation}
Upon solving Eq.~(\ref{eq:in_cond_in_out_field}), the matrix in Eq.~(\ref{eq:CO_in_out_field_0}) can be written as
\begin{equation}
   \renewcommand{\arraystretch}{1.9}{
\begin{array}{lll}
\rho^{1,1}_S(t)&=&\displaystyle\langle\phi_2^{\text{out}}(t)\phi_1^{\text{in}}(0)\rangle{\rho}^{(N^{\text{out}}\!=0, N^{\text{in}}\!=0, N=0)}_{\vec{n}^{\text{out}}=\vec{0},n^\text{in}=0,\vec{n}=\vec{0}}\\
&&\displaystyle - \sum_{\vec{n}^{\text{out}}:N^{{\text{out}}}=1}{\rho}^{(N^{\text{out}}\!=1, N^{\text{in}}\!=1, N=0)}_{\vec{n}^{\text{out}}\!,n^\text{in}=1,\vec{n}=\vec{0}}\;.
\end{array}}
\end{equation}
where $N^\text{out}=\sum_{\alpha,k}n^\text{out}_{\alpha k}$ so that the last sum is over all terms labeled by an index $\vec{n}^\text{out}$ having only one non-zero element, which is equal to $1$.

In the next section, we analyze the case where pairs of fields are used to define both the input and the output.
\subsubsection{Quadratic Case}
\label{sec:quadratic_case}
We now consider the case where the output consists of two  fields $\phi_j^{\text{out}}(t)$, $j=1,2$, and, similarly, the input consists of two fields $\phi_j^{\text{in}}(0)$, $j=1,2$, so that $m_\text{in}=2$ and $m_\text{out}=2$ in Eq.~(\ref{eq:CO_in_out}) and $m=m_\text{out}+m_\text{in}=4$. This can be used to model a situation in which a single photon  wave-packet is injected into a thermal bath and observables quadratic in the bosonic environmental fields are measured at a final time. In this setting, we want to compute
\begin{equation}
\label{eq:rho22}
\rho_S^{2,2}(t)=\text{Tr}_B\left[\mathcal{T}\phi_2^{\text{out}}(t)\phi_1^{\text{out}}(t)\phi_2^{\text{in}}(0)\phi_1^{\text{in}}(0)\rho(t)\right]\;,
\end{equation}
To do this, we can use Eq.~(\ref{eq:rhomSt_main}) and the notation defined in the previous sections to write
      \begin{equation}
   \label{eq:CO_in_out_field_0_quadratic}
   \begin{array}{lll}
\rho^{2,2}_S(t)=\left\langle\phi_1^{\text{out}}(t)\phi_2^{\text{out}}(t)\phi_1^{\text{in}}(0)\phi_2^{\text{in}}(0)\right\rangle\rho_S(t)\\

-\displaystyle\mathcal{T}_S\left[\left\langle {\phi}^{\text{out}}_{1}(t){\phi}^{\text{out}}_{2}(t)\right\rangle \mathcal{Y}^{\text{in};1}_t\mathcal{Y}^{\text{in};2}_t\right.\\

\displaystyle\phantom{\mathcal{T}_S[~~}+\left\langle {\phi}^{\text{out}}_{1}(t){\phi}^{\text{in}}_{1}(0)\right\rangle  \sum_{\alpha,k}\mathcal{Y}^{\text{out};2 \alpha k}_t\mathcal{Y}^{\text{in};2}_t\\

\displaystyle\phantom{\mathcal{T}_S[~~}+\left\langle {\phi}^{\text{out}}_{1}(t){\phi}^{\text{in}}_{2}(0)\right\rangle \sum_{\alpha,k}\mathcal{Y}^{\text{out};2\alpha k}_t\mathcal{Y}^{\text{in};1}_t\\

\displaystyle\phantom{\mathcal{T}_S[~~}+\left\langle {\phi}^{\text{out}}_{2}(t){\phi}^{\text{in}}_{1}(0)\right\rangle  \sum_{\alpha,k}\mathcal{Y}^{\text{out};1\alpha k}_t\mathcal{Y}^{\text{in};2}_t\\

\displaystyle\phantom{\mathcal{T}_S[~~}+\left\langle {\phi}^{\text{out}}_{2}(t){\phi}^{\text{in}}_{2}(0)\right\rangle\sum_{\alpha,k}\mathcal{Y}^{\text{out};1 \alpha k}_t\mathcal{Y}^{\text{in};1}_t\\

\displaystyle\phantom{\mathcal{T}_S[~~}+\left\langle {\phi}^{\text{in}}_{1}(0){\phi}^{\text{in}}_{2}(0)\right\rangle  \sum_{\alpha,k}\sum_{\alpha',k'}\mathcal{Y}^{\text{out};1\alpha k}_t\mathcal{Y}^{\text{out};2\alpha'k'}_t\\

\phantom{\mathcal{T}_S[~~}-\displaystyle\left.  \mathcal{Y}^{\text{in};1}_t\mathcal{Y}^{\text{in};2}_t\sum_{\alpha,k}\sum_{\alpha',k'}\mathcal{Y}^{\text{out};1 \alpha k}_t\mathcal{Y}^{\text{out};2\alpha'k'}_t\right]\rho_S(t)\;,

\end{array}
   \end{equation}
   where, explicitly, we used the definitions in Eq.~(\ref{eq:out_sup_quadr}) and Eq.~(\ref{eq:in_sup_quadr}), and the ansatz in Eq.~(\ref{eq:spectr_ansatz}) for the output fields, i.e.,
   \begin{equation}
   \label{eq:ans_oo}
      \langle{{\phi}}_j^{\text{out}}(t)\chi^\alpha_\tau\rangle=\sum_{k}c^{j\alpha k} e^{-\gamma^{j\alpha k}(t-\tau)}\;,
  \end{equation}
  for $j=1,2$. The expansion in Eq.~(\ref{eq:CO_in_out_field_0_quadratic}) let us appreciate the amount of information packed in the compact Eq.~(\ref{eq:rhomSt_main}). Importantly, the quantities  in Eq.~(\ref{eq:CO_in_out_field_0_quadratic}) can be written as
\begin{equation}
   \renewcommand{\arraystretch}{1.9}{
\begin{array}{lll}
\mathcal{T}_S \mathcal{Y}^{\text{in};1}_t\mathcal{Y}^{\text{in};2}_t\rho_S(t)&=&\displaystyle\rho^{(0, 2,0)}_{\vec{0},\vec{0},1,1,\vec{0}}\\
\displaystyle\mathcal{T}_S\mathcal{Y}^{\text{out};1\alpha k}_t\mathcal{Y}^{\text{in};1}_t\rho_S(t)&=&\displaystyle\rho^{(1, 1,0)}_{\vec{e}(\alpha,k),\vec{0},1,0,\vec{0}}\\
\displaystyle\mathcal{T}_S\mathcal{Y}^{\text{out};1\alpha k}_t\mathcal{Y}^{\text{in};2}_t\rho_S(t)&=&\displaystyle\rho^{(1, 1,0)}_{\vec{e}(\alpha,k),\vec{0},0,1,\vec{0}}\\
\displaystyle\mathcal{T}_S\mathcal{Y}^{\text{out};2\alpha k}_t\mathcal{Y}^{\text{in};1}_t\rho_S(t)&=&\displaystyle\rho^{(1, 1,0)}_{\vec{0},\vec{e}(\alpha,k),1,0,\vec{0}}\\
\displaystyle\mathcal{T}_S\mathcal{Y}^{\text{out};2\alpha k}_t\mathcal{Y}^{\text{in};2}_t\rho_S(t)&=&\displaystyle\rho^{(1, 1,0)}_{\vec{0},\vec{e}(\alpha,k),0,1,\vec{0}}\\
\displaystyle\mathcal{T}_S\mathcal{Y}^{\text{out};1\alpha k}_t\mathcal{Y}^{\text{out};2\alpha' k'}_t\rho_S(t)&=&\displaystyle\rho^{(2, 0,0)}_{\vec{e}(\alpha,k),\vec{e}(\alpha'\!,k'),0,0,\vec{0}}\;,
 \end{array}}
\end{equation}
 in terms of the auxiliary density matrices
            \begin{equation}
   \label{eq:LrhoNnt_main_in_out_field_quadratic}
   \begin{array}{lll}
\rho^{(N^{\text{out}}\!\!, N^{\text{in}}\!\!, N)}_{\vec{n}_1^{\text{out}}\!\!,\vec{n}_2^{\text{out}}\!\!,n_1^\text{in}\!\!,n_2^\text{in}\!\!,\vec{n}}\!\!\!\!&\equiv&\!\!\! \displaystyle\mathcal{T}_S\!\!\!\!\prod_{\alpha\alpha'\!kk'}
\! \! \left[\mathcal{Y}^{\text{out};1\alpha'k'}_t\right]^{n^{\text{out}}_{1\alpha'k'}}\!\! \left[\mathcal{Y}^{\text{out};2\alpha k}_t\right]^{n^{\text{out}}_{2\alpha k}}\\
&\times&\displaystyle\left[\mathcal{Y}^{\text{in};1}_t\right]^{n_1^{\text{in}}}\left[\mathcal{Y}^{\text{in};2}_t\right]^{n_2^{\text{in}}}\prod_{\sigma}\left[\Theta^{\sigma}_t\right]^{n_\sigma}\rho_S.
\end{array}
   \end{equation} 
Following the same arguments as in the sections above, these matrices satisfy the extended HEOM
           \begin{equation}
   \label{eq:HEOM_extended_main_in_out_field_quadratic}
   \renewcommand{\arraystretch}{1.9}{
\begin{array}{l}\dot{\rho}^{(N^{\text{out}}\!\!, N^{\text{in}}\!\!, N)}_{\vec{n}_1^{\text{out}}\!,\vec{n}_2^{\text{out}}\!,n_1^\text{in}\!,n_2^\text{in}\!,\vec{n}}=\displaystyle \text{HEOM}_{0}\left[\rho^{(N^{\text{out}}\!\!, N^{\text{in}}\!\!, N)}_{\vec{n}_1^{\text{out}}\!,\vec{n}_2^{\text{out}}\!,n_1^\text{in}\!,n_2^\text{in}\!,\vec{n}}\right]\\

+n_1^{\text{in}}\mathcal{G}^1_{t}\rho^{(N^{\text{out}}\!\!, N^{\text{in}}-1, N)}_{\vec{n}_1^{\text{out}}\!,\vec{n}_2^{\text{out}}\!,n_1^\text{in}-1\!,n_2^\text{in}\!,\vec{n}} +n_2^{\text{in}}\mathcal{G}^2_{t}\rho^{(N^{\text{out}}\!\!, N^{\text{in}}-1, N)}_{\vec{n}_1^{\text{out}}\!,\vec{n}_2^{\text{out}}\!,n_1^\text{in}\!,n_2^\text{in}-1\!,\vec{n}}\\

-\displaystyle\sum_{\alpha,k} \left[n^{{{\text{out}}}}_{1\alpha k} \gamma^{1\alpha k}+n^{{{\text{out}}}}_{2\alpha k} \gamma^{2\alpha k}\right]{\rho}^{(N^{\text{out}}\!\!, N^{\text{in}}\!\!, N)}_{\vec{n}_1^{\text{out}}\!,\vec{n}_2^{\text{out}}\!,n_1^\text{in}\!,n_2^\text{in}\!,\vec{n}}\\
+\displaystyle\sum_{\alpha,k} c^{1\alpha k}\mathcal{S}_t^{\alpha}{\rho}^{(N^{\text{out}}-1, N^{\text{in}}\!\!, N)}_{\vec{n}_1^{\text{out}}-\vec{e}{(\alpha,k)},\vec{n}_2^{\text{out}}\!,n_1^\text{in}\!,n_2^\text{in}\!,\vec{n}}\\
+\displaystyle\sum_{\alpha,k} c^{2 \alpha k}\mathcal{S}_t^{\alpha}{\rho}^{(N^{\text{out}}-1, N^{\text{in}}\!\!, N)}_{\vec{n}_1^{\text{out}}\!,\vec{n}_2^{\text{out}}-\vec{e}{(\alpha,k)},n_1^\text{in}\!,n_2^\text{in}\!,\vec{n}}\;,
\end{array}}
\end{equation}
   with the  initial condition
  \begin{equation}
  \label{eq:in_cond_in_out_field_quadratic}
\rho^{(N^{\text{out}}\!\!, N^{\text{in}}\!\!, N)}_{\vec{n}_1^{\text{out}}\!\!,\vec{n}_2^{\text{out}}\!\!,n_1^\text{in}\!\!,n_2^\text{in}\!\!,\vec{n}}(0)=\delta_{\vec{n}_1^{{{\text{out}}}},\vec{0}}\delta_{\vec{n}_2^{{{\text{out}}}},\vec{0}}\delta_{{n_1^\text{in}},{0}}\delta_{{n_2^\text{in}},{0}}\delta_{\vec{n},\vec{0}}\;\rho_S(0)\;,
\end{equation}
and where
        \begin{equation}
   \label{eq:Yprop_1_in_out_field_new}
      \mathcal{G}^j_{t}\equiv   \sum_\alpha\langle{\phi}_j^{\text{in}}(0)\chi^{\alpha}_{t}\rangle \mathcal{S}^{\alpha}_t\;,
   \end{equation}
for $j=1,2$.  In Appendix \ref{sec:app:ioHEOM_ex} we provide an explicit expression for these HEOM for a specific instance of an open quantum system characterized by a single pair of system and bath coupling operators and then characterize it even more by focusing on a single-mode  bath.

As a summary, in this section we analyzed specific cases involving input and output modeled by one or two fields. In the next section, we provide a formal analysis which can be used to model more general cases. We are going to do this by distinguishing dynamical and static fields which include the input and output cases considered here.
\section{General case}
\label{sec:GeneralCase}
In this section, we analyze the derivation for an extended version of the HEOM for the general reduced correlation in Eq.~(\ref{eq:redCO}). In fact, while introducing input and output fields might be physically motivated, a conceptually more precise distinction can be obtained defining ``dynamic'' fields (evaluated at the dynamics time) and ``static'' field (evaluated at fixed times). In fact, this is the main feature leading to the qualitative different terms in the HEOM, which we introduced in the previous sections.

Following these considerations, we distinguish between``dynamical'' fields, ${{\phi}}^\text{dyn}_j(t)$, $j\in J_\text{dyn}=1,\dots,m_\text{dyn}$ (evaluated at the same time $t$ as the dynamics) and  ``static'' fields, ${{\phi}}^\text{stat}_{j'}(t^\text{stat}_{j'})$,  $j'\in J_\text{stat}=m_\text{dyn}+1,\dots,m_\text{dyn}+m_\text{stat}$ (evaluated at fixes times $t^\text{stat}_{j'}$)  so that $m_\text{stat}+m_\text{dyn}=m$. Importantly, this distinction includes two relevant cases, i.e., output fields (evaluated at the final time) and input fields (evaluated at the initial, fixed time). 

Leaving a more detailed derivation to Appendix \ref{sec:HEOMEnvCorr}, here we aim at highlighting the main logic which just consists in defining a formalism as parallel as possible to the one used to derive the regular HEOM. In fact, all the steps which are going to follow are the direct generalization of the techniques used in the previous sections.

To start, we simply impose an exponential ansatz for the correlations depending on the time $t$, i.e, involving dynamical fields so that
    \begin{equation}
   \label{eq:ansatzOchi}
   \langle{{\phi}}^{\text{dyn}}_{j}\chi^\alpha_\tau\rangle=\sum_{k}c^{j\alpha k} e^{-\gamma^{j\alpha k}(t-\tau)}\;,
   \end{equation}
   for $j\in J_\text{dyn}$. No specific ansatz is required for static fields. At this point, it is convenient to define the multi-index $\eta\equiv(\eta_1,\eta_2,\eta_3)\equiv(j,\alpha,k)$ and the sets $\eta_\text{dyn}$ and $\eta_\text{stat}$ such that $\eta\in\eta_\text{dyn}$ if $\eta_1=j\in J_\text{dyn}$ and $\eta\in\eta_\text{stat}$ otherwise, i.e., if $\eta_1=j\in J_\text{stat}$. With this notation, we can define
       \begin{equation}
   \label{eq:defY_main2}
   \begin{array}{lll}
     \mathcal{Y}^{\eta}_t\overset{\eta\in \eta_{\text{dyn}}}{\equiv} \mathcal{Y}^{j\alpha k}_t\overset{j\in J_{\text{dyn}}}{\equiv}c^{j\alpha k}\int_0^t d\tau  e^{-\gamma^{j\alpha k}(t-\tau)}\mathcal{S}^{\alpha}_\tau\\
  \mathcal{Y}^{\eta}_t\overset{\eta\in \eta_{\text{stat}}}{\equiv} \mathcal{Y}^{j\alpha k}_{t}\overset{j\in J_{\text{stat}}}{\equiv} \delta_{\alpha,0}\delta_{k,0}\int_0^t d\tau  \;\tilde{\mathcal{Y}}^{j}_{\tau}\;.
   \end{array}
   \end{equation}
   where
      \begin{equation}
   \label{eq:Ytilde_main}
   \tilde{\mathcal{Y}}^{j}_{\tau}=\sum_{\alpha'} \langle{{\phi}_{j}}(t_j)\chi^{\alpha'}_\tau\rangle\mathcal{S}^{\alpha'}_\tau\;.
   \end{equation}
   We note that, in the expression for the static fields, the indexes $\alpha$ and $k$ were artificially introduced in order to achieve a more uniform and compact formulation in the following of this section.
   The derivatives of the expressions in Eq.~(\ref{eq:Ytilde_main}) satisfy the identities
      \begin{equation}
   \label{eq:Yprop}
       \dot{\mathcal{Y}}^{\eta}_t\overset{\eta\in \eta_{\text{dyn}}}{\equiv}    c^\eta \mathcal{S}^{\eta_2}_t-\gamma^\eta{\mathcal{Y}}^{\eta}_t\; \text{, and}~~\dot{\mathcal{Y}}^{\eta}_t\overset{\eta\in \eta_{\text{stat}}}{\equiv} \tilde{\mathcal{Y}}^{j}_{t}\;,
   \end{equation}
   corresponding to the one in Eq.~(\ref{eq:temptemp2}) for the standard HEOM. 
The redundancy in the notation in the definition for the static fields in Eq.~(\ref{eq:defY_main2}) is introduced to allow a more homogeneous treatment of two classes of fields in the following. In fact, we can now already write out main quantities of interest, i.e., those in Eq.~(\ref{eq:Sat_main}) as a function of the $\mathcal{Y}$ operators as
       \begin{equation}
   \label{eq:Sat_main_last}
   \rho_S^{\mathfrak{a}}(t)=\prod_{j\in \mathfrak{a}_{\text{c}} } \sum_{\alpha,k}  \mathcal{Y}_t^{j\alpha k}\rho_S(t)\;,
   \end{equation}
   where we note that the complement of the set $\mathfrak{a}$ is taken, as required by Eq.~(\ref{eq:Sat_main}).
To compute these quantities, we now follow the  formal similarities to the regular HEOM in order to define the following extended auxiliary density matrices 
      \begin{equation}
   \label{eq:LrhoNnt_main}
   \begin{array}{lll}
\rho_{\vec{n}^{{{\phi}}},\vec{n}}(t)&\equiv&   \alpha_0^N\mathcal{T}\prod_{\eta} \left[\mathcal{Y}^{\eta}_t\right]^{n^{{{\phi}}}_{\eta}}\prod_{\sigma}\left[\Theta^{\sigma}_t\right]^{n_\sigma}\rho_S(t)\;,
\end{array}
   \end{equation}
   where $n^{{\phi}}_\eta\equiv n^{{\phi}}_{j k\alpha}\in \mathbb{N}$.  These quantities encode information about the reduced system dynamics (as in the regular HEOM) alongside information about the statistics of the fields present in the correlation. In fact, knowledge of these auxiliarty density matrices allows to recover $\rho^\mathfrak{a}_S(t)$ as
      \begin{equation}
   \label{eq:rhoSat_main}
     \rho_S^{\mathfrak{a}}(t)= \sum_{\vec{n}^{{{\phi}}}: \vec{n}^{\text{tot}}_{j\in \mathfrak{a}_{\text{c}}}=1,\vec{n}^{\text{tot}}_{j\not\in \mathfrak{a}_{\text{c}}}=0}\rho_{\vec{n}^{{{\phi}}},\vec{n}=\vec{0}}(t)\;.
   \end{equation}
   where $   \vec{n}^{\text{tot}}_j\equiv\sum_{\alpha k}n^{{{\phi}}}_{j\alpha k}$. 
   Now, the  exponential ansatz in Eq.~(\ref{eq:ansatzOchi}) and the property in  Eq.~(\ref{eq:Yprop}), allow to follow parallel arguments to the ones used  in section \ref{sec:HEOM_main_rsd} for the standard HEOM to write
      \begin{equation}
 \label{eq:HEOM_extended_main_short}
\begin{array}{l}\dot{\rho}^{(N^{{\phi}},N)}_{\vec{n}^{{{\phi}}},\vec{n}}=\displaystyle\text{HEOM}_0[{\rho}^{(N^{{\phi}},N)}_{\vec{n}^{{{\phi}}},\vec{n}}]-\sum_{\eta\in\eta_{\text{out}}} n^{{{\phi}}}_\eta \gamma^{\eta}\rho^{(N^{{\phi}},N)}_{\vec{n}^{{{\phi}}},\vec{n}}\\
\displaystyle+\sum_{\eta\in\eta_{\text{dyn}}}n^{{{\phi}}}_\eta a^{\eta}\mathcal{S}_t^{\eta_2}\rho^{(N^{{\phi}}-1,N)}_{\vec{n}^{{{\phi}}}-\vec{e}{(\eta)},\vec{n}}+\sum_{\eta\in\eta_{\text{stat}}}n^{{{\phi}}}_\eta\tilde{\mathcal{Y}}^{\eta_1}_{t}\rho^{(N^{{\phi}}-1,N)}_{\vec{n}^{{{\phi}}}-\vec{e}{(\eta)},\vec{n}}\;,
\end{array}
\end{equation}
or, in full,
   \begin{equation}
   \label{eq:HEOM_extended_main}
\begin{array}{l}\dot{\rho}^{(N^{{\phi}},N)}_{\vec{n}^{{{\phi}}},\vec{n}}=\displaystyle -\sum_{\sigma} n_\sigma b^{\sigma}\rho^{(N^{{\phi}},N)}_{\vec{n}^{{{\phi}}},\vec{n}}+\alpha_0\sum_{\sigma}n_\sigma a^{\sigma}\mathcal{B}^{\sigma}\rho^{(N^{{\phi}},N-1)}_{\vec{n}^{{{\phi}}},\vec{n}-\vec{e}{(\sigma)}}\\
\displaystyle +\alpha^{-1}_0\sum_{\sigma}\mathcal{A}^{\sigma}\rho^{(N^{{\phi}},N+1)}_{\vec{n}^{{{\phi}}},\vec{n}+\vec{e}(\sigma)} -\sum_{\eta\in\eta_{\text{out}}} n^{{{\phi}}}_\eta \gamma^{\eta}\rho^{(N^{{\phi}},N)}_{\vec{n}^{{{\phi}}},\vec{n}}\\
\displaystyle+\sum_{\eta\in\eta_{\text{dyn}}}n^{{{\phi}}}_\eta a^{\eta}\mathcal{S}_t^{\eta_2}\rho^{(N^{{\phi}}-1,N)}_{\vec{n}^{{{\phi}}}-\vec{e}{(\eta)},\vec{n}}+\sum_{\eta\in\eta_{\text{stat}}}n^{{{\phi}}}_\eta\tilde{\mathcal{Y}}^{\eta_1}_{t}\rho^{(N^{{\phi}}-1,N)}_{\vec{n}^{{{\phi}}}-\vec{e}{(\eta)},\vec{n}}\;,
\end{array}
\end{equation}
where $N^{{\phi}}=\sum_{\eta}\vec{n}^{{{\phi}}}$. These extended HEOM are associated to the initial condition
  \begin{equation}
  \label{eq:in_cond}
\rho_{\vec{n}^{{{\phi}}},\vec{n}}(0)=\delta_{\vec{n}^{{{\phi}}},\vec{0}}\delta_{\vec{n},\vec{0}}\;\rho_S(0)\;.
\end{equation}
  It is interesting to note that the extension of the HEOM operator on the new indexes is bounded, i.e., there is no ``process'' which increases $N^{\phi}$. 
  
  In summary, the solution of Eq.~(\ref{eq:HEOM_extended_main}) with initial condition in Eq.~(\ref{eq:in_cond}) can be used in Eq.~(\ref{eq:rhoSat_main}) for all the set $a$ required in Eq.~(\ref{eq:rhomSt_main}). In turn, this allows to compute the quantities $\rho^{m}_S(t)$ from the knowledge of the cross correlations involving the full set of superoperators  made by the interaction operators $\chi^\alpha_t$ and the fields $\phi_j(t_j)$. 
  
We conclude with a remark on the scaling of the HEOM indexes. To do this, we note that, from Eq.~(\ref{eq:rhoSat_main}), we are always interested in evaluating the auxiliary density matrices for $\vec{n}^\text{tot}_{j}=1$ for each $j$ labeling a field present in the correlation which we need to compute. Together with the fact that, as already noted,  Eq.~(\ref{eq:HEOM_extended_main_short})
does not contain any  ``process'' which increases $N^{\phi}$, this allows us to provide a faithful estimate on the complexity which is added to the regular HEOM. In fact,  if we set $N^\phi_\text{exp}$ to be the scale for the number of elements in the  spectral representation  approximating the cross correlations between dynamical fields and environmental interaction operators, the extended HEOM can be interpreted as effectively requiring, with respect to the regular version, an extra index taking $2^{N_\text{tot}}$ possible values where
\begin{equation}
\label{eq:scaling}
N_\text{tot}={2 N^H_{\text{I}}N^\phi_{\text{exp}}m_{\text{dyn}}+m_{\text{stat}}}\;.
\end{equation}
Here,  $m_{\text{dyn}}$ and $m_{\text{stat}}$ are the number of dynamical and static field operators considered, and $N^H_{\text{I}}$ the number of terms in the interaction Hamiltonian.

  \section{Input-output extension of the Lindblad equation}
  \label{sec:io-Lindblad}
  As an exemplicative application of this formalism, in this section we derive an input-output extension of the Lindblad equation. In fact, while the HEOM are usually used to analyze parameter regimes beyond the Markovian approximation, here we show that the input-output HEOM can also be used to extend the Markovian Lindblad equation to an input-output version. 
  \subsection{A Markovian Model}
For the sake of specificity, we are going to consider the input-output hierarchical equations of motion for a 1-dimensional environment interacting with a quantum system within the Markovian limit. 
  To do this, the main setting consists of a collection of environmental mode densities $b_\mathfrak{r}(p)$, written as a function of a momentum variable $p$ and a label $\mathfrak{r}=\mathfrak{p},\mathfrak{n}$, characterizing positive and negative energy so that $[b_{\mathfrak{r}}(p),b^\dagger_{\mathfrak{r}'}(p')]=\delta(p-p')\delta_{\mathfrak{r} \mathfrak{r}'}$ in terms of the Dirac and Kronecker deltas, respectively. The inclusion of negative-energy modes is one of the standard avenues to introduce the Markovian approximation by means of white noise as analyzed in \cite{Gardiner}. In fact,  the extension of  the energy domain to the negative axis corresponds to the flat spectral density approximation defining the Markovian limit.  While the existence of these modes allows the simplification of the effective description of the system dynamics into a Lindblad form, the physical validity of the corresponding Markovian limit requires the population of the negative energy modes to remain negligible throughout the environmental dynamics. 

To keep the notation explicit, we further describe the system as a two-level system with raising and lowering Pauli operators $\sigma_\pm$ and a zero-temperature underlying environmental state $\rho_{\beta_\infty}$.
In this setting, we  assume the following system+bath Hamiltonian
\begin{equation}
\label{eq:1D}
\begin{array}{lll}
    H_\text{1D}&=&\displaystyle H_S+\sum_{\mathfrak{r}=\mathfrak{p},\mathfrak{n}}  \int_{-\infty}^\infty dp\; \omega_\mathfrak{r}(p) b_\mathfrak{r}^\dagger(p) b_\mathfrak{r}(p)\\
    &&\displaystyle+\sum_{\mathfrak{r}=\mathfrak{p},\mathfrak{n}}\int_{-\infty}^\infty dp\;\left[g_\mathfrak{r}(p)\sigma_+ b_\mathfrak{r}(p)+\bar{g}_r(p) b_\mathfrak{r}^\dagger(p)\sigma_-\right]\;,
    \end{array}
\end{equation}
in terms of a system Hamiltonian $H_S=\omega_s/2 \;\sigma_z$ with splitting $\omega_S$, a dispersion $\omega_\mathfrak{r}(p)$ [such that $\omega_\mathfrak{n}(p)=-\omega_\mathfrak{p}(p)$], and coupling parameters $g_\mathfrak{r}(p)$ characterizing a rotating-wave-like interaction. As shown in \cite{Gardiner}, the white noise approximation  corresponds to imposing a flat interaction profile in energy which, using Eq.~(\ref{eq:1D}), corresponds to a flat spectral density
\begin{equation}
J_{\mathfrak{r},j}(E)\equiv\frac{|g^2_\mathfrak{r}(p_j^E)|}{|\frac{d\omega_\mathfrak{r}}{dp}(p_j^E)|}\equiv \Gamma/n\pi\;,
\end{equation}
in terms of the effective decay rate $\Gamma$. Here,  $p_j^E$ are the momenta such that $\omega(p_j^E)=E$, so that the index $j=1,\cdots,n$ labels the $n-$degeneracy in energy of the dispersion.  Assuming a Ohmic bath with dispersion $\omega(p)= c |p|$ in terms of the speed of light $c$ in the bath, opposite momenta corresponds to the same energy so that $n=2$, ultimately requiring that
\begin{equation}
\label{eq:sp_flat_white_noise}
g_\mathfrak{r}(p)\equiv \sqrt{\Gamma c/2\pi}\;.
\end{equation}
In the notation of Eq.~(\ref{eq:sup_int_H}), the model in Eq.~(\ref{eq:1D}) corresponds to an interaction superoperator written in the interaction picture as
\begin{equation}
\mathcal{H}^I_t=\sum_{\alpha}\mathcal{S}^\alpha_t\chi_t^\alpha\;,
\end{equation}
where $\alpha=(p,q)$ specifies left/right  ($p=\text{l}/\text{r}$) and raising/lowering ($q=\pm$)  indexes, i.e.,
\begin{equation}
\label{eq:chiSdef}
\renewcommand{\arraystretch}{1.4}{
\begin{array}{lllll}
\chi_t^{\text{l},\pm}&=&X_t^\pm[\cdot],~~~~~~~\mathcal{S}_t^{\text{l},\pm}&=&\sigma_\mp[\cdot]e^{\mp i\omega_S t}\\
\chi_t^{\text{r},\pm}&=&[\cdot]X_t^\pm,~~~~~~~\mathcal{S}_t^{\text{r},\pm}&=&-[\cdot]\sigma_\mp e^{\mp i\omega_S t}\;,
\end{array}}
\end{equation}
in terms of the operators
\begin{equation}
\renewcommand{\arraystretch}{1.4}{
\begin{array}{lll}
X_t^{-}&=&\displaystyle\sum_{\mathfrak{r}=\mathfrak{p},\mathfrak{n}}\int_{-\infty}^\infty dp\;g_\mathfrak{r}(p)b_\mathfrak{r}(p) e^{-i\omega_\mathfrak{r}(p) t}\\
X_t^{+}&=&\displaystyle\sum_{\mathfrak{r}=\mathfrak{p},\mathfrak{n}}\int_{-\infty}^\infty dp\;\bar{g}_\mathfrak{r}(p)b^\dagger_\mathfrak{r}(p)e^{i\omega_\mathfrak{r}(p) t}\;.
\end{array}}
\end{equation}
As a consequence, the reduced system dynamics is completely determined by the $4\times 4$  correlation matrix $C^{\alpha_2\alpha_1}_{t_2,t_1}$ (for $t_2>t_1$) in Eq.~(\ref{eq:Corr}) which, thanks to the rotating-wave form of the interaction and the zero-temperature assumption has only four non-zero elements which read
\begin{equation}
\renewcommand{\arraystretch}{1.4}{
\begin{array}{lll}
C^{\alpha_2=(\text{l},-),\alpha_1=(\text{l},+)}_{t_2,t_1}&=&C^{\alpha_2=(\text{r},-),\alpha_1=(\text{l},+)}_{t_2,t_1}=\langle X_{t_2}^-X_{t_1}^+\rangle_\infty\\
&=&\!2\Gamma\delta(t_2-t_1)\\
C^{\alpha_2=(\text{l},+),\alpha_1=(\text{r},-)}_{t_2,t_1}&=&C^{\alpha_2=(\text{r},+),\alpha_1=(\text{r},-)}_{t_2,t_1}=\langle X_{t_1}^-X_{t_2}^+\rangle_\infty\\
&=&2\Gamma\delta(t_2-t_1)\;,
\end{array}}
\end{equation}
where the expectation value $\langle\cdot\rangle_\infty$ is defined in terms of the trace over the state $\rho_{\beta_\infty}$. We note that the possibility to write these correlations as delta functions relies on the Ohmic structure of the dispersion \emph{and} the presence of negative energy modes which are necessary to achieve actual spectral flatness.
The resulting expressions for the correlations can now be inserted in Eq.~(\ref{eq:F_rep}) to show that, using Eq.~(\ref{eq:chiSdef}) and $\int_0^t dt'~\delta(t-t')=1/2$, see Eq.~(5.3.12) in \cite{Gardiner}, the influence superoperator  simplifies to write the reduced system dynamics  as the Lindblad equation
\begin{equation}
\label{eq:F_rep_lind}
\renewcommand{\arraystretch}{1.4}{
\begin{array}{lll}
\dot{\rho}_S&=&L_0\rho_S\\
&=&\!\!-i[H_S,\rho_S]+\Gamma\left(2\sigma_-\rho_S\sigma_+-\sigma_+\sigma_-\rho_S-\rho_S\sigma_+\sigma_-\right),
    \end{array}}
\end{equation}
back in the Shr\"{o}dinger picture for the system. So far, our considerations constitute a version of the regular derivation of the Lindblad equation. 

However, our goal is, within this Markovian model, to reframe the formalism developed in this article to extend the Lindblad equation to allow both the presence of wave-packets propagating in the environment and the possibility to probe bath observables. Specifically, we are going to look for an extension of Eq.~(\ref{eq:F_rep_lind}) to describe a bath initially prepared in the state
\begin{equation}
\label{eq:in_cond_markov}
\rho_B\equiv\phi^\text{in}_1(0)\phi^\text{in}_2(0)\rho_{\beta_\infty}=\varphi_1^\text{in}\rho_{\beta_\infty}\varphi_2^\text{in}\;,
\end{equation}
where, following the notation in section \ref{sec:input_output}, we defined the supeoperators $\phi_1^\text{in}(0)=\varphi_1^\text{in}[\cdot]$ and $\phi_2^\text{in}(0)=[\cdot]\varphi_2^\text{in}$, in terms of the input fields
\begin{equation}
\renewcommand{\arraystretch}{1.4}{
\begin{array}{lll}
\varphi_1^\text{in}&=&\int_{-\infty}^\infty dp\; \sqrt{g_\text{in}(p)}e^{-i p x_\text{in}}b_{\mathfrak{p}}^\dagger(p)\\
\varphi_2^\text{in}&=&\int_{-\infty}^\infty dp\; \sqrt{g_\text{in}(p)}e^{i p x_\text{in}}b_{\mathfrak{p}}(p)\;.
\end{array}}
\end{equation}
Here,  the initial condition \emph{only} only depends on the positive-energy modes $b_{\mathfrak{p}}(p)$ to represent a phsyical single-boson  wave-packet with a space-bias $x_\text{in}$ and a Gaussian profile
\begin{equation}
\label{eq:gin}
g_\text{in}(p)\equiv g(p;p_\text{in},\sigma_\text{in})\equiv e^{-(p-p_\text{in})^2/2\sigma_\text{in}^2}/\sqrt{2\pi}\sigma_\text{in}\;,
\end{equation}
characterized by an initial momentum-bias $p_\text{in}$ and momentum-width $\sigma_\text{in}$. We can check the actual meaning of these definitions by evaluating some of the expected properties of the initial state. For example, we can define the Fourier transformed operators 
\begin{equation}
b_{\mathfrak{r}}(x)=\frac{1}{\sqrt{2\pi}}\int_{-\infty}^\infty dp\; e^{i p x} b_{\mathfrak{r}}(p)\;,
\end{equation}
so that
\begin{equation}
\renewcommand{\arraystretch}{1.4}{
\begin{array}{lll}
\text{Tr}[b_{\mathfrak{p}}^\dagger(p) b_{\mathfrak{p}}(p)\rho_B]&=& g(p;p_\text{in},\sigma_\text{in})\\
\text{Tr}[b_{\mathfrak{p}}^\dagger(x) b_{\mathfrak{p}}(x)\rho_B]&=& g(x;x_\text{in},\zeta_\text{in})\;,
\end{array}}
\end{equation}
in terms of the the standard deviation of the initial wave-packet in space $\zeta_\text{in}=1/(2\sigma_\text{in})$.

We are now going to model the output by explicitly defining what observable we are interested to model. This raises the rather subtle question about whether negative-energy modes should be included in the definition. To motivate a possible choice, in this section we are going to impose a white-noise structure to the observable, similarly at what we did to derive the master equation itself.  In other words, we are looking for the simplest input-output Markovian extension of the Lindlblad equation, through the definition
\begin{equation}
\label{eq:Obs_x}
O^x(t_\text{out})=\Delta x ~b_{+}^\dagger(x_\text{out}) b_{+}(x_\text{out})\;,
\end{equation}
for a chosen position $x_\text{out}$, a space-discretization $\Delta x$, and where 
\begin{equation}
b_{\pm}^\dagger(x_\text{out})=b_{\mathfrak{p}}^\dagger(x_\text{out}) \pm b_{\mathfrak{n}}^\dagger(x_\text{out})\;,
\end{equation}
 Equivalently, we can write
\begin{equation}
\label{eq:Obs_+}
O^x(t_\text{out})=\varphi^\text{out}_{1}(t_\text{out})\varphi^\text{out}_{2}(t_\text{out}),
\end{equation}
 in terms of the fields
   \begin{equation}
      \label{eq:obs_pos_neg}
     \renewcommand{\arraystretch}{1.4}{
   \begin{array}{lll}
  \varphi^\text{out}_{1}(t_\text{out})&=&\displaystyle\sum_{\mathfrak{r}={\mathfrak{p}},{\mathfrak{n}}}\int_{-\infty}^\infty dp \;g_\text{out}(p) b_\mathfrak{r}^\dagger(p)e^{i\omega_\mathfrak{r}(p) t_\text{out}}\\
    \varphi^\text{out}_{2}(t_\text{out})&=&\displaystyle\sum_{\mathfrak{r}={\mathfrak{p}},{\mathfrak{n}}}\int_{-\infty}^\infty dp \;\bar{g}_\text{out}(p) b_\mathfrak{r}(p)e^{-i\omega_\mathfrak{r}(p) t_\text{out}}\;.
  \end{array}}
   \end{equation} 

We note that the expression in Eq.~(\ref{eq:Obs_x}) treats positive and negative energy modes on equal grounds, thereby corresponding to the ``output-Markovian'' choice for the observable which was mentioned above. More concretely, this qualification can be understood  from the fact that the integrals defining the cross correlations between this observable and the environmental coupling operators are now going to be effectively performed over the full energy axis, corresponding to the flat Markovian assumption.

We finish this section by noting that other choices, such as limiting the focus on positive energy modes only, is also possible within this formalism by simply computing the corresponding cross-correlations. Here, the ``output-Markovian'' choice serves to present the simplest of such possibilities. We point out that, at this level, it might be tempting to argue that a focus on positive-energy modes is the more valuable from a physical point of view. While following this temptation, it is important to consider that any choice is always going to rely on the very same underlying Markovian assumption i.e.,  the existence of the negative energy modes. As a consequence, even by imposing those modes not to be populated in the initial state (as we do), they are still going to be involved in the dynamics, whether we decide to add them to the output observable or not. In turn, this points towards considering the discrepancy among the results obtained from their inclusion or exclusion in the output observable as simply a measure of the validity of the underlying Markovian assumption, rather than physically meaningful. 

In other words, whenever the output population of negative energy modes becomes comparable to the positive energy ones, the validity of the original Markov approximation should be questioned or, at least, limiting the resolution of its predictions. In all other cases, the discrepancy should be negligible, making the original choice a matter of convenience.

This completes the general analysis of the Markovian case. In the following, we are going to apply the input-output HEOM formalism to this setup.

  \subsection{An input-output Lindblad equation}
  To proceed gradually, we begin by deriving the input version of the Lindblad equation. 
To do this, we follow the formalism in section \ref{subsec:i}, we only need to compute the key quantity, which defines the corresponding time-dependent input-HEOM,
       \begin{equation}
       \label{eq:Gin}
       \mathcal{G}^{\text{in},j}_t =   \sum_\alpha\text{Tr}[\mathcal{T}_B{\phi}_j^{\text{in}}(0)\chi^{\alpha}_{t}\rho_{\beta_\infty}]\mathcal{S}^{\alpha}_t\;,
   \end{equation}
  for $j=1,2$. Using Eq.~(\ref{eq:chiSdef}), we find that
         \begin{equation}
         \renewcommand{\arraystretch}{1.4}{
         \begin{array}{llllll}
       \mathcal{G}^{\text{in},1}_t &=&\text{Tr}[{X^{-}_{t}\varphi}_1^{\text{in}}\rho_{\beta_\infty}] [\sigma_+,\cdot]e^{i\omega_S t}&\equiv& D^\text{in}_+(t)e^{ i\omega_S t}\\
       
              \mathcal{G}^{\text{in},2}_t &=&\text{Tr}[{X^{+}_{t}\rho_{\beta_\infty}\varphi}_2^{\text{in}}] [\sigma_-,\cdot]e^{-i\omega_S t}&\equiv&D^\text{in}_-(t)e^{- i\omega_S t},
       \end{array}}
   \end{equation}
   where we introduced the driving superoperator terms
   \begin{equation}
   \label{eq:Din}
   D^\text{in}_\pm(t)=-i\Omega^\text{in}_\pm(t) [\sigma_\pm,\cdot]\;,
   \end{equation}
  in terms of the time-dependent complex frequencies
  \begin{equation}
  \Omega^\text{in}_\pm(t)\equiv i\sqrt{{\Gamma c}/{2\pi}}\int_{-\infty}^\infty dp\;\sqrt{g_\text{in}(p)}e^{\mp i[ p x_\text{in}+ \omega(p) t]}\;,
  \end{equation}
  which, using Eq.~(\ref{eq:gin}), becomes
    \begin{equation}
  \renewcommand{\arraystretch}{1.6}{
  \begin{array}{lll}
  \Omega^\text{in}_\pm(t)
  \!\!&=&\!\!i K_\text{in}\left\{e^{-x^2_t/4\zeta^2_{\text{in}}\mp i{p}_\text{in} x_t}\left[1\mp i\text{Erfi}\left(\frac{x_t}{2\zeta_\text{in}}\pm i\frac{p_\text{in}}{2\sigma_\text{in}}\right)\right]\right.\\
    \!\!&+&\!\!\left.e^{-x^2_{-t}/4\zeta^2_{\text{in}}\mp i{p}_\text{in} x_{-t}}\left[1\pm i\text{Erfi}\left(\frac{x_{-t}}{2\zeta_\text{in}}\pm i\frac{ p_\text{in}}{2\sigma_\text{in}}\right)\right]\right\}\;,
\end{array}}
  \end{equation}
in terms of  the effective frequency $K_\text{in}=(2\pi)^{-1/4}\sqrt{\Gamma c/\zeta_\text{in}}/2$, $x_t=x_\text{in}+c t$, and the imaginary error function $\text{Erfi}(z)=-2i\int_0^{iz} dq ~e^{-q^2}/\sqrt{\pi}$. In the equations above, the extra imaginary constant was only introduced to recover some of the standard notation for the infinitesimal evolution in  quantum mechanics in Eq.~(\ref{eq:Din}).
We now have set up all the formalism which allows us to define the input version of the Lindlbad equation in Eq.~(\ref{eq:F_rep_lind}) by simply reading it from Eq.~(\ref{eq:HEOM_extended_main_1_in_field_quadr}) as
  \begin{equation}
  \label{eq:inputShr}
  \dot{\Psi}^\text{in}(t)=L^\text{in}\Psi^\text{in}(t)\;,
  \end{equation}
where
  \begin{equation}
  \label{eq:inputL}
  L^\text{in}=\mathbb{I}^{\text{in}}_2\otimes\mathbb{I}^{\text{in}}_2\otimes L_0+ \mathbb{I}^{\text{in}}_2\otimes\tau^{\text{in}}_-\otimes D^\text{in}_++ \tau^{\text{in}}_-\otimes\mathbb{I}^{\text{in}}_2\otimes D^\text{in}_-\;,
  \end{equation}
   in the system Shr\"{o}dinger picture, written terms of the 2-dimensional identity $\mathbb{I}^{\text{in}}_2$ and the  lowering Pauli operator  $\tau^{\text{in}}_-$, and the state-vector $\Psi=(\Psi_0,\Psi_1,\Psi_2,\Psi_3)=(\rho_{00},\rho_{10},\rho_{01},\rho_{11})^T$, where $T$ denotes transposition. 
   Following Eq.~(\ref{eq:rho20}), this input Lindblad-equation 
   is defined with an initial condition given by $\Psi(0)=(\rho_S(0),0,0,0)^T$ and it characterizes the overall system dynamics as
   \begin{equation}
   \rho_S(t)=  \langle\phi_1^\text{in}\phi_2^\text{in}\rangle_\infty\Psi_0(t)-\Psi_3(t)=\Psi_0(t)-\Psi_3(t)\;,
   \end{equation}
   since $  \langle\phi_1^\text{in}\phi_2^\text{in}\rangle_\infty=\int_{-\infty}^\infty dp\; g_\text{in}(p)=1$.

We are now going to complement this analysis to allow the further evaluation of the output environmental observable given in Eq.~(\ref{eq:Obs_+}).
Interestingly, there are two possible routes on how to treat this output observables. One, is to follow the output analysis given in section \ref{subsec:o}. This route relies on the exponential ansatz of the kind given in Eq.~(\ref{eq:spectr_ansatz}). While this is the standard way forward for numerical implementations, to obtain the simplest possible result, here we are going to follow the general analysis in section \ref{sec:GeneralCase} instead. In fact, as mentioned there, the exponential ansatz for the HEOM can always be relaxed 
by fixing the evaluation time $t_\text{out}$ throughout the dynamics. In other words, by fixing the time at which the fields are evaluated all fields can be treated in the same way as we did for the input ones above. From this point of view, as we just did for the input fields, we just need to compute the quantity
       \begin{equation}
       \label{eq:Gout}
       \mathcal{G}^{\text{out},j}_t =   \sum_\alpha\text{Tr}[\mathcal{T}_B{\phi}_j^{\text{out}}(t_\text{out})\chi^{\alpha}_{t}\rho_{\beta_\infty}]\mathcal{S}^{\alpha}_t\;.
   \end{equation}
By comparing this expression to the corresponding input one in Eq.~(\ref{eq:Gin}), we can appreciate that the action of the time-ordering superoperator qualitatively depends on whether $t<t_\text{out}$ or $t>t_\text{out}$. Given the definition of the correlations in Eq.~(\ref{eq:CO_in_out}), the meaningful quantity to analyze is the one in which the dynamical time $t$ is at least as big as the observation time $t_\text{out}$, so that a minimal requirement is to analyze the regime up to times $t=t_\text{out}$. However, it is essential to note that this condition is only to be imposed after the dynamics has been calculated. Equivalently,  $t_\text{out}$ is to be fixed \emph{before} solving for the dynamics whose solution is going to be meaningful for times $t\geq t_\text{out}$. We now can analyze Eq.~(\ref{eq:Gout}) more explicitly as
       \begin{equation}
\label{eq:G1_ex}
  \renewcommand{\arraystretch}{1.4}{
\begin{array}{lll}
       \mathcal{G}^{\text{out},1}_t &\stackrel{t_\text{out}<t}{=}& \langle X^-_t\varphi^\text{out}_1(t_\text{out})\rangle_\infty (\sigma_+[\cdot]-[\cdot]\sigma_+)e^{ i\omega_S t}\\       
       &\stackrel{t_\text{out}>t}{=} &-\langle X^-_t\varphi^\text{out}_1(t_\text{out})\rangle_\infty [\cdot]\sigma_+\equiv D_+^{\text{out}}(t)e^{ i\omega_S t},
       \end{array}}
   \end{equation}
   where
   \begin{equation}
   \label{eq:D+}
D_+^{\text{out}}(t)=i\Omega^\text{out}_+(t)[\cdot]\sigma_+\;,
   \end{equation}
   in terms of 
   \begin{equation}
   \label{eq:omegaoutp}
     \renewcommand{\arraystretch}{1.4}{
   \begin{array}{lll}
    \Omega^\text{out}_+(t)&\equiv&     i\langle X^-_t\varphi^\text{out}_1(t_\text{out})\rangle_\infty\\
    &=&i\sqrt{\Gamma c/2\pi}\int_{-\infty}^\infty dp\; g_\text{out}(p) e^{i\omega(p) (t_{\text{out}}-t)}\;.
    \end{array}}
   \end{equation}
   The result in Eq.~(\ref{eq:G1_ex}) can be intuitively justified by simply noting that, for $j=1$ only the $\alpha=(\text{l},-)$ and $\alpha=(\text{r},-)$ terms give non-zero contributions. For $t_\text{out}<t$, the field always acts first on the vacuum because of the time ordering, thereby allowing both cases to contribute. On the other hand, when $t_\text{out}>t$, the interaction operator acts first, thereby annihilating the case $\alpha=(\text{l},-)$. Similarly, for $j=2$, we have
          \begin{equation}
\label{eq:G2_ex}
  \renewcommand{\arraystretch}{1.4}{
\begin{array}{lll}
       \mathcal{G}^{\text{out},2}_t &\stackrel{t_\text{out}<t}{=}& 0\\       
       &\stackrel{t_\text{out}>t}{=} &\langle \varphi^\text{out}_2(t_\text{out})X^+_t\rangle_\infty\sigma_- [\cdot]\equiv D^\text{out}_-(t)e^{- i\omega_S t}\;,
       \end{array}}
   \end{equation}
   where
      \begin{equation}
      \label{eq:D-}
  D_-^{\text{out}}(t)=-i\Omega^\text{out}_-(t)\sigma_-[\cdot]\;,
   \end{equation}
   in terms of 
   \begin{equation}
      \label{eq:omegaoutm}
     \renewcommand{\arraystretch}{1.4}{
   \begin{array}{lll}
     \Omega^\text{out}_-(t)&\equiv &  i \langle \varphi^\text{out}_2(t_\text{out})X^+_t\rangle_\infty\\
     &=&i\sqrt{\Gamma c/2\pi}\int_{-\infty}^\infty dp\; \bar{g}_\text{out}(p) e^{-i\omega(p) (t_{\text{out}}-t)}\;.
            \end{array}}
   \end{equation}
We consider that, if we are interested in the dynamical solution at $t_\text{out}$, we can restrict our attention to the case $t<t_\text{out}$.
   We  further note that the sign and the imaginary constant in Eq.~(\ref{eq:D+}) and Eq.~(\ref{eq:D-}), are introduced to align the notation in analogy to the standard one for infinitesimal time-evolution in which a $(-i)$ corresponds to the Hamiltonian acting on the left and a $(+i)$ to the action on the right.
   Because of the presence of both negative- and positive-energy modes as in Eq.~(\ref{eq:obs_pos_neg}), Eq.~(\ref{eq:omegaoutp}) and Eq.~(\ref{eq:omegaoutm}) take the form
   \begin{equation}
     \renewcommand{\arraystretch}{1.4}{
   \begin{array}{lll}
    \Omega^\text{out}_{+}&\equiv&  i\sqrt{\Gamma c/2\pi}\sum_{\mathfrak{r}=\mathfrak{p},\mathfrak{n}}\int_{-\infty}^\infty dp\; g_\text{out}(p) e^{i\omega_\mathfrak{r}(p) \Delta t}\\
         \Omega^\text{out}_{-}&\equiv &i\sqrt{\Gamma c/2\pi}\sum_{\mathfrak{r}=\mathfrak{p},\mathfrak{n}}\int_{-\infty}^\infty dp\; \bar{g}_\text{out}(p) e^{-i\omega_\mathfrak{r}(p) \Delta t}\;,
    \end{array}}
   \end{equation}
where $\Delta t = t_\text{out} - t$ and which, in a typical Markovian fashion, simplifies as 
\begin{equation}
\label{eq:Omega_delta}
    \Omega^\text{out}_{\pm}(t)=i\sqrt{\Gamma \Delta x/c}\left[\delta(t-t^+_\text{out})+\delta(t-t^-_\text{out})\right]\;,
\end{equation}
for $t_\text{out}^\pm=t_\text{out}\pm x_\text{out}/c$ and in terms of the Dirac delta.
We note that, considering other observables in the same limit, such as the ones involving only positive-energy modes, would require a further use of the Sokhotski–Plemelj formula resulting in rescaling the delta contribution above and in the presence of an additional term dependent on the Cauchy principal value.

To proceed, if we were interested in just computing the output without the input, we could now simply write the equation corresponding to Eq.~(\ref{eq:inputShr}) with the replacement $\text{in}\rightarrow\text{out}$, i.e., defined in terms of an extended Lindlbad operator in the form
  \begin{equation}
  \label{eq:outputShr}
L^\text{out}=L_0\otimes\mathbb{I}^{\text{out}}_2\otimes\mathbb{I}^{\text{out}}_2+ D^\text{out}_+\otimes\mathbb{I}^{\text{out}}_2\otimes\tau^{\text{out}}_-+ D^\text{out}_-\otimes\tau^{\text{out}}_-\otimes\mathbb{I}^{\text{out}}_2.
  \end{equation}
It turns out that, while being more interesting, the input-output case does not present further conceptual complications. In fact,  we can proceed recursively by defining
\begin{equation}
\label{Psi_inout}
\Psi^\text{in-out}\equiv(\Psi_{00},\Psi_{10},\Psi_{01},\Psi_{11})^T\;,
\end{equation}
where the indexes in $\Psi_{ij}$, $i,j=0,1$ identify the input and where
\begin{equation}
 \begin{array}{lll}
\Psi_{00}&\equiv&(\rho_{00,00},\rho_{00,10},\rho_{00,01},\rho_{00,11})\\
\Psi_{10}&\equiv&(\rho_{10,00},\rho_{10,10},\rho_{10,01},\rho_{10,11})\\
\Psi_{01}&\equiv&(\rho_{01,00},\rho_{01,10},\rho_{01,01},\rho_{01,11})\\
\Psi_{11}&\equiv&(\rho_{11,00},\rho_{11,10},\rho_{11,01},\rho_{11,11})\;.
    \end{array}
\end{equation}
so that the vector components can be written as $\Psi_{ab,ij}\equiv\rho_{ab,ij}$, where, in $\rho_{ab,ij}$, the first two indexes $a,b=0,1$ identify the input, while the last  two $i,j=0,1$ the output as in the general formalism developed in the previous sections. The input-output Lindblad equation can now be written by simply using $L^\text{in}$ in place of $L_0$ in Eq.~(\ref{eq:outputShr}), i.e., 
\begin{equation}
\label{eq:inoutLindblad}
\dot{\Psi}^\text{in-out}(t)=L^\text{in-out}\Psi^\text{in-out}(t)\;,
\end{equation}
where
     \begin{equation}
     \label{eq:L_in_out}
       \renewcommand{\arraystretch}{1.4}{
     \begin{array}{lll}
L^\text{in-out}&\!\!\!\!=&\!\!\!\mathbb{I}^{\text{in}}_4\otimes\mathbb{I}^{\text{out}}_4\otimes L_0\\
  
  &\!\!\!\!+&\!\!\! \mathbb{I}^{\text{in}}_2\otimes\tau^{\text{in}}_-\otimes\mathbb{I}^{\text{out}}_4\otimes D^\text{in}_++ \tau^{\text{in}}_-\otimes\mathbb{I}^{\text{in}}_2\otimes\mathbb{I}^{\text{out}}_4\otimes D^\text{in}_-\\

  &\!\!\!\!+&\!\!\! \mathbb{I}^{\text{in}}_4\otimes\mathbb{I}^{\text{out}}_2\otimes\tau^{\text{out}}_-\otimes D^\text{out}_+\!\!+\! \mathbb{I}^{\text{in}}_4\otimes\tau^{\text{out}}_-\otimes\mathbb{I}^{\text{out}}_2\otimes D^\text{out}_- \!.
       \end{array}}
  \end{equation}
     The input-output Lindblad equation in Eq.~(\ref{eq:inoutLindblad}) has initial condition given simply by a vector $\Psi(0)$ characterized by $\Psi_{00,00}=\rho_S(0)$ and all other elements zero in Eq.~(\ref{Psi_inout}).  
     
     Thanks to the explicit expression given in Eq.~(\ref{eq:Omega_delta}), it is possible to manipulate this equation further. In fact, the effect of the Dirac deltas in Eq.~(\ref{eq:Omega_delta}) is to provide a ``kick'' at the times $t^\pm_\text{out}$ when their constraint are satisfied. Intuitively, these are the times when the information has to leave the system in order to influence the value of the observable at the position $x_\text{out}$ at the later time $t_\text{out}$. In fact, when $|x_\text{out}/c|>t_\text{out}$, the two deltas do not contribute, inutuitively corresponding to the impossibility for the information to propagate that far from the system.  For $|x_\text{out}/c|=t_\text{out}$, then one of the deltas is satisfied at $t=0$, intuitively corresponding to the fact that, in order to arrive at those location in time, information has to leave the system immediately.  When $0<|x_\text{out}/c|\leq  t_\text{out}$, the contraints are satified, respectively at the intermediate times $0<t^\pm_\text{out}<t_\text{out}$.
       \begin{center}\begin{figure*}[t!]
\includegraphics[clip,width=2\columnwidth]{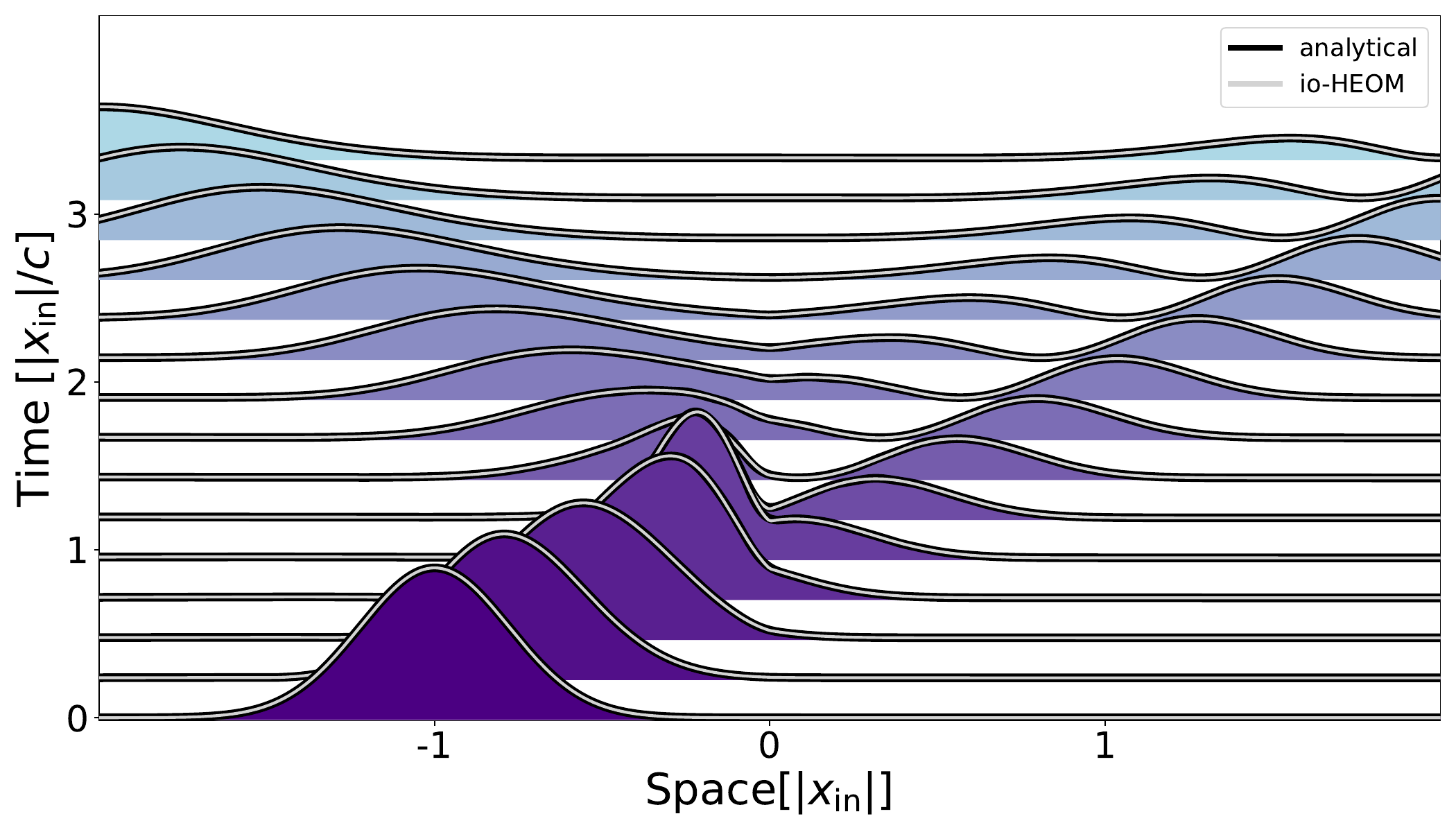}
\caption{Ridgeline plot of the bosonic occupation density, i.e. $\langle O^x\rangle(t_\text{out})/\Delta x$, along the $x$-direction of a 1-dimensional Markovian environment.   Here, in the y-axis, the density scale is not shown in favor of the offset time, i.e., the evolution time $t_\text{out}$ at which each curve is evaluated. 
At the beginning of the dynamics (zero offset), the environment is prepared in a 1-photon wave packet centered at $x_\text{in}$, and  propagating towards the system,  located at the position $x=0$. The bath occupation density in black corresponds to the analytical solution in Eq.~(\ref{eq:inoutLindblad}), and it is overlayed (in light-grey) by the solution of the input-output Lindlbad master equation in Eq.~(\ref{eq:analytical_res}) inspired by the io-HEOM formalism. The filling of the curves to the offset highlights the shape of the initial, transmitted, forward-, and back-scattered waves as the environment and the system interact. Other parameters are: $\omega_S = 4.5 c / |x_\text{in}|$, $\Gamma = 0.4 \omega_S$, $\sigma_\text{in}=p_\text{in}/2$, $p_\text{in}=\omega_S/c$, $\zeta_\text{in}=1/p_\text{in}$. The code used to generate this figure is available at \cite{io-HEOM-Markovian}.}
\label{fig:markovian}
\end{figure*}
\end{center} 
Taking this analysis into account, the input and output effects in Eq.~(\ref{eq:inoutLindblad}) effectively ``decouple'' to 
\begin{equation}
\label{eq:inoutLindblad2}
\dot{\Psi}^\text{in-out}(t)=[\tilde{L}_0+\tilde{L}^\text{in}(t)]\Psi^\text{in-out}(t)\;,
\end{equation}
where $\tilde{L}_0=\mathbb{I}^{\text{in}}_4\otimes\mathbb{I}^{\text{out}}_4\otimes L_0$ and where
     \begin{equation}
     \label{eq:L_in_out2}
       \renewcommand{\arraystretch}{1.4}{
     \begin{array}{lll}
\tilde{L}^\text{in}(t)&=&\!\! \mathbb{I}^{\text{in}}_2\otimes\tau^{\text{in}}_-\otimes\mathbb{I}^{\text{out}}_4\otimes D^\text{in}_++ \tau^{\text{in}}_-\otimes\mathbb{I}^{\text{in}}_2\otimes\mathbb{I}^{\text{out}}_4\otimes D^\text{in}_-.
       \end{array}}
  \end{equation}
  Whenever, $0<|x_\text{out}/c|\leq  t_\text{out}$, this dynamics is further combined with a single, extra ``kick'' at the time $t^*$, i.e.,
  \begin{equation}
  \label{eq:kick}
  \Psi^\text{in-out}(t^*+dt)=  K(t^*)\Psi^\text{in-out}(t^*)
\end{equation}
  where, omitting the explicit tensor products,   
  \begin{equation}
 K(t^*)=\exp{\left[\sqrt{\frac{\Gamma\Delta x}{c}}\mathbb{I}^{\text{in}}_4\left(\tau^{\text{out}}_-\mathbb{I}^{\text{out}}_2\sigma_-[\cdot]-\mathbb{I}^{\text{out}}_2\tau^{\text{out}}_-[\cdot]\sigma_+\right)\right]}\;,
 \end{equation}
for $t^*>0$ and with an extra factor $1/2$ in the exponent for $t^*=0$. Here, $t^*=t^+_\text{out}$ for $x_\text{out}/c<0$ and $t^*=t^-_\text{out}$ for $x_\text{out}/c>0$.

     The solution to Eq.~(\ref{eq:inoutLindblad2}), combined with the kick in Eq.~(\ref{eq:kick}), can then be used to compute the expectation value of the corresponding observable  conditioned on the initial state in Eq.~(\ref{eq:in_cond_markov}) as
    \begin{equation}
    \label{eq:Obreconstruction}
      \renewcommand{\arraystretch}{1.4}{
   \begin{array}{lll}
\langle O^x(t_\text{out})\rangle_\text{in}
&=&\left\langle\phi_1^{\text{out}}\phi_2^{\text{out}}\phi_1^{\text{in}}\phi_2^{\text{in}}\right\rangle\Psi_{00,00}+\Psi_{11,11}\\

&-&\displaystyle\left\langle {\phi}^{\text{out}}_{1}{\phi}^{\text{out}}_{2}\right\rangle\Psi_{11,00}-\left\langle {\phi}^{\text{out}}_{1}{\phi}^{\text{in}}_{1}\right\rangle  \Psi_{01,01}\\

&-&\displaystyle\left\langle {\phi}^{\text{out}}_{1}{\phi}^{\text{in}}_{2}\right\rangle \Psi_{10,01}-\left\langle {\phi}^{\text{out}}_{2}{\phi}^{\text{in}}_{1}\right\rangle  \Psi_{01,10}\\

&-&\displaystyle\left\langle {\phi}^{\text{out}}_{2}{\phi}^{\text{in}}_{2}\right\rangle\Psi_{10,10}-\left\langle {\phi}^{\text{in}}_{1}{\phi}^{\text{in}}_{2}\right\rangle  \Psi_{00,11}\;,
\end{array}}
   \end{equation}
in the system Shr\"{o}dinger picture and where we omitted the fields time dependencies.
   Using the rotating-wave form of the interaction and the zero-temperature assumption impose $\langle {\phi}^{\text{out}}_{1}{\phi}^{\text{in}}_{1}\rangle=\langle {\phi}^{\text{out}}_{2}{\phi}^{\text{in}}_{2}\rangle=0$ and $\langle {\phi}^{\text{out}}_{1}{\phi}^{\text{out}}_{2}\rangle=0$, respectively. 
As a consequence, the only non-zero free correlations are 
\begin{equation}\label{eq:CO_in_out_field_0_quadratic_second}
     \renewcommand{\arraystretch}{1.6}{
   \begin{array}{lll}
\left\langle {\phi}^{\text{out}}_{1}{\phi}^{\text{in}}_{2}\right\rangle &=&\displaystyle\int_{-\infty}^\infty dp \sqrt{g_\text{in}(p)}g_\text{out}(p)e^{ipx_\text{in}+i\omega(p) t_\text{out}}\\

\left\langle {\phi}^{\text{out}}_{2}{\phi}^{\text{in}}_{1}\right\rangle&=&\overline{\left\langle {\phi}^{\text{out}}_{1}{\phi}^{\text{in}}_{2}\right\rangle }\;,
\end{array}}\;,
   \end{equation}
   together with the normalization of the initial state $\left\langle {\phi}^{\text{in}}_{1}{\phi}^{\text{in}}_{2}\right\rangle=1$, and 
\begin{equation}
\left\langle\phi_1^{\text{out}}\phi_2^{\text{out}}\phi_1^{\text{in}}\phi_2^{\text{in}}\right\rangle=\left\langle {\phi}^{\text{out}}_{1}{\phi}^{\text{in}}_{2}\right\rangle \left\langle {\phi}^{\text{out}}_{2}{\phi}^{\text{in}}_{1}\right\rangle=|\left\langle {\phi}^{\text{out}}_{1}{\phi}^{\text{in}}_{2}\right\rangle|^2\;,
\end{equation}
obtained using Eq.~(\ref{eq:CO_in_out_field_0_quadratic_second}) and Wick's theorem. Using the specific expressions for the input and output fields, it is possible to write, more explicitly,
       \begin{equation}
  \renewcommand{\arraystretch}{1.6}{
  \begin{array}{l}
\left\langle {\phi}^{\text{out}}_{1}{\phi}^{\text{in}}_{2}\right\rangle =\displaystyle\sqrt{\frac{{\Delta x}}{\zeta_\text{in}}}\frac{1}{2(2\pi)^{1/4}}\times\\
\left\{e^{-{\Delta x}^2_{-t}/4\zeta^2_{\text{in}}- i{p}_\text{in} \Delta x_{-t}}\left[1- i\text{Erfi}\left(\frac{{\Delta x}_{-t}}{2\zeta_\text{in}}+i\frac{p_\text{in}}{2\sigma_\text{in}}\right)\right]\right.\\
+\left.e^{-{\Delta x}^2_{t}/4\zeta^2_{\text{in}}- i{p}_\text{in} \Delta x_{t}}\left[1+ i\text{Erfi}\left(\frac{\Delta x_{t}}{2\zeta_\text{in}}+i\frac{p_\text{in}}{2\sigma_\text{in}}\right)\right]\right\}\;,
\end{array}}
  \end{equation}
      with ${\Delta x}_t=x_\text{out}-x_\text{in}+c t_\text{out}$
  Using these results, Eq.~(\ref{eq:Obreconstruction}) simplifies to
      \begin{equation}
         \label{eq:ObsLind}\renewcommand{\arraystretch}{1.6}{
   \begin{array}{lll}
\langle O^x(t_\text{out})\rangle_\text{in}&=&\left|\langle {\phi}^{\text{out}}_{1}{\phi}^{\text{in}}_{2}\rangle\right|^2\Psi_{00,00}+\Psi_{11,11}- \Psi_{00,11}\\

\displaystyle
&-&\left\langle {\phi}^{\text{out}}_{1}{\phi}^{\text{in}}_{2}\right\rangle \Psi_{10,01}-\overline{\left\langle {\phi}^{\text{out}}_{1}{\phi}^{\text{in}}_{2}\right\rangle} \Psi_{01,10}\;,
\end{array}}
   \end{equation}
   which completes our derivation. 
   
As a summary, the input-output version of the Lindblad equation presented in this section can be used to analyze a bath prepared in the state $\rho_B=\varphi_1^\text{in}\rho_{\beta_\infty}\varphi_2^\text{in}$, consisting of a single-photon wave packet $\varphi_1^\text{in}$ injected upon an underlying zero-temperature equilibrium state $\rho_{\beta_\infty}$. In this setting, the goal is to compute the population  $O^x(t_\text{out})$ of the bath at position $x_\text{out}$ at the time $t_\text{out}$. For simplicity, we have assumed the system to be located at $x=0$. The bath population's dynamics can be evaluated by extending the Lindblad operator $L_0$ (corresponding to the underlying zero-temperature Markovian open system) to its input-output version 
   \begin{equation}
   \label{eq:L_in_out}
   L^\text{in-out}(t)=\tilde{L}_0+\tilde{L}^\text{in}(t)\;,
   \end{equation}
   where
        \begin{equation}
       \renewcommand{\arraystretch}{1.4}{
     \begin{array}{lll}
     \tilde{L}_0&=&\mathbb{I}^{\text{in}}_4\otimes\mathbb{I}^{\text{out}}_4\otimes L_0\\
\tilde{L}^\text{in}(t)&=&\!\! \mathbb{I}^{\text{in}}_2\otimes\tau^{\text{in}}_-\otimes\mathbb{I}^{\text{out}}_4\otimes D^\text{in}_++ \tau^{\text{in}}_-\otimes\mathbb{I}^{\text{in}}_2\otimes\mathbb{I}^{\text{out}}_4\otimes D^\text{in}_-.
       \end{array}}
  \end{equation}
  As it can be seen from the expression above, the input-output Lindblad operator acts the tensor product of a 4-dimensional input space, a 4-dimensional output space, and the Liouville space for the system, in this specific case, taken to be a two-level system. Interestingly, the infinitesimal dynamics described by the propagator  in Eq.~(\ref{eq:L_in_out}) is trivial in the output space. In fact, the effects in such a space are fully encoded in a single ``kick'', to be operated at a specific time $t^*$ which depends on the time $t_\text{out}$ and space $x_\text{out}$  of the output observation. In other words, assuming $0<|x_\text{out}/c|\leq  t_\text{out}$, the input output Lindblad equation reads
   \begin{equation}
   \begin{array}{llllll}
\label{eq:inoutLindblad3}
\dot{\Psi}^\text{in-out}(t)&=  &L^\text{in-out}(t)~\Psi^\text{in-out}(t)&\text{for}&t\neq t^*\\
  \Psi^\text{in-out}(t+dt)&=  &K(t)~\Psi^\text{in-out}(t)&\text{for}&t= t^*\;,
\end{array}
\end{equation}
where $t^*=t_\text{out}+x_\text{out}/c$, for $x_\text{out}<0$ and  $t^*=t_\text{out}-x_\text{out}/c$, for $x_\text{out}>0$.
The kick operator above is defined as
\begin{equation}
\begin{array}{llllll} K(t)&=&\exp{\left[\sqrt{\Gamma\Delta x/c}\left(\sigma_-[\cdot]-[\cdot]\sigma_+\right)/2\right]}&\text{for}&t=0\\
&=&\exp{\left[\sqrt{\Gamma\Delta x/c}\left(\sigma_-[\cdot]-[\cdot]\sigma_+\right)\right]}&\text{for}&t>0\;.
\end{array}
 \end{equation}
For $|x_\text{out}/c|>  t_\text{out}$, no additional kicks are needed to complement the dynamics. The solution of Eq,~(\ref{eq:inoutLindblad3}) contains all the information needed to reproduce the value of the population $\langle O^ x\rangle$ through Eq.~(\ref{eq:ObsLind}). 

It is important to point out that  this model can be analytically solved within the single-excitation sector of the open system, see Appendix \ref{sec:analytical}. In fact, the main purposes of this derivation is to exemplify the overall formalism, to show the possibility to use it to derive an input-output extension of the regular Lindblad equation, and to provide a numerical benchmark.  However, we also note that the formalism presented in this article, including this Markovian case, is designed to go beyond several of the approximations upon which such an analytical treatment critically relies such as, for example, allowing Rabi-like system-bath interactions, colored, i.e. non-white, spectral properties,  multiple initial wave-packets, arbitrary initial condition on the system, and general output bath-observables. 

In Fig.~\ref{fig:markovian}, we provide a further illustration to the potential applications of the method and also benchmark its validity by presenting an explicit numerical solution of the input-output Lindblad equation in Eq.~(\ref{eq:inoutLindblad}). Specifically, we plot the dynamics of the real-space probability density, see Eq.~(\ref{eq:Obs_+}), for a 1-dimensional  Markovian environment injected with the single-photon wave-packet, as in Eq.~(\ref{eq:in_cond_markov}). We compare the input-output HEOM result with the analytical result, see Appendix \ref{sec:analytical} for a derivation. The plot shows that both models agree in predicting the scattering distribution of the single-photon wave-packet as it interacts with the system. This model can be used to model open systems beyond the Markovian approximation by simply revisiting the definitions of the spectral density and dispersion of the bath which, in turn, corresponds to an upgrade both the expressions for the correlations involving the environmental coupling operator (to replace the Lindblad operator $L_0$ in Eq.~(\ref{eq:L_in_out}) with the regular-HEOM kernel) and the expressions for the correlations involving the environmental coupling operator and the input-output fields. 
To generalize the model further to allow, for example, Rabi-like system-bath interactions and multiple initial wave-packets, it is always possible to consider the input-output HEOM in Eq.~(\ref{eq:HEOM_extended_main_short}).

   This section presented an input-output version of the Lindblad equation which allows to compute both the system dynamics and bath observables when a  Markovian environment is prepared in a non-equilibrium input state. Because of the reliance on the same formalism as the input-output HEOM, this construction can be generalized to different cases including Rabi-like system-bath interactions, colored spectral interactions, or iteratively applied to more complex input states of output observables.  While the explicit expression are provided at zero temperature, the same formalism can be used in terms of a finite temperature Lindblad equation and by evaluating the correlations on the corresponding finite temperature underlying equilibrium state for the bath.
 
\section{Conclusions}
\label{sec:Conclusions}
We showed that input and output properties of a bosonic open quantum system encoded in environmental correlations can be formally decomposed, without approximation, in terms of a finite series whose evaluation can be computed by solving an extended version of the regular HEOM.  In other words, these input-output HEOM can
be used to track the expectation values of bath observables as the environment interacts with a system in a non-Markovian regime  while also allowing to analyze 
 the case in which the environment is prepared in a non-Gaussian initial state. Combining these two settings, it is possible to  compute bath observables as a quantum system interacts with a bath prepared in a non-Gaussian state. 
We further showed that, in the Markovian limit, these HEOM correspond to an input-output extension of the Lindblad equation. As a consequence, this formalism can find applications to upgrade the modeling of general Lindblad systems whenever the inclusion of wave-packets in the bath or the computation of the dynamics of bath observables is desirable.
The complexity of the extended HEOM depends on whether the environmental property under analysis is to be tracked along the dynamics or at a fixes time. For dynamical properties characterized by environmental fields (such as those defining output observables), the range of the indexes extending the HEOM depends on  the number of exponentials in the spectral representation of cross-correlations between the fields and bath coupling operators describing the open system. For  static properties (such as those defining a non-Gaussian initial state  or an output observable at a fixed time), no spectral ansatz is required and the HEOM is extended with additional binary index and  a time-dependent contribution. 
 Specifically, the dynamics of output observables can be modeled either dynamically, i.e., as a function of time while imposing a spectral ansatz, or statically, i.e., for a fixed time, without additional spectral assumptions, leading to an interesting trade-off between computational resources.

Overall, the extended HEOM acquire an extra complexity (with respect to their regular version) scaling with the intricacy of the input-output task, such as the deviation of the statistics of the initial state from Gaussianity. 

As an outlook, it would be interesting to compare this approach with other methodologies used to model non-Gaussian open quantum systems such as \cite{PhysRevLett.132.170402}.
In parallel, as noted in Appendix \ref{app:a_generalization}, the formalism presented here can be adapted to compute correlation among fields with non-trivial support on the system. This might allow to optimize the modeling of time-delays as different subsystems are coupled to the same environment.

\section{Acknowledgments.}
M.C. acknowledges support from NSFC (Grant No. 11935012) and NSAF (Grant No. U2330401).
N.L. is supported by the RIKEN Incentive Research Program and by MEXT KAKENHI Grants No. JP24H00816 and No. JP24H00820.

\newpage
\appendix

In this appendix, we present details about the derivations in the main text. 
We start by reviewing the most relevant symbols used throughout the article in the table \ref{table:symbols}.
\renewcommand{\arraystretch}{1.9}{
\begin{table}[h!]
\begin{tabularx}{.5\textwidth}{|c|X|}
\hline
\textbf{Symbol}          & {\thead{\textbf{Description}}} \\
\hline
$\text{Tr}$, $\text{Tr}_S$, $\text{Tr}_B$& Trace over the system+bath, the system, and the bath, respectively.\\
\hline
$\mathcal{T}$, $\mathcal{T}_S$, $\mathcal{T}_B$& Time-ordering for superoperators (in the overall system+bath, the system, and the bath, respectively) in the interaction picture.\\
\hline
$H^{\text{I}}(t)$& Interaction Hamiltonian operator.\\
\hline
$N^H_{\text{I}}$& Number of terms defining the interaction Hamiltonian operator.\\
\hline
$\mathcal{H}^{\text{I}}_t$&Interaction Hamiltonian superoperator.\\
\hline
$N_{\text{I}}$&Number of terms defining the interaction Hamiltonian superoperator.\\
\hline
$\mathcal{S}^\alpha_t$&  System interaction superoperator, $\alpha=1,\cdots,N_{\text{I}}$.\\
\hline
$\chi^\alpha_t$&  Bath interaction superoperator, $\alpha=1,\cdots,N_{\text{I}}$.\\
\hline
$\mathcal{F}_t$&  Influence superoperator.\\
\hline
$\sigma$&  Multi-index $\sigma=(\alpha,\beta,k)$, for $\alpha,\beta=1,\cdots,N_{\text{I}}$, and $k=1,\cdots,N_{\text{exp}}$.\\
\hline
${\phi}_j(t_j)$&Field superoperator in the interaction picture, with   $j\in J=\{1,\cdots,m\}$. \\
\hline
${{\phi}_j^{\text{out}}}(t^{\text{out}})$&Output field superoperator: ${{\phi}_j^{\text{out}}}(t^{\text{out}})\equiv{\phi}_j(t)$ for $j\in J_{\text{out}}$.\\
\hline
${{\phi}_j^{\text{in}}}(t^{\text{in}})$&Input field superoperator: ${{\phi}_j^{\text{in}}}(t^{\text{in}})\equiv{\phi}_j(0)$ for $j\in J_{\text{in}}$.\\
\hline
$\tilde{\mathcal{Y}}^{j}_{\tau}$&HEOM superoperator for $j\not\in J_{\text{out}}$, $\tilde{\mathcal{Y}}^{j}_{\tau}=\sum_{\alpha'} \langle{{\phi}_{j}}(t_j)\chi^{\alpha'}_\tau\rangle\mathcal{S}^{\alpha'}_\tau$\\
\hline
$\langle[\cdot]\rangle$&Trace over the environment:  $\langle[\cdot]\rangle\equiv\text{Tr}_B\left\{\mathcal{T}_B[\cdot]\rho_B\right\}$.\\
\hline
$M$& Set $M=\{1,\cdots,m\}$, $m> 0$.\\
\hline
$[m/2]$, $m>0$ &Closest integer to $m/2$ from below: $[m/2]=m/2$ for even $m$ and $(m-1)/2$ for odd $m$.\\
\hline
 $\mathfrak{m}(m,2k)$& Collection of sets $\mathfrak{a}$ made by choosing $2k$ elements from $M$.\\
 \hline
 $\mathfrak{a}_{\text{c}}$& Complement of a set $\mathfrak{a}$ in $M$.\\
 \hline
 $\vec{n}^{\text{out}}$( $\vec{n}^{\text{in}}$)& Vector of $n^{\text{out}}_{j \alpha k}$($n^{\text{in}}_{j}$) augmenting the indexes of the auxiliary density matrices for output (input) fields, whose total number is $N^{{\text{out}}}=\sum_{j,\alpha,k}n^{\text{out}}_{j\alpha k}$ ($N^{{\text{in}}}=\sum_{j}n^{\text{in}}_{j}$).\\
 \hline
\end{tabularx}
\caption{List of symbols \label{table:symbols}}
\end{table}
}

\section{Open Quantum Systems}
\label{sec:app:openquantumsystems}
Here, we provide details about the formal considerations presented in section \ref{sec:Introduction}.  For an open system described by the interaction Hamiltonian superoperator in Eq.~(\ref{eq:sup_int_H}), the influence superoperator can be written as
\begin{equation}
\label{eq:basedone}
\begin{array}{lll}
    \mathcal{F}&=&-\displaystyle\int_0^t dt_2\int_0^{t_2} dt_1\sum_{\alpha_2,\alpha_1}\text{Tr}_B\left[\chi^{\alpha_2}_{t_2}\chi^{\alpha_1}_{t_1}\rho_B\right]\mathcal{S}^{\alpha_2}_{t_2}\mathcal{S}^{\alpha_1}_{t_1}\\

&=&\displaystyle-\int_0^t dt_2\sum_{\alpha_2}\mathcal{S}^{\alpha_2}_{t_2}\int_0^{t_2} dt_1\sum_{\alpha_1}C^{\alpha_2\alpha_1}_{t_2,t_1}\mathcal{S}^{\alpha_1}_{t_1}\\

&\equiv&\displaystyle\int_0^t dt_2\sum_{\alpha_2}\mathcal{A}^{\alpha_2}_{t_2}\int_0^{t_2} dt_1\sum_{\alpha_1}C^{\alpha_2\alpha_1}_{t_2,t_1}\mathcal{B}^{\alpha_2\alpha_1}_{t_1}\;,
    \end{array}
\end{equation}
which corresponds to Eq.~(\ref{eq:F_rep}).
It is interesting to further find the explicit dependence of this notation in terms of the system and bath operators defining the interaction Hamiltonian for the open system.  In this case, the index $\alpha$ in the expression above can be interpreted as a double-index $\alpha\mapsto(p,q)$, in which $p=1,2$ (corresponding to  $p=1=\text{l}, p=2=\text{r}$ in the notation of section \ref{sec:Introduction}) and $q=1,\cdots,N_I^H$ as in Eq.~\ref{eq:HI}, so that
\begin{equation}
\label{eq:S_chi_def}
\begin{array}{lllllll}
\chi_t^{(1,q)}[\cdot]&=&X^q_t[\cdot]\;,&\mathcal{S}_t^{(1,q)}[\cdot]&=& s^q_t[\cdot]\\
\chi_t^{(2,q)}[\cdot]&=&[\cdot]X^q_t\;,&\mathcal{S}_t^{(2,q)}[\cdot]&=&-[\cdot] s^q_t\;.
\end{array}
\end{equation}
Using these definitions, we can write
\begin{equation}
\label{eq:full_corr}
\begin{array}{lll}
C^{\alpha_2\alpha_1}_{t_2,t_1}&=&\text{Tr}_B\left[\chi^{(p_2,q_2)}_{t_2}\chi^{(p_1,q_1)}_{t_1}\rho_B\right]\\
&=&\displaystyle\delta_{p_1,1} \tilde{C}^{q_2q_1}_{t_2,t_1}+\delta_{p_1,2}  \tilde{C}^{q_1q_2}_{t_2,t_1}\;,
\end{array}
\end{equation}
where
\begin{equation}
    \tilde{C}^{q_2q_1}_{t_2,t_1}=\text{Tr}_B\left[X^{q_2}{(t_2)}X^{q_1}{(t_1)}\rho_B\right]\;,
\end{equation}
is correlation matrix written for the original operators $X^q(t)$.
We can now note the formal asymmetry between the indexes $p_1$ and $p_2$ on the right hand-side of the expression in Eq.~(\ref{eq:full_corr}). This is a direct consequence of the fact that the order in which the operators $X^q$ are applied critically depends on whether the left-most superoperator $\chi$ on the left hand-side, acts on the left of right of $\rho_B(0)$. With this notation, we can write Eq.~(\ref{eq:basedone}) as
\begin{equation}
\label{eq:basedone_2}
\begin{array}{lll}
    \mathcal{F}&=&\displaystyle-\int_0^t dt_2\sum_{q_2}\left[\sum_{p_2}\mathcal{S}^{p_2,q_2}_{t_2}\right]\sum_{j_1}\int_0^{t_2} dt_1\\
    &&\displaystyle
    \left[ \tilde{C}^{q_2q_1}_{t_2,t_1}\mathcal{S}^{1,j_1}_{t_1}+ \tilde{C}^{q_1q_2}_{t_2,t_1}\mathcal{S}^{2,q_1}_{t_1}\right]\;.
    \end{array}
\end{equation}
For the specific case in which $N^H_{\text{I}}=1$, i.e., only a single term is present in the interaction Hamiltonian in Eq.~(\ref{eq:HI}), there is no need to introduce the extra index $q$ in Eq.~(\ref{eq:S_chi_def}) which then becomes
\begin{equation}
    \begin{array}{lllllll}
        \chi^{p=1}_t&=&X(t)\cdot,&~~~~&\mathcal{S}^{p=1}_t&=& s_t\cdot,\\
        \chi^{p=2}_t&=&\cdot X(t),&~~~~&\mathcal{S}^{p=2}_t&=&-\cdot  s_t.\\
    \end{array}
\end{equation}
and
\begin{equation}
\begin{array}{lll}
C^{p_1,p_2}_{t_2,t_1}&=&\text{Tr}_B\left[\chi^{p_2}_{t_2}\chi^{p_1}_{t_1}\rho_B\right]\\
&=&\delta_{p_1,1} \tilde{C}(t_2,t_1)+\delta_{p_1,2} \tilde{C}(t_1,t_2)\\

&=&\delta_{p_1,1} \tilde{C}(t)+\delta_{p_1,2} \tilde{C}(-t)\;,
\end{array}
\end{equation}
where 
\begin{equation}
\tilde{C}(t_2,t_1)=\text{Tr}_B[X_{t_2} X_{t_1}\rho_B]\;,
\end{equation}
and where we further assumed stationarity, i.e., $\tilde{C}(t_2,t_1)=\tilde{C}(t)$, in terms of the time difference $t=t_2-t_1$.

 In the following section, we are going to use these results to explicitly write two equivalent representations of the influence superoperator which, in turn, will lead to different representations for the HEOM.

\section{HEOM representations}
\label{sec:HEOMrepresentations}
In this section, we consider two explicit representation for the HEOM in Eq.~(\ref{eq:sX}) for the open system described by the Hamiltonian in Eq.~(\ref{eq:sX}), i.e., involving a single pair $s_t$ and $X_t$ of Hermitian coupling operators for the system and the environment, respectively. 
In this case, it is possible to write the influence superoperator in Eq.~(\ref{eq:F_rep}) in, for example, two different ways to highlight either the real-imaginary structure of the correlation, see Eq.~(\ref{eq:F2}), or its causal structure, see Eq.~(\ref{eq:F1}). These representations can then be used to derive alternative forms for the HEOM as in Eq.~(\ref{eq:HEOM_Neill}) and Eq.~(\ref{eq:HEOM_Yan_2}), which are analyzed in depth  in, for example,  \cite{Lambert_Bofin} and \cite{Yan_HEOM,FreePoles}. We present the derivation for these two cases in the next two subsections.
\subsection{``Causal'' representation}
Here, we use the term ``causal'', to denote a representation which highlights the difference between positive- and negative-time contributions. 
Using the formalism and the assumptions presented in the previous section, we can write the influence superoperator  in Eq.~(\ref{eq:basedone_2}) as 
\begin{equation}
\label{eq:F1}
\begin{array}{lll}
\mathcal{F}_t&=&\displaystyle-\int_0^t dt_2\left[\sum_{\alpha}\mathcal{S}^\alpha_{t_2}\right]\int_0^{t_2} dt_1\\
&&\times\left[C(t_2-t_1)\mathcal{S}^1_{t_1}+\bar{C}(t_2-t_1)\mathcal{S}^2_{t_1}\right],
\end{array}
\end{equation}
where we used that $C(-\tau)=\bar{C}(\tau)$ for all $\tau$ (in which the overbar implies complex conjugation) which follows from the fact that both $ s$ and $X$ are Hermitian (for $N=1$). 
Comparing Eq.~(\ref{eq:F1}) with Eq.~(\ref{eq:F_rep}), we have $\alpha\rightarrow1$, i.e., $\alpha$ no longer represents an index,  $\beta\rightarrow\pm$, $\sigma=(\alpha,\beta,k)\rightarrow(\beta,k)$, $\mathcal{B}^{\alpha\beta}\rightarrow\mathcal{B}^\beta$, $N_{\alpha\beta}\rightarrow N_{\beta}$, and
\begin{equation}
\begin{array}{lll}
\mathcal{A}^{\alpha=1}_t&\rightarrow&-\sum_{\alpha_2=1}^2\mathcal{S}^{\alpha_2}\equiv- s^X_t[\cdot]\\
\mathcal{B}^+_t&\rightarrow&\mathcal{S}^{\alpha_1=1}\\
\mathcal{B}^-_t&\rightarrow&\mathcal{S}^{\alpha_1=2}\\
D^{\alpha\beta}_{t_2,t_1}&\rightarrow&D^\beta(t_2,t_1)\\
D^{\beta=+}(t_2,t_1)&\equiv&C(t_2-t_1)\equiv\sum_{k}c_k^{+}e^{-\gamma_k^{+}t}\\
D^{\beta=-}(t_2,t_1)&\equiv&\bar{C}(t_2-t_1)\equiv\sum_{k}c_k^{-}e^{-\gamma_k^{-}t}\;,
\end{array}
\end{equation}
where $ s^X_t[\cdot]\equiv s_t[\cdot]-[\cdot] s_t$ and such that, consistently, $c_k^{+}=\bar{c}_k^{-}$, and $\gamma_k^{+}=\bar{\gamma}_k^{-}$. Using this formalism in Eq.~(\ref{eq:HEOM_Yan}), and imposing $\alpha_0\rightarrow -i$, we obtain
\begin{equation}
\label{eq:HEOM_Yan_2}
    \begin{array}{lll}
\dot{\rho}^{(N)}_{n^{ + }_{1},\cdots,n^{ + }_{N_ + },n^{ - }_{1},\cdots,n^{ - }_{N_ - }}(t)\\

=\displaystyle -\left(\sum_{k=1}^{N_ + } n^ + _k \gamma^ + _k+\sum_{k=1}^{N_ - } n^ - _k \gamma^ - _k\right)\rho^{(N)}_{n^{ + }_{1},\cdots,n^{ + }_{N_ + },n^{ - }_{1},\cdots,n^{ - }_{N_ - }}(t)\\

\displaystyle-i\left[\sum_{k}^{N_ + } \sqrt{n^ + _k c^ + _k}  s_t\rho^{(N)}_{n^{ + }_{1},\cdots,n^{ + }_{k}-1,\cdots,n^{ + }_{N_ + },n^{ - }_{1},\cdots,n^{ - }_{N_ - }}(t)\right]\\ 

\displaystyle--i\left[\sum_{k}^{N_ - } \sqrt{n^ - _k c^ - _k} \rho^{(N)}_{n^{ + }_{1},\cdots,n^{ + }_{N_ + },n^{ - }_{1},\cdots,n^{ - }_{k}-1,\cdots,n^{ - }_{N_ - }}(t) s_t\right]\\

\displaystyle-i\sum_{k}^{N_ + } \sqrt{(n^ + _k +1) c^ + _k}  s_t^X\rho^{(N)}_{n^{ + }_{1},\cdots,n^{ + }_{k}+1,\cdots,n^{ + }_{N_ + },n^{ - }_{1},\cdots,n^{ - }_{N_ - }}(t)\\

\displaystyle-i\sum_{k}^{N_ - } \sqrt{(n^ - _k + 1) c^ - _k}  s_t^X\rho^{(N)}_{n^{ + }_{1},\cdots,n^{ + }_{N_ + },n^{ - }_{1},\cdots,n^{ - }_{k}+1,\cdots,n^{ - }_{N_ - }}(t),
    \end{array}
\end{equation}
as in Eq.~(4) in the supplemental of \cite{FreePoles}.

\subsection{``Real-structure'' representation}
Here, we use the term ``real-structure'', to denote a representation which highlights a real structure in the correlations complex plane. Specifically, we write the influence superoperator in Eq.~(\ref{eq:basedone_2}) as
\begin{equation}
\label{eq:F2}
\begin{array}{lll}
F&=&-\displaystyle\int_0^t dt_2\left[\sum_{\alpha}\mathcal{S}^\alpha_{t_2}\right]\int_0^{t_2} dt_1\left\{C_{\text{R}}(t_2-t_1)\left[\mathcal{S}^1_{t_1}+\mathcal{S}^2_{t_1}\right]\right.\\
&&+\displaystyle\left. i C_{\text{I}}(t_2-t_1)\left[\mathcal{S}^1_{t_1}-\mathcal{S}^2_{t_1}\right]\right\}\;,
\end{array}
\end{equation}
 where the indexes R and I indicate real an imaginary part, respectively. 
Comparing Eq.~(\ref{eq:F2}) with Eq.~(\ref{eq:F_rep}), we have $\alpha\rightarrow1$, i.e., $\alpha$ no longer represents an index,  $\beta\rightarrow\text{R},\text{I}$, $\sigma=(\alpha,\beta,k)\rightarrow(\beta,k)$,  $\mathcal{B}^{\alpha\beta}\rightarrow\mathcal{B}^\beta$, $N_{\alpha\beta}\rightarrow N_{\beta}$, and
\begin{equation}
\label{eq:ri}
\begin{array}{lll}
\mathcal{A}^{\alpha=1}_t&\rightarrow&-\sum_{\alpha_2=1}^2\mathcal{S}^{\alpha_2}\equiv- s^X_t[\cdot]\\
\mathcal{B}^{\text{R}}_t&\rightarrow&\mathcal{S}^1(t_1)+\mathcal{S}^2(t_1)\\
\mathcal{B}^{\text{I}}_t&\rightarrow&i\left[\mathcal{S}^1(t_1)-\mathcal{S}^2(t_1)\right]\\
D^{\alpha\beta}_{t_2,t_1}&\rightarrow&D^\beta(t_2,t_1)\\
D^{\beta=\text{R}}(t_2,t_1)&\equiv&C_{\text{R}}\equiv\sum_{k}c_k^{\text{R}}e^{-\gamma_k^{\text{R}}t}\\
D^{\beta=\text{I}}(t_2,t_1)&\equiv&C_{\text{I}}\equiv\sum_{k}c_k^{\text{I}}e^{-\gamma_k^{\text{I}}t}\;.
\end{array}
\end{equation}
Using Eq.~(\ref{eq:ri}) into Eq.~(\ref{eq:HEOM}) for $\alpha=-i$, we get
\begin{equation}
\label{eq:HEOM_Neill}
    \begin{array}{lll}
\dot{\rho}^{(N)}_{n^{\text{R}}_{1},\cdots,n^{\text{R}}_{N_{\text{R}}},n^{\text{I}}_{1},\cdots,n^{\text{I}}_{N_{\text{I}}}}(t)\\

=\displaystyle -\left(\sum_{k=1}^{N_{\text{R}}} n^{\text{R}}_k \gamma^{\text{R}}_k+\sum_{k=1}^{N_{\text{I}}} n^{\text{I}}_k \gamma^{\text{I}}_k\right)\rho^{(N)}_{n^{\text{R}}_{1},\cdots,n^{\text{R}}_{N_{\text{R}}},n^{\text{I}}_{1},\cdots,n^{\text{I}}_{N_{\text{I}}}}(t)\\

\displaystyle-i\left[\sum_{k}^{N_{\text{R}}} n^{\text{R}}_k c^{\text{R}}_k  s^X(t)\rho^{(N)}_{n^{\text{R}}_{1},\cdots,n^{\text{R}}_{k}-1,\cdots,n^{\text{R}}_{N_{\text{R}}},n^{\text{I}}_{1},\cdots,n^{\text{I}}_{N_{\text{I}}}}(t)\right]\\ 

\displaystyle-ii\left[\sum_{k}^{N_{\text{I}}} n^{\text{I}}_k c^{\text{I}}_k  s^\circ(t)\rho^{(N)}_{n^{\text{R}}_{1},\cdots,n^{\text{R}}_{N_{\text{R}}},n^{\text{I}}_{1},\cdots,n^{\text{I}}_{k}-1,\cdots,n^{\text{I}}_{N_{\text{I}}}}(t)\right]\\

\displaystyle-i\sum_{k}^{N_{\text{R}}}  s^X\rho^{(N)}_{n^{\text{R}}_{1},\cdots,n^{\text{R}}_{k}+1,\cdots,n^{\text{R}}_{N_{\text{R}}},n^{\text{I}}_{1},\cdots,n^{\text{I}}_{N_{\text{I}}}}(t)\\

\displaystyle-i\sum_{k}^{N_{\text{I}}}  s^X\rho^{(N)}_{n^{\text{R}}_{1},\cdots,n^{\text{R}}_{N_{\text{R}}},n^{\text{I}}_{1},\cdots,n^{\text{I}}_{k}+1,\cdots,n^{\text{I}}_{N_{\text{I}}}}(t)\;,
    \end{array}
\end{equation}
as in Eq.~(11) in \cite{Lambert_Bofin}.

\section{Environmental correlations: formal series}
\label{sec:BathCorrelations}
In this section, we consider a set of $m$ environmental superoperators ${{\phi}}_1(t_1), \cdots, {{\phi}}_m(t_m)$ written in the interaction picture. In the following, we will also use the notation ${{\phi}}_j(t_j)$ with  $j\in J=\{1,\cdots,m\}$. We assume these superoperators to be defined using operators with support on the bath which are linear in the bosonic creation and annihilation bath operators. In other words, we assume that these superoperators satisfy the superoperator version of the Wick's theorem. Our goal is to compute
\begin{equation}
\rho^m_S(t)=\text{Tr}_B\left\{\mathcal{T}\left[\prod_{j=1}^m{{\phi}}_j(t_j) \right] e^{-i\int_0^t d\tau\;\mathcal{H}_\tau^{\text{I}}}\rho_S(0)\rho_B\right\}\;.
\end{equation}
Using Eq.~(\ref{eq:sup_int_H}), we can write
\begin{equation}
\label{eq:rhoSm}
\rho^{m}_S(t)=\sum_{n=0}^\infty\frac{(-i)^n}{n!}\rho_n^{m}\;,
\end{equation}
where
\begin{equation}
\label{eq:rhomn}
\rho_n^{m}\equiv \int_0^t d\tau_1\cdots\int_0^t d\tau_n\;\sum_{\vec{\alpha}}C^{m}_{\vec{\alpha};n} \mathcal{T}_S\mathcal{S}^{\alpha_1}_{\tau_1}\cdots\mathcal{S}^{\alpha_n}_{\tau_n}\rho_S(0)\;,
\end{equation}
in terms of the indexes $\vec{\alpha}=(\alpha_1,\cdots,\alpha_n)$, and
\begin{equation}
\label{eq:rhomn_corr}
C^{m}_{\vec{\alpha};n}=\text{Tr}_B\left\{\mathcal{T}_B{{\phi}}_1(t_1) \cdots{{\phi}}_m(t_m)  \chi^{\alpha_1}_{\tau_1}\cdots\chi^{\alpha_n}_{\tau_n}\rho_B\right\}\;.
\end{equation}
Importantly, we have introduced $\mathcal{T}_S$ and $\mathcal{T}_B$ as the time ordering acting on superoperators on the system and the bath, respectively. Because of these assumptions, computing a more explicit version of this expression is just a matter of an iterative application of Wick's theorem, i.e., a combinatorial problem. To simplify the notation, we now define 
\begin{equation}
\langle[\cdot]\rangle\equiv\text{Tr}_B\left\{\mathcal{T}_B[\cdot]\rho_B\right\}\;.
\end{equation}
We also assume that $\rho_B$ is quadratic so that all expectations containing an odd number of creation or annihilation bosonic operators are zero. This implies that $C^m_{\vec{\alpha};n}=0$ whenever $n+m$ is odd. Assuming then $(n+m)$ to be even, we introduce a parameter $k\geq 0$ to label the number of contractions between the field superoperators ${{\phi}}_j(t_j)$ originating when applying Wick's theorem to  $C^m_{\vec{\alpha};n}$. Given $m$, the idea is to decompose  the expansion of $C^m_{\vec{\alpha};n}$ into (i) components containing $k$ contractions between $2k$ superoperators ${{\phi}}$, (ii) $(m-2k)$ contractions between $(m-2k)$ superoperators ${{\phi}}$ and $(m-2k)$ superoperators $\chi$, and (iii) $[n-(m-2k)]/2$ contractions between $[n-(m-2k)]$  superoperators $\chi$. Graphically, we can represent the terms in this decomposition as
\begin{equation}
\label{eq:graphical}
C^{m}_{\vec{\alpha};k;n}=\langle\underbrace{{{\phi}}_1(t_1) \cdots}_{2k}\tallvdots\underbrace{\cdots{{\phi}}_m(t_m)}_{m-2k}  \underbrace{\chi^{\alpha_1}_{\tau_1}\cdots}_{m-2k}\tallvdots\underbrace{\cdots\chi^{\alpha_n}_{\tau_n}}_{n-(m-2k)}\rangle\;.
\end{equation}
It is important to note that the quantity graphically represented above, is actually a sum of all terms compatible with the decomposition. To gather more intuition on these terms, we note that the decomposition above implies the constraint $(m-2k)\geq 0$. If $n\geq m$, this is the only constraint. However, if $n<m$, then there might not be enough $\chi$ to contract with, when $k$ is too large. In this case, we need to further  impose $(m-2k)\leq n$. Since this equation is always satisfied when $n\geq m$ because $k\geq 0$, we can write, in all cases
\begin{equation}
n\geq m-2k\geq 0\;.
\end{equation}
This implies that
\begin{equation}
\begin{array}{llll}
k&=&\text{max}[0,(m-n)/2],\cdots,m/2\;,&\text{for}\;m\;\text{even}\\
k&=&\text{max}[0,(m-n)/2],\cdots,(m-1)/2\;, &\text{for}\;m\;\text{odd}\;,
\end{array}
\end{equation}
which we condense into
\begin{equation}
k=\text{max}[0,(m-n)/2],\cdots,[m/2]\;,
\end{equation}
where $[m/2]\equiv m/2$ for even $m$ and $(m-1)/2$ otherwise. We can now use these constraints to push forward a logic in which, after fixing $m$, we choose $k$ and only then choose $n$. This can be done by writing 
\begin{equation}
\label{eq:Cmn}
C^{m}_{\vec{\alpha};n}=\sum_{k=0}^{[m/2]}\theta[n-(m-2k)]C^{m}_{\vec{\alpha};k;n}\;,
\end{equation}
and the corresponding integrated version which is obtained by inserting Eq.~(\ref{eq:Cmn}) into Eq.~(\ref{eq:rhomn}). We note that the presence of the theta function in the expressions above, allows  to treat the sum over $k$ as independent from the sum over $n$, in turn introducing the possibility to conceptually fix $k$ before $n$. In the context of this section, this is an important point, since we are ultimately interested to formally \emph{re-sum over $n$ given $m$ and $k$}. Furthermore, we also note that we stressed the computation of the integrated version of the coefficients because the corresponding presence of dummy variables will introduce non-trivial simplifications.

We now go back to Eq.~(\ref{eq:graphical}), to give a more precise quantification of the process we are considering. Mainly, the number of elements in the sum implicit in Eq.~(\ref{eq:graphical}), can be calculated as follows. First, as mentioned, we choose a set of $2k$ elements from the list of $m$ superoperators ${{\phi}}$. This corresponds to $\binomial{m}{2k}$ choices. For each of these choices, the correlation involving $2k$ superoperators ${{\phi}}$ can be further fully contracted leading to an extra $(2k-1)!!$ terms. We are then left with $m-2k$ superoperators ${{\phi}}$ which needs to be contracted with the superoperators $\chi$. We then have $\#_{{{\phi}}\chi}(m;k)$ ways to choose $(m-2k)$ elements from the product $\chi_{\tau_1}^{\alpha_1}\cdots\chi_{\tau_n}^{\alpha_n}$. Note that such a choice is ordered because it corresponds to the order in which the superoperators ${{\phi}}$ appear. We are then left with a product of $n-(m-2k)$ superoperators $\chi$ which can be contracted in $\#_{{{\phi}}{{\phi}}}(m;k)$ ways. Explicitly, these combinatorial numbers are
\begin{equation}
\begin{array}{lll}
\#_{{{\phi}}\chi}(m;k)&=&\displaystyle\frac{n!}{(n-m+2k)!}\\
\#_{\chi\chi}(m;k)&=&\displaystyle(n-m+2k-1)!!\\
&=&\displaystyle\frac{(n-m+2k)!}{2^{(n-m+2k)/2}[(n-m+2k)/2]!}\;.
\end{array}
\end{equation}
As a check for these combinatorial expressions, we can consider that the sum over $k$ of the product of all of them must corresponds to the total number of contractions for the full set of $m+n$ operators. We then need to check that
\begin{equation}
(m+n-1)!!=\sum_{k=k_0}^{[m/2]}\binomial{m}{2k}\frac{(2k)!}{2^k k!}\#_{{{\phi}}\chi}(m;k)\#_{\chi\chi}(m;k)\;,
\end{equation}
where $k_0=\text{max}[0,(m-n)/2]$ and where we used that $(2k-1)!!=\frac{(2k)!}{2^k k!}$. The identity above can be checked numerically.
To proceed, we define $\mathfrak{m}(m,2k)$ as the set of all possible lists of $2k$ elements chosen from $\{1,\dots,m\}$.  For each of the lists $\mathfrak{a}\in \mathfrak{m}(m,2k)$, we define $\mathfrak{a}_{\text{c}}$ as its complement in $\mathfrak{m}(m,2k)$, i.e., the list made by the elements in $\mathfrak{m}(m,2k)$ which are not in $a$. With these definitions, we can finally write Eq.~(\ref{eq:rhoSm}) as

   \begin{equation}
   \begin{array}{l}
\rho^{m}_S  =\displaystyle \sum_{k=0}^{[m/2]}\sum_{n=0}^{\infty}\theta[(n-(m-2k))](-i)^{(m-2k)}\sum_{\mathfrak{a}\in \mathfrak{m}(m,2k)}\\
   \displaystyle\left\langle\prod_{i\in \mathfrak{a}}{{\phi}}_i(t_i)\right\rangle\mathcal{T}_S\prod_{j\in \mathfrak{a}_{\text{c}} }\int_0^t d\tau\sum_{\alpha}\langle{{\phi}}_{j}(t_j)\chi^\alpha_\tau\rangle\mathcal{S}^\alpha_\tau\\
   \displaystyle{(-1)}^{(n-m+2k)/2}\frac{\mathcal{G}_t^{(n-m+2k)/2}\rho_S(0)}{2^{(n-m+2k)/2}[(n-m+2k)/2]!}\;,
      \end{array}
   \end{equation}
   where
   \begin{equation}
   \mathcal{G}_t=\int_0^t d t_2 \int_0^t d t_1 \sum_{\alpha_2,\alpha_1}\langle \chi^{\alpha_2}_{t_2}\chi^{\alpha_1}_{t_1}\rangle\mathcal{S}_{t_2}^{\alpha_2}\mathcal{S}_{t_1}^{\alpha_1}\;,
   \end{equation}
   and   where we used that 
   \begin{equation}
      \begin{array}{lll}
(-i)^n&=&[(-i)^2]^{[n/2-(m-2k)/2]+(m-2k)/2}\\
&=&  -1^{[n-(m-2k)]/2}(-i)^{(m-2k)}\;.
      \end{array}
   \end{equation}
   It is important to note that the presence of the sums over $\alpha$ and the integrals are what allowed, once $k$ has been fixed, all the remaining choices to lead to equivalent factors so that we could factorize them. Even further, the full sum over $n$ can now be factorized in front. In fact, we can now define $\tilde{n}=(n-m+2k)/2$ which is an integer since $n+m$ is even. Furthermore,  the theta function imposes $\tilde{n}=0,\cdots,\infty$ so that
      \begin{equation}
      \label{eq:rhomS_app}
   \begin{array}{lll}
\rho^{m}_S(t)  &=&\displaystyle \sum_{k=0}^{[m/2]}\sum_{\mathfrak{a}\in \mathfrak{m}(m,2k)}\left\langle\prod_{i\in \mathfrak{a}}{{\phi}}_i(t_i)\right\rangle(-i)^{(m-2k)}\\
     &&\displaystyle\times\mathcal{T}_S\prod_{j\in \mathfrak{a}_{\text{c}} }\int_0^t d\tau\sum_{\alpha}\left\langle{{\phi}}_{j}(t_j)\chi^\alpha_\tau\right\rangle\mathcal{S}^\alpha_\tau e^{\mathcal{F}_t}\rho_S(0)\;,
      \end{array}
   \end{equation}
   where
      \begin{equation}
   \mathcal{F}_t=-\int_0^t d t_2 \int_0^{t_2} d t_1 \sum_{\alpha_2,\alpha_1}\langle\chi^{\alpha_2}_{t_2}\chi^{\alpha_1}_{t_1}\rangle\mathcal{S}_{t_2}^{\alpha_2}\mathcal{S}_{t_1}^{\alpha_1}\;.
   \end{equation}
   As a check, we can note that the extra factor $(-i)^{(m-2k)}$ is justified directly from Eq.~(\ref{eq:SE_SO}) and Eq.~(\ref{eq:sup_int_H}). In fact, the final expression should have a $(-i)$ for each of the superoperators $\mathcal{S}$ present in the expression. Those present in the exponents always come in pair, thereby giving rise to the minus sign in the definition of $\mathcal{F}_t$ above, which leads to Eq.~(\ref{eq:F_rep}) through Eq.~(\ref{eq:basedone}). However, we also have an extra $(m-2k)$ superoperators $\mathcal{S}$ for each choice of $2k$ superoperators ${{\phi}}$ to contract among each other. The expression above highlights the beauty of the time ordering operator in allowing to analyze all fields on the same footing even when evaluated at different times. This flexibility is formal as the time ordering power is hidden in its evaluation complexity. For this reason, in the next section, we consider a HEOM to iteratively solve Eq.~(\ref{eq:rhomS_app}). Before doing that, we consider the possibility to relax the restriction of using field operators with support limited to the environmental space.
   \subsection{A generalization}
   \label{app:a_generalization}
   We note that this derivation can be generalized to the case where the superoperators  $\phi_j(t_j)$ involve fields which have non-trivial support also on the system. We can see this, for example, by imposing $\phi_j(t_j)\rightarrow \sum_{{\alpha'}}\phi^{{\alpha'}}_j(t_j)\otimes\phi^{{\alpha'}}_{S;j}(t_j)$ in which  $\phi^{{\alpha'}}_j(t_j)$ and $\phi^{{\alpha'}}_{S;j}(t_j)$ have support on the environment and the system, respectively, and intended in the interaction picture.  In this case, the terms in Eq.~(\ref{eq:rhomn}) and Eq.~(\ref{eq:rhomn_corr}) would require a slight redefinition as
   \begin{equation}
   \begin{array}{lll}
\rho_n^{m}&\equiv&\displaystyle \int_0^t d\tau_1\cdots\int_0^t d\tau_n\;\sum_{\vec{\alpha},\vec{\alpha}'}C^{m}_{\vec{{\alpha}}',\vec{\alpha};n} \\
&&\displaystyle\times\mathcal{T}_S\phi^{\alpha'_1}_{S;1}(\tau_1)\cdots\phi^{\alpha'_n}_{S;n}(\tau_n)\mathcal{S}^{\alpha_1}_{\tau_1}\cdots\mathcal{S}^{\alpha_n}_{\tau_n}\rho_S(0)\;,
\end{array}
\end{equation}
in terms of the indexes $\vec{\alpha}=(\alpha_1,\cdots,\alpha_n)$, $\vec{\alpha}'=(\alpha'_1,\cdots,\alpha'_n)$, and
\begin{equation}
C^{m}_{\vec{\alpha}',\vec{\alpha};n}=\text{Tr}_B\left\{\mathcal{T}_B{{\phi}}^{\alpha_1'}_1(t_1) \cdots{{\phi}}^{\alpha_m'}_m(t_m)  \chi^{\alpha_1}_{\tau_1}\cdots\chi^{\alpha_n}_{\tau_n}\rho_B\right\}\;.
\end{equation}
As a consequence, one can then reproduce all the arguments presented after Eq.~(\ref{eq:rhomn_corr}) to get a new version of Eq.~(\ref{eq:rhomS_app}) involving the fields $\phi_{j}\rightarrow\phi^{\alpha'_j}_{j}(t_j)$, $\phi_{S;j}\rightarrow\phi^{\alpha'_j}_{j}(t_j)$, and the sum over $\alpha'_j$. In fact, this sum over $ \alpha'_j$ can be recomposed leading to an expression corresponding to Eq.~(\ref{eq:rhomS_app}), i.e.,
      \begin{equation}
   \begin{array}{lll}
\rho^{m}_S(t)  &=&\displaystyle \sum_{k=0}^{[m/2]}\sum_{\mathfrak{a}\in \mathfrak{m}(m,2k)}\mathcal{T}_S\left\langle\prod_{i\in \mathfrak{a}}{{\phi}}_i(t_i)\right\rangle(-i)^{(m-2k)}\\
     &&\displaystyle\times\prod_{j\in \mathfrak{a}_{\text{c}} }\int_0^t d\tau\sum_{\alpha}\left\langle{{\phi}}_{j}(t_j)\chi^\alpha_\tau\right\rangle\mathcal{S}^\alpha_\tau e^{\mathcal{F}_t}\rho_S(0)\;,
      \end{array}
   \end{equation}
where the only formal difference with Eq.~(\ref{eq:rhomS_app}) originates from the different position of the time ordering due to the redefinition of the fields $\phi_j(t_j)\rightarrow \sum_{{\alpha'}}\phi^{{\alpha'}}_j(t_j)\otimes\phi^{{\alpha'}}_{S;j}(t_j)$.

  \section{Environmental correlations: an analysis}
  \label{sec:app:environmentalcorrelationsanalysis}
In section \ref{sec:environmentalcorrelations}, we defined our main quantity of interest as
\begin{equation}
\label{eq:rhoms_app}
\rho_S^m(t_j;t)=\text{Tr}_B\left[\mathcal{T}\prod_{j=1}^m{{\phi}}_j(t_j)\rho(t)\right]\;,
\end{equation}
whose knowledge allows to compute the correlations in Eq.~(\ref{eq:CO}) as
\begin{equation}
\label{eq:rhoms_app_trace}
\Phi(t_j;t)=\text{Tr}_S\left[\rho_S^m(tj;t)\right]\;.
\end{equation}
By including system observables in this expression, it is also possible to compute other interesting quantities, such as how properties of the system are affected by preparing the bath in a non-Gaussian state.

For the sake of completeness, here we provide some further general analysis on the reduced correlation in Eq.~(\ref{eq:rhoms_app}), which is written in the interaction picture. In order to gain more intuition, it is possible  to write its expression more explicitly and interpret it in the Shr\"{o}dinger picture. In fact, the quantities involved in Eq.~(\ref{eq:rhoms_app}) are explicitly defined as
\begin{equation}
    \begin{array}{lll}
        \rho(t)&=&\mathcal{T} e^{-i\int_0^t d\tau\;\mathcal{H}^I_\tau}\rho(0)=U_I(t)\rho(0) U_I^\dagger(t)\\
        U_I(t)&=&U^\dagger_0(t)U(t)\\
        \varphi_j(t)&=&U^\dagger_0(t)\varphi_j U_0(t)\;,
    \end{array}
\end{equation}
where $U(t)=\exp{[-iH^\text{tot}t]}$ and $U_0(t)=\exp{[-iH^{(0)}t]}$, in terms of the total  $H^\text{tot}=H^{(0)}+H^I$ and free Hamiltonian $H^{(0)}=H_S+H_B$. Here,  $H_S$,  $H_B$, and $H^I$ are, respectively, the  system,  bath, and interaction  Hamiltonians, written in the Shr\"{o}dinger picture. We remark that the field $\varphi_j$ is the operator underlying the definition of the superoperators $\phi_j$. By using these expressions in Eq.~(\ref{eq:rhoms_app}), and imposing $t_1\leq\cdots\leq t_m\leq t$, we can write
\begin{equation}
\label{eq:rhoms_app2}
\rho_S^m(t_j;t)=\mathcal{U}_S(t)\text{Tr}_B\left[\mathcal{U}(t,t_m)\cdots\mathcal{U}(t_2,t_1)\phi_1 \mathcal{U}(t_1,0)\rho(0)\right]\;,
\end{equation}
where $\mathcal{U}(t_b,t_a)=\exp\{-i(t_b-t_a)[H^\text{tot},\cdot]\}=U(t_{ba})[\cdot] U^\dagger(t_{ba})$, with $t_{ba}=t_b-t_a$. The presence of the last $\mathcal{U}_S(t)=\exp\{-i t[H_S,\cdot]\}$ is to change frame for the overall reduced correlation, which is a matrix in the system basis. As mentioned, we further considered that the superoperators $\phi_j(t)$ can be written in terms of a set of operators $\varphi(t)$ in the interaction picture and acting either on the left or on the right. 

The reduced correlation in Eq.~(\ref{eq:rhoms_app2}) can be interpreted, in the Shr\"{o}dinger picture, as the evolution of the full system interrupted, at each time $t_j$, by the action of the superoperators $\phi_j$. This is a direct consequence of the presence of the time-ordering in the definition in Eq.~(\ref{eq:rhoms_app}). This intuitive interpretation, is valid when $t_j\leq t$.

Furthermore, we note that Eq.~(\ref{eq:rhoms_app2}) can be used to compute expectation values for operators in the interaction picture of the full system+environment, i.e., in the Heisemberg representation for the system once the bath has been traced out. When we only require the computation of correlations involving only bath operators, i.e., Eq.~(\ref{eq:rhoms_app}), we can directly use Eq.~(\ref{eq:rhoms_app_trace}) to simply take the trace of Eq.~(\ref{eq:rhoms_app2}) over the system to obtain
\begin{equation}
\label{eq:rhoms_app3}
\begin{array}{lll}
\Phi(t_j;t)&=&\text{Tr}_S\left[\mathcal{U}(t,t_m)\cdots\mathcal{U}(t_2,t_1)\phi_1 \mathcal{U}(t_1,0)\rho(0)\right]\\

&=&\text{Tr}_S\left[\phi_m\cdots\mathcal{U}(t_2,t_1)\phi_1 \mathcal{U}(t_1,0)\rho(0)\right]\\
&=&\Phi(t_j;t_j)\;.
\end{array}
\end{equation}
This shows that, when $t_j\leq t$ the correlation $\Phi^m(t_j;t)$ is constant in $t$, i.e., it fulfills the identity
\begin{equation}
\label{eq:to_prove_app}
    \Phi(t_j;t)=\Phi(t_j;t_j) \;\text{for}\; t\geq t_j\;.
\end{equation}
It is interesting to check that this relation is also encoded in the representation given in Eq.~(\ref{eq:rhomSt_main}) and Eq.~(\ref{eq:Sat_main}). In fact, in the regime where $t_j\leq t$, the quantities

      \begin{equation}
      \label{eq:temp_app_3}
   \rho_S^{\mathfrak{a}}(t_j;t)=\mathcal{T}_S\prod_{j\in \mathfrak{a}_{\text{c}} } \sum_{\alpha}\int_0^t d\tau \; \langle{{\phi}}_{j}(t_j)\chi^\alpha_\tau\rangle\mathcal{S}^{\alpha}_\tau\rho_S(t)\;,
   \end{equation}
   can be simplified. This is a consequence of the fact that the Hamiltonian in Eq.~(\ref{eq:sup_int_H}) represents a commutator, so that the index $\alpha$ always comes in pairs, let us call them $\alpha$ and $\bar{\alpha}$ to represent a system operator acting on the left, and the same operator acting on the right, with a minus sign. While this is in general not enough to further simplify the expression for the correlation, the time ordering implicit in the definition 
   $\langle{{\phi}}_{j}(t_j)\chi^\alpha_\tau\rangle=\text{Tr}_B[\mathcal{T}_B {\phi}_{j}(t_j)\chi^\alpha_\tau\rho_B]$ implies that, for $t_j\leq t$, 
   \begin{equation}
\langle{{\phi}}_{j}(t_j)\chi^\alpha_\tau\rangle=\langle{{\phi}}_{j}(t_j)\chi^{\bar{\alpha}}_\tau\rangle\;,
   \end{equation}
   for all $\tau\geq t_j$,    since $\phi_j(t_j)$ is always pushed on the very right. We can use this in Eq.~(\ref{eq:temp_app_3}) where, by splitting the integral into a sum for $\tau\leq t_j$ and for $\tau>t_j$. Since the latter is zero by the argument above, we are left with 
         \begin{equation}
   \rho_S^{a}(t_j;t)=\mathcal{T}_S\prod_{j\in \mathfrak{a}_{\text{c}} } \sum_{\alpha}\int_0^{t_j} d\tau \; \langle{{\phi}}_{j}(t_j)\chi^\alpha_\tau\rangle\mathcal{S}^{\alpha}_\tau\rho_S(t)\;,
   \end{equation}
   so that $\rho_S^{\mathfrak{a}}(t_j;t)=\rho_S^{\mathfrak{a}}(t_j;t_j)$ for $t\geq t_j$. Inserting this result in Eq.~(\ref{eq:rhomSt_main}), directly reproduces Eq.~(\ref{eq:to_prove_app}). This concludes the check that, indeed, both the original definition in Eq.~\ref{eq:rhoms_app} and the 
 equivalent representation in Eq.~(\ref{eq:rhomSt_main}), lead to the same conclusion.

   \section{HEOM for environmental correlations: general case}
   \label{sec:HEOMEnvCorr}
   In this section, we specify the details of the derivation of an extended version of the  HEOM to analyze  reduced environmental correlations, i.e., computing the quantity $\rho^m_S(t)$ in Eq.~(\ref{eq:rhomS_app}). 
   This task is going to critically depend on whether the fields are required to be tracked along the dynamics or just for a fixed time. We take into account this distinction by  introducing two classes of fields. One is constituted by ``dynamical'' fields, ${{\phi}}^\text{dyn}_j(t)$, $j\in J_\text{dyn}=1,\dots,m_\text{dyn}$, evaluated at the same time $t$ as the dynamics. The other class describes ``static'' fields, ${{\phi}}^\text{stat}_{j'}(t^\text{stat}_{j'})$,  $j'\in J_\text{stat}=m_\text{dyn}+1,\dots,m_\text{dyn}+m_\text{stat}$, evaluated at fixes times $t^\text{stat}_{j'}$, so that $m_\text{stat}+m_\text{dyn}=m$. Importantly, this distinction includes two relevant cases, i.e., output fields (evaluated at the final time) and input fields (evaluated at the initial, fixed time). 

Before presenting the details of the derivation, we first gather some more intuition on the meaning of the time dependence in the quantities we want to compute. In fact, the solution of a set of differential equations in the time-variable $t$ for  $\rho^m_S(t)$ would allow, for example, to compute  full correlations in the form given  in Eq.~(\ref{eq:CO}), and which we report for convenience here as
\begin{equation}
\label{eq:eq:CO_app}
\Phi(t_j;t)=\text{Tr}_S[\rho^m_S(t)]=\text{Tr}\left\{\mathcal{T}\prod_{j=1}^m{{\phi}}_j(t_j)\rho(t)\right\}\;.
\end{equation}
From the equation above, it is already intuitive that the  ``dynamical'' fields are going to be treated differently than the other fields since their argument is the same as the time-variable of the differential equation we are seeking. 
We further note that,  for $t_j\leq t$, the quantities $\Phi(t_j;t)$ have a direct operative interpretation obtained by simply translating their expression to the Shr\"{o}dinger picture. However, whenever Eq.~(\ref{eq:eq:CO_app}) is the solution of a differential equation in the time $t$ with initial condition at $t=0$, there will always exist a regime where $t_j>t$, whose interpretation is less evident. In other words, in the presence of ``static'' fields, the resulting correlation might not have a clear meaning when $t< t_j$ for some $j$.

Following these intuitive consideration, as mentioned, we are now going to explicitly distinguish bewteen dynamical and static fields. We then start by assuming the typical HEOM ansatz
   \begin{equation}
   \langle{{\phi}_{ j }}(t)\chi^\alpha_\tau\rangle=\sum_{k}c^{ j \alpha k} e^{-\gamma^{ j \alpha k}(t-\tau)}~\text{for}~ j\in J_{\text{dyn}}\;,
   \end{equation}
   for the ``dynamical'' fields. The correlation for the ``static'' fields depends only on the time $\tau$, so that we do not need to impose the corresponding ansatz.
Inspired by the standard HEOM proof, we now define
   \begin{equation}
   \label{eq:defY}
   \mathcal{Y}^{j\alpha k}_{t}=c^{j\alpha k}\int_0^t d\tau  e^{-\gamma^{j\alpha k}(t-\tau)}\mathcal{S}^{\alpha}_\tau\;
   \end{equation}
   for $j\in J_{\text{dyn}}$ and
      \begin{equation}
   \label{eq:defY}
   \mathcal{Y}^{j\alpha k}_{t}\equiv \mathcal{Y}^{j}_{t}\delta_{\alpha,0}\delta_{k,0}\;,
   \end{equation}
   for    $j\in J_{\text{stat}}$ , with
   \begin{equation}
    \mathcal{Y}^{j}_{t}=\sum_{\alpha'}\int_0^t d\tau \; \langle{{\phi}_{j}}(t_j)\chi^{\alpha'}_\tau\rangle\mathcal{S}^{\alpha'}_\tau\equiv\int_0^t d\tau  \;\tilde{\mathcal{Y}}^{j}_{\tau}\;,
   \end{equation}
   where we defined 
   \begin{equation}
   \label{eq:Ytilde}
   \tilde{\mathcal{Y}}^{j}_{\tau}=\sum_{\alpha'} \langle{{\phi}_{j}}(t_j)\chi^{\alpha'}_\tau\rangle\mathcal{S}^{\alpha'}_\tau\;,
   \end{equation}
    for later convenience.
We note that the artificial values for the indexes $\alpha$ and $k$ in Eq.~(\ref{eq:defY}) introduces a  notational redundancy aimed at providing a unified formalism for all cases in the following calculations.
   In terms of these assumptions and definitions, Eq.~(\ref{eq:rhomS_app}) can be written as 
         \begin{equation}
         \label{eq:rhomSt}
   \begin{array}{lll}
\rho^{m}_S(t)  &=&\displaystyle \sum_{k=0}^{[m/2]}\sum_{\mathfrak{a}\in \mathfrak{m}(m,2k)}\left\langle\prod_{i\in \mathfrak{a}}{{\phi}}_i(t_i)\right\rangle(-i)^{(m-2k)}\rho_S^{\mathfrak{a}}(t)\;,
      \end{array}
   \end{equation}
   where 
   \begin{equation}
   \label{eq:Sat}
   \rho_S^{\mathfrak{a}}(t)=\prod_{j\in \mathfrak{a}_{\text{c}} } \sum_{\alpha,k}  \mathcal{Y}_t^{j\alpha k}\rho_S(t)\;,
   \end{equation}
   constitutes the main quantity of interest in this section. We now proceed by further defining
   \begin{equation}
   \label{eqLrhoNnt}
   \begin{array}{lll}
\rho_{\vec{n}^{{{\phi}}},\vec{n}}(t)&\equiv&  \mathcal{T}\prod_{j\alpha k} \left[\mathcal{Y}^{j\alpha k}_t\right]^{n^{{{\phi}}}_{j\alpha k}}\prod_{\sigma}\left[\Theta^{\sigma}_t\right]^{n_\sigma}\rho_S(t)\\
&=&  \mathcal{T}\prod_{\eta} \left[\mathcal{Y}^{\eta}_t\right]^{n^{{{\phi}}}_{\eta}}\prod_{\sigma}\left[\Theta^{\sigma}_t\right]^{n_\sigma}\rho_S(t)\;,
\end{array}
   \end{equation}
   where $n^{{\phi}}_{j\alpha k}\equiv n^{{\phi}}_{\eta} \in\mathbb{N}$ in terms of the multi-index  $\eta\equiv(\eta_1,\eta_2,\eta_2)\equiv (j,\alpha,k)$. We define $\eta_\text{dyn}$ and $\eta_\text{stat}$ as the set of $\eta$ having its first $j-$component is either in $J_\text{dyn}$ or $J_\text{stat}$. As a comment, we note that $\rho_{\vec{n}^{\phi},\vec{n}}(t)$ contains all possible contributions from the superoperators $\Theta^\sigma$ necessary to define the HEOM for the reduced system dynamics. Alongside them, it also contains all the information about the superoperators $\mathcal{Y}_t^{j\alpha k}$ for all values of their indexes. In other words, it encodes all the information required to reconstruct $\rho_S^{\mathfrak{a}}(t)$ for all $\mathfrak{a}$. As a reminder, here $\mathfrak{a}$ is a subset of $M=\{1,\cdots,m\}$ since, in fact, it is a subset of $\mathfrak{m}(m,2k)$, itself a subset made out of $2k$ elements of $M$. To decode this information, i.e., to recover $\rho_S^{\mathfrak{a}}(t)$ from $\rho_{\vec{n}^{{{\phi}}},\vec{n}}(t)$ we can simply note that each of the terms in the sum in  Eq.~(\ref{eq:Sat}) contains a product of $\mathcal{Y}_t^{j\alpha k}$ containing one, and only one, operator for each $j\in \mathfrak{a}_{\text{c}}$. This leads to defining  
   \begin{equation}
   \vec{n}^{\text{tot}}_j\equiv\sum_{\alpha k}n^{{{\phi}}}_{j\alpha k}\;,
   \end{equation}
which allows to write
   \begin{equation}
   \label{eq:rhoSat}
     \rho_S^{\mathfrak{a}}(t)= \sum_{\vec{n}^{{{\phi}}}: \vec{n}^{\text{tot}}_{j\in \mathfrak{a}_{\text{c}}}=1,\vec{n}^{\text{tot}}_{j\not\in \mathfrak{a}_{\text{c}}}=0}\rho_{\vec{n}^{{{\phi}}},\vec{n}=\vec{0}}(t)\;.
   \end{equation}
   In other words, we need to sum over all contributions in which the tensor $n^{{{\phi}}}_{j\alpha k}$ has one, and only one, element different from zero for all $j\in \mathfrak{a}_{\text{c}}$. In other words, from Eq.~(\ref{eqLrhoNnt}), this corresponds to a sum of a product of operators $\prod_{j\in \mathfrak{a}_{\text{c}}}\mathcal{Y}^{j\alpha_j k_j}_t$ over the indexes $\alpha_j$ and $k_j$.
   
Thanks to these definitions, we can now obtain the HEOM for the quantity in Eq.~(\ref{eqLrhoNnt}). This can be done thanks to the identities
\begin{equation}
\label{eq:Y_diff_app1}
    \dot{\mathcal{Y}}^{\eta}_t=    c^\eta \mathcal{S}^{\eta_2}_t-\gamma^\eta{\mathcal{Y}}^{\eta}_t\;,
\end{equation}
for $\eta\in \eta_{\text{dyn}}$, and
\begin{equation}
\label{eq:Y_diff_app2}
    \dot{\mathcal{Y}}^{\eta}_t=    \tilde{\mathcal{Y}}^{j}_{t}\;,
\end{equation}
for $\eta\in \eta_{\text{stat}}$, following Eq.~(\ref{eq:Ytilde}).
These expressions imply that taking a derivative of Eq.~(\ref{eqLrhoNnt}) will simply reproduce the same effects as in the original HEOM in Eq.~(\ref{eq:HEOM}) with the replacements  $a^\sigma, b^\sigma\rightarrow c^\eta,\gamma^\eta$, $\mathcal{B}^{\sigma_2}_t \rightarrow\mathcal{S}^{\eta_2}_t$, and ${\Theta}^{\sigma}_t\rightarrow {\mathcal{Y}}^{\eta}_t$. In other words
\begin{equation}
\label{eq:gen_HEOM_app}
\begin{array}{l}\dot{\rho}^{(N^{{\phi}},N)}_{\vec{n}^{{{\phi}}},\vec{n}}=\displaystyle\text{HEOM}_0[\rho_{\vec{n}^{{{\phi}}},\vec{n}}] -\sum_{\eta\in\eta_{\text{dyn}}} n^{{{\phi}}}_\eta \gamma^{\eta}\rho^{(N^{{\phi}},N)}_{\vec{n}^{{{\phi}}},\vec{n}}\\
\displaystyle+\sum_{\eta\in\eta_{\text{dyn}}}n^{{{\phi}}}_\eta c^{\eta}\mathcal{S}_t^{\eta_2}\rho^{(N^{{\phi}}-1,N)}_{\vec{n}^{{{\phi}}}-\vec{e}{(\eta)},\vec{n}}
+\sum_{\eta\in\eta_{\text{stat}}}n^{{{\phi}}}_\eta\tilde{\mathcal{Y}}^{\eta_1}_{t}\rho^{(N^{{\phi}}-1,N)}_{\vec{n}^{{{\phi}}}-\vec{e}{(\eta)},\vec{n}}\;,
\end{array}
\end{equation}
where $\vec{e}(\eta)$ is the tensor in the multi-index $\eta$ space with all entries equal to zero apart from a unit entry at position $\eta$, i.e., $[\vec{e}(\eta)]_{\eta_0}=\delta_{\eta,\eta_0}$. Here, $\text{HEOM}_0$ is the operator acting on the indexes $\vec{n}$ as defined in Eq.~(\ref{eq:HEOM}). We note that, in the extended indexes there is no term dependent on  $\mathcal{A}$ since the derivative on the system dynamics is taken once, and encoded in the HEOM part. This corresponds to the fact that the HEOM operator in the new indexes is bounded, i.e., there is no ``process'' which increases the total $N^{{\phi}}=\sum_{\eta}{n}^{{{\phi}}}_\eta$. We note that, while time-dependent factors might appear in Eq.~(\ref{eq:gen_HEOM_app}), there is no more explicit reference to the time ordering, which is the ultimate goal of the HEOM.
The initial condition for this differential equation is 
\begin{equation}
\rho_{\vec{n}^{{{\phi}}},\vec{n}}(0)=\delta_{\vec{n}^{{{\phi}}},\vec{0}}\delta_{\vec{n},\vec{0}}\;\rho_S(0)\;.
\end{equation}
We can finish this section by writing the whole HEOM explicitly as
\begin{equation}
\begin{array}{lll}\dot{\rho}^{(N^{{\phi}},N)}_{\vec{n}^{{{\phi}}},\vec{n}}&=&\displaystyle -\sum_{\sigma} n_\sigma b^{\sigma}\rho^{(N^{{\phi}},N)}_{\vec{n}^{{{\phi}}},\vec{n}}+\sum_{\sigma}n_\sigma a^{\sigma}\mathcal{B}^{\sigma_2}\rho^{(N^{{\phi}},N-1)}_{\vec{n}^{{{\phi}}},\vec{n}-\vec{e}{(\sigma)}}\\
&&\displaystyle +\sum_{\sigma}\mathcal{A}^{\sigma}\rho^{(N^{{\phi}},N+1)}_{\vec{n}^{{{\phi}}},\vec{n}+\vec{e}(\sigma)} -\sum_{\eta\in\eta_{\text{dyn}}} n^{{{\phi}}}_\eta \gamma^{\eta}\rho^{(N^{{\phi}},N)}_{\vec{n}^{{{\phi}}},\vec{n}}\\
&&\displaystyle+\sum_{\eta\in\eta_{\text{dyn}}}n^{{{\phi}}}_\eta c^{\eta}\mathcal{S}_t^{\eta_2}\rho^{(N^{{\phi}}-1,N)}_{\vec{n}^{{{\phi}}}-\vec{e}{(\eta)},\vec{n}}\\
&&\displaystyle+\sum_{\eta\in\eta_{\text{stat}}}n^{{{\phi}}}_\eta\tilde{\mathcal{Y}}^{\eta_1}_{t}\rho^{(N^{{\phi}}-1,N)}_{\vec{n}^{{{\phi}}}-\vec{e}{(\eta)},\vec{n}}\;,
\end{array}
\end{equation}
which constitutes the main result of this article. In fact, a solution to these equations can be used in Eq.~(\ref{eq:rhoSat}) for all the set $a$ required in Eq.~(\ref{eq:rhomSt}) which can then be computed explicitly from the knowledge of the correlations involving the operators ${{\phi_j}}$.

\section{Input-Output HEOM: a specific instance in full form}
\label{sec:app:ioHEOM_ex}
In this section, we present a specific exemplification of the extended HEOM in section \ref{sec:input_output} to write them in a full explicit form. For clarity, we present the output, input, and input-output in independent sections. 
\subsection{Output}
 Here, we analyze the output HEOM in the case where the system-bath interaction can be written in terms of one system and one environmental interaction operator. We further assume the two output fields to be written as $\phi^\text{out}_1(t)=\varphi^\dagger(t)[\cdot]$ and  $\phi^\text{out}_2(t)=\varphi(t)[\cdot]$ in terms of a non-Hermitian underlying field $\varphi$. This time,  the cross-correlations satisfy
       \begin{equation}
       \langle{{\phi}}_j^{\text{out}}(t)\chi^\alpha_\tau\rangle=\overline{\langle{{\phi}}_{{j_\text{c}}}^{\text{out}}(t)\chi^{{\alpha_\text{c}}}_\tau\rangle}\;,
   \end{equation}
   where, since in this case, $\alpha=\text{l}, \text{r}$, and $j=1,2$, we denoted by $\alpha_\text{c}$ and by $j_\text{c}$ their complementary  value, i.e., for example, if $\alpha=\text{l}$, then ${\alpha_\text{c}}=\text{r}$. The constraints above implies that
   \begin{equation}
  c^{j\alpha k}=\bar{c}^{{j_\text{c}}{\alpha_\text{c}} k}\;, \gamma^{{j}{\alpha} k}=\bar{\gamma}^{{j_\text{c}}{\alpha_\text{c}} k}\;.
 \end{equation}
 The extended HEOM in Eq.~(\ref{eq:HEOM_extended_main_in_out_field_quadr}) can then be written, more explicitly, as
         \begin{equation}
         \label{eq:outout_ex}
 \begin{array}{ll}\dot{\rho}^{(N^{\text{out}}\!\!,  N)}_{\vec{n}^{\text{out}}\!,\vec{n}}&=\displaystyle \text{HEOM}_{0}\!\!\left[\rho^{(N^{\text{out}}\!\!, N)}_{\vec{n}^{\text{out}}\!,\vec{n}}\right] \\
 &\displaystyle+\sum_{k} n^{{{\text{out}}}}_{1\text{l} k}\left[c_\text{l}^{k}s_t\rho^{(N^{{\text{out}}}\!-1,N)}_{\vec{n}^{{{\text{out}}}}-\vec{e}{(1,\text{l},k)},\vec{n}}-\gamma_\text{l}^{k}\rho^{(N^{\text{out}}\!\!,  N)}_{\vec{n}^{\text{out}}\!,\vec{n}}\right]\\

 &\displaystyle+\sum_{k} n^{{{\text{out}}}}_{2\text{r} k}\left[-\bar{c}_\text{l}^{k}\rho^{(N^{{\text{out}}}\!-1,N)}_{\vec{n}^{{{\text{out}}}}-\vec{e}{(2,\text{r},k)},\vec{n}}s_t-\bar{\gamma}_\text{l}^{k}\rho^{(N^{\text{out}}\!\!,  N)}_{\vec{n}^{\text{out}}\!,\vec{n}}\right]\\

 &\displaystyle+\sum_{k} n^{{{\text{out}}}}_{2\text{l} k}\left[\bar{c}_\text{r}^{k}s_t\rho^{(N^{{\text{out}}}\!-1,N)}_{\vec{n}^{{{\text{out}}}}-\vec{e}{(2,\text{l},k)},\vec{n}}-\bar{\gamma}_\text{r}^{k}\rho^{(N^{\text{out}}\!\!,  N)}_{\vec{n}^{\text{out}}\!,\vec{n}}\right]\\

 &\displaystyle+\sum_{k} n^{{{\text{out}}}}_{1\text{r} k}\left[-{c}_\text{r}^{k}\rho^{(N^{{\text{out}}}\!-1,N)}_{\vec{n}^{{{\text{out}}}}-\vec{e}{(1,\text{r},k)},\vec{n}}s_t-{\gamma}_\text{r}^{k}\rho^{(N^{\text{out}}\!\!,  N)}_{\vec{n}^{\text{out}}\!,\vec{n}}\right],
 \end{array}
 \end{equation}
 in terms of the decompositions
 \begin{equation}
 \begin{array}{lll}
 \langle\varphi^\dagger(t)X(\tau)\rangle&=&\displaystyle\sum_{k}c_\text{l}^{k} e^{-\gamma_\text{l}^{k}(t-\tau)}\\
 \langle X(\tau)\varphi^\dagger(t)\rangle&=&\displaystyle\sum_{k}c_\text{r}^{k} e^{-\gamma_\text{r}^{k}(t-\tau)}\;.
 \end{array}
 \end{equation}
 where we defined $c^{1\text{l}k}\rightarrow c_\text{l}^{k}$, $\gamma^{1\text{l} k}\rightarrow \gamma_\text{l}^{k}$, $c^{1\text{r}k}\rightarrow c_\text{r}^{k}$, and $\gamma^{1\text{r} k}\rightarrow \gamma_\text{r}^{k}$.
\subsection{Input}
We start from the pure-input analysis done in section \ref{subsec:i} and specialize it for the case of an open system characterized by the interaction Hamiltonian in Eq.~(\ref{eq:sX}), and for input fields written as
\begin{equation}
\phi^\text{in}_{j=1}(0)=\varphi^\dagger[\cdot]\;,~\phi^\text{in}_{j=2}(0)=[\cdot]\varphi\;,
\end{equation}
in terms of an underlying field $\varphi$ which could specify a wave-packet injected in the bath. In this case, we have
     \begin{equation}
     \mathcal{G}^{j=1}_t=\langle X_t \varphi^\dagger\rangle[s_t,\cdot]\;,~\mathcal{G}^{j=2}_t=\langle \varphi X_t \rangle[s_t,\cdot]\;,
   \end{equation}
   where we used the implicit presence of a time-ordering in the definition of $\langle[\cdot]\rangle$.
   In this way, the HEOM in Eq.~(\ref{eq:HEOM_extended_main_1_in_field_quadr}) becomes
         \begin{equation}
         \label{eq:exHEOM_inp2} 
            \renewcommand{\arraystretch}{1.9}{
\begin{array}{l}\dot{\rho}^{(N^{\text{in}}\!\!,N)}_{\vec{n}^{\text{in}}\!,\vec{n}}=\displaystyle \text{HEOM}_{0}\left[\rho^{(N^{\text{in}}\!\!,N)}_{\vec{n}^{\text{in}}\!,\vec{n}}\right]\\
+ n_1^{\text{in}}\langle X_t \varphi^\dagger\rangle[s_t,\rho^{(N^{{\text{in}}}\!-1,N)}_{\vec{n}^{{{\text{in}}}}\!-(1,0),\vec{n}}]+ n_2^{\text{in}}\langle \varphi X_t \rangle[s_t,\rho^{(N^{{\text{in}}}\!-1,N)}_{\vec{n}^{{{\text{in}}}}\!-(0,1),\vec{n}}]\;.
\end{array}}
\end{equation}
It is interesting to further increase in specificity and write the HEOM considered in this section in a pure-dephasing limit where the system coupling operators commute with the free system dynamics.  We do this in Appendix \ref{sec:PureDephasing}. Here, we focus on fully expanding Eq.~(\ref{eq:exHEOM_inp2})  on the input indexes to write
  \begin{equation}
  \label{eq:pure_input}
     \renewcommand{\arraystretch}{1.9}{
\begin{array}{lll}
\dot{\rho}^{(0,N)}_{0,0,\vec{n}}&=&\!\!\displaystyle \text{HEOM}_{0}\left[\rho^{(0,N)}_{0,0,\vec{n}}\right]\\
\dot{\rho}^{(1,N)}_{1,0,\vec{n}}&=&\!\!\displaystyle \text{HEOM}_{0}\left[\rho^{(1,N)}_{1,0,\vec{n}}\right]\!+{f}_t[s_t,\rho^{(0,N)}_{0,0,\vec{n}}]\\
\dot{\rho}^{(1,N)}_{0,1,\vec{n}}&=&\!\!\displaystyle \text{HEOM}_{0}\left[\rho^{(1,N)}_{0,1,\vec{n}}\right]\!+\bar{f}_t[s_t,\rho^{(0,N)}_{0,0,\vec{n}}]\\
\dot{\rho}^{(2,N)}_{1,1,\vec{n}}&=&\!\!\displaystyle \text{HEOM}_{0}\left[\rho^{(2,N)}_{1,1,\vec{n}}\right]\!+{f}_t[s_t,\rho^{(1,N)}_{0,1,\vec{n}}]+\bar{f}_t[s_t,\rho^{(1,N)}_{1,0,\vec{n}}],
\end{array}}
\end{equation}
where $f_t\equiv\langle X_t \varphi^\dagger\rangle$ and $\bar{f_t}=\langle \varphi X_t \rangle$. The initial condition for this differential equation is
  \begin{equation}
       \renewcommand{\arraystretch}{1.9}{
  \begin{array}{lll}
{\rho}^{(0,0)}_{0,0,\vec{n}}(0)&=&\rho_S(0)\delta_{\vec{n},\vec{0}}\\
{\rho}^{(1,N)}_{1,0,\vec{n}}(0)&=&{\rho}^{(1,N)}_{0,1,\vec{n}}(0)={\rho}^{(2,N)}_{1,1,\vec{n}}(0)=0\;.
\end{array}}
\end{equation}
The density matrix at time $t$ can then be computed as
\begin{equation}
{\rho}_S(t)={\rho}^{(0,0)}_{0,0,\vec{0}}-{\rho}^{(2,N)}_{1,1,\vec{0}}\;.
\end{equation}
We can now build on this result to present an exemplification for the input-output HEOM in section \ref{subsec:io}.
\subsection{Input-Output}
As in the previous subsection, we consider a special case for the results above for an open system characterized by the interaction Hamiltonian in Eq.~(\ref{eq:sX}), written in terms of a Hermitian environmental coupling operator $X_t$. In this case, we can write the input fields written as
\begin{equation}
\phi^\text{in}_{j=1}(0)=\varphi_\text{in}^\dagger[\cdot]\;,~\phi^\text{in}_{j=2}(0)=[\cdot]\varphi_\text{in}\;,
\end{equation}
in terms of an underlying field $\varphi_\text{in}$ which could specify a wave-packet injected in the bath. In this case, we have, as in section \ref{sec:quadraticInput},
     \begin{equation}
     \begin{array}{lll}
     \mathcal{G}^{j=1}_t&=&\langle X_t \varphi_\text{in}^\dagger\rangle[s_t,\cdot]\equiv f_t s^X\\\mathcal{G}^{j=2}_t&=&\langle \varphi_\text{in} X_t \rangle[s_t,\cdot]\equiv \bar{f}_t s^X\;,
     \end{array}
   \end{equation}
   where $f_t\equiv\langle X_t \varphi_\text{in}^\dagger\rangle$, $\bar{f_t}=\langle \varphi_\text{in} X_t \rangle$, and where $s^X\cdot\equiv[s,\cdot]$. For output fields, we further impose
   \begin{equation}
   \phi^\text{out}_1(t)=\varphi_\text{out}^\dagger(t)[\cdot],~~\phi^\text{out}_2(t)=\varphi_\text{out}(t)[\cdot]\;,
   \end{equation}
  in terms of an underlying field $\varphi_\text{out}$. In this case,  the cross-correlations are
      \begin{equation}
      \label{eq:phiphiphiphi}
         \renewcommand{\arraystretch}{1.9}{
      \begin{array}{lllll}
      \langle{{\phi}}_1^{\text{out}}(t)\chi^\text{l}_\tau\rangle&=&\!\langle \varphi^\dagger_\text{out}(t)X_\tau\rangle,\langle{{\phi}}_1^{\text{out}}(t)\chi^\text{r}_\tau\rangle&=&\!\langle X_\tau \varphi^\dagger_\text{out}(t)\rangle\\
      
            \langle{{\phi}}_2^{\text{out}}(t)\chi^\text{l}_\tau\rangle&=&\!\langle \varphi_\text{out}(t)X_\tau\rangle,\langle{{\phi}}_2^{\text{out}}(t)\chi^\text{r}_\tau\rangle&=&\!\langle X_\tau \varphi_\text{out}(t)\rangle,
      \end{array}}
  \end{equation}
as in section \ref{sec:quadratic_output}. In this notation, the coefficients in the ansatz in Eq.~(\ref{eq:ans_oo}) can be written in terms of $c^{k}_{\text{l/r}}$ and $\gamma^k_{\text{l/r}}$, as
\begin{equation}
\label{eq:simplified}
\begin{array}{lll}
\langle\varphi_\text{out}^\dagger(t)X_\tau\rangle&=&\displaystyle\sum_{k}c_\text{l}^{k} e^{-\gamma_\text{l}^{k}(t-\tau)}\\
\langle X_\tau\varphi_\text{out}^\dagger(t)\rangle&=&\displaystyle\sum_{k}c_\text{r}^{k} e^{-\gamma_\text{r}^{k}(t-\tau)}\;,
\end{array}
\end{equation}
alongside the fact that correlations involving $\varphi_\text{out}$ can be computed by complex conjugation. This allows to write the input-output HEOM as
   \begin{equation}
   \label{eq:HEOMioquad}
   \renewcommand{\arraystretch}{1.9}{
\begin{array}{l}\dot{\rho}^{(N^{\text{out}}\!\!, N^{\text{in}}\!\!, N)}_{\vec{n}_1^{\text{out}}\!,\vec{n}_2^{\text{out}}\!,n_1^\text{in}\!,n_2^\text{in}\!,\vec{n}}=\displaystyle \text{HEOM}_{0}\!\!\left[\rho^{(N^{\text{out}}\!\!, N^{\text{in}}\!\!, N)}_{\vec{n}_1^{\text{out}}\!,\vec{n}_2^{\text{out}}\!,n_1^\text{in}\!,n_2^\text{in}\!,\vec{n}}\right] +\sum_{k}\biggl\{\\

\displaystyle n^{{{\text{out}}}}_{1\text{l} k}\left[c_\text{l}^{k}s_t\rho^{(N^{\text{out}}-1, N^{\text{in}}\!\!, N)}_{\vec{n}_1^{\text{out}}-\vec{e}{(\text{l},k)}\!,\vec{n}_2^{\text{out}}\!,n_1^\text{in}\!,n_2^\text{in}\!,\vec{n}}-\gamma_\text{l}^{k}\rho^{(N^{\text{out}}\!\!, N^{\text{in}}\!\!, N)}_{\vec{n}_1^{\text{out}}\!,\vec{n}_2^{\text{out}}\!,n_1^\text{in}\!,n_2^\text{in}\!,\vec{n}}\right]\\

\displaystyle-n^{{{\text{out}}}}_{1\text{r} k}\left[{c}_\text{r}^{k}\rho^{(N^{\text{out}}-1, N^{\text{in}}\!\!, N)}_{\vec{n}_1^{\text{out}}-\vec{e}{(\text{r},k)}\!,\vec{n}_2^{\text{out}}\!,n_1^\text{in}\!,n_2^\text{in}\!,\vec{n}}s_t+{\gamma}_\text{r}^{k}\rho^{(N^{\text{out}}\!\!, N^{\text{in}}\!\!, N)}_{\vec{n}_1^{\text{out}}\!,\vec{n}_2^{\text{out}}\!,n_1^\text{in}\!,n_2^\text{in}\!,\vec{n}}\right]\\

\displaystyle+ n^{{{\text{out}}}}_{2\text{l} k}\left[\bar{c}_\text{r}^{k}s_t\rho^{(N^{\text{out}}-1, N^{\text{in}}\!\!, N)}_{\vec{n}_1^{\text{out}}\!,\vec{n}_2^{\text{out}}-\vec{e}{(\text{l},k)}\!,n_1^\text{in}\!,n_2^\text{in}\!,\vec{n}}-\bar{\gamma}_\text{r}^{k}\rho^{(N^{\text{out}}\!\!, N^{\text{in}}\!\!, N)}_{\vec{n}_1^{\text{out}}\!,\vec{n}_2^{\text{out}}\!,n_1^\text{in}\!,n_2^\text{in}\!,\vec{n}}\right]\\

\displaystyle- n^{{{\text{out}}}}_{2\text{r} k}\left[\bar{c}_\text{l}^{k}\rho^{(N^{\text{out}}-1, N^{\text{in}}\!\!, N)}_{\vec{n}_1^{\text{out}}\!,\vec{n}_2^{\text{out}}-\vec{e}{(\text{r},k)}\!,n_1^\text{in}\!,n_2^\text{in}\!,\vec{n}}s_t+\bar{\gamma}_\text{l}^{k}\rho^{(N^{\text{out}}\!\!, N^{\text{in}}\!\!, N)}_{\vec{n}_1^{\text{out}}\!,\vec{n}_2^{\text{out}}\!,n_1^\text{in}\!,n_2^\text{in}\!,\vec{n}}\right]\!\!\biggl\}\\

+ n_1^{\text{in}}f_t s^X_t\rho^{(N^{\text{out}}\!\!, N^{\text{in}}-1, N)}_{\vec{n}_1^{\text{out}}\!,\vec{n}_2^{\text{out}}\!,n_1^\text{in}-1\!,n_2^\text{in}\!,\vec{n}}
\!\!
+ n_2^{\text{in}}\bar{f}_ts^X_t\rho^{(N^{\text{out}}\!\!, N^{\text{in}}-1, N)}_{\vec{n}_1^{\text{out}}\!,\vec{n}_2^{\text{out}}\!,n_1^\text{in}\!,n_2^\text{in}-1\!,\vec{n}}.
\end{array}}
\end{equation}
These equations are simply the composition of the output and input equations in Eq.~(\ref{eq:outout_ex}) and Eq.~(\ref{eq:exHEOM_inp2}). In particular, because of its time-independence  the output part of these equations, i.e., Eq.~(\ref{eq:outout_ex}), can be interpreted as
providing additional indexes to the original $\text{HEOM}_0$, on top of which, the time-dependent input contribution, i.e., Eq.~(\ref{eq:exHEOM_inp2}), can be built upon, leading to Eq.~(\ref{eq:HEOMioquad}).

To make this construction even more explicit, it is possible to further simplify the equations by considering the underlying equilibrium state of the bath to be at zero temperature and for an output field which is a single bath mode, i.e.,  $\varphi_\text{out}\rightarrow b_0$. This allows to drop from the equations the sum over $k$ and also the left index $\alpha=\text{l}$ for $\phi_1^\text{out}=\varphi^\dagger_\text{out}\rightarrow b^\dagger_{0}$ (since, in this case, $\langle{{\phi}}_1^{\text{out}}(t)\chi^\text{l}_\tau\rangle=\langle \varphi_\text{out}^\dagger(t) X_\tau\rangle\rightarrow 0$) and the right index
 $\alpha=\text{r}$ for $\phi_2^\text{out}=\varphi_\text{out}\rightarrow b_{0}$ (since, in this case, $\langle{{\phi}}_2^{\text{out}}(t)\chi^\text{r}_\tau\rangle=\langle  X_\tau\varphi_\text{out}(t)\rangle\rightarrow 0$). In this case, the ansatz in Eq.~(\ref{eq:simplified}) can be written in terms of the parameters $c_0$ and $\gamma_0$ as
 \begin{equation}
 \label{eq:output_example_exp}
 \langle X_\tau\varphi_\text{out}^\dagger(t)\rangle\rightarrow c_0 e^{-\gamma_0(t-\tau)}\;,
 \end{equation}
so that the extended HEOM simplify to
  \begin{equation}
   \label{eq:HEOMioquad_new_in_full}
   \renewcommand{\arraystretch}{1.9}{
\begin{array}{l}\dot{\rho}^{(N^{\text{out}}\!\!, N^{\text{in}}\!\!, N)}_{{n}_1^{\text{out}}\!,{n}_2^{\text{out}}\!,n_1^\text{in}\!,n_2^\text{in}\!,\vec{n}}=\displaystyle \text{HEOM}_{0}\!\!\left[\rho^{(N^{\text{out}}\!\!, N^{\text{in}}\!\!, N)}_{{n}_1^{\text{out}}\!,{n}_2^{\text{out}}\!,n_1^\text{in}\!,n_2^\text{in}\!,\vec{n}}\right] \\

\displaystyle+n^{{{\text{out}}}}_{1}\left[-{c}_0\rho^{(N^{\text{out}}-1, N^{\text{in}}\!\!, N)}_{{n}_1^{\text{out}}-1,{n}_2^{\text{out}}\!,n_1^\text{in}\!,n_2^\text{in}\!,\vec{n}}s_t-{\gamma}_0\rho^{(N^{\text{out}}\!\!, N^{\text{in}}\!\!, N)}_{{n}_1^{\text{out}}\!,{n}_2^{\text{out}}\!,n_1^\text{in}\!,n_2^\text{in}\!,\vec{n}}\right]\\

\displaystyle+ n^{{{\text{out}}}}_{2}\left[\bar{c}_0s_t\rho^{(N^{\text{out}}-1, N^{\text{in}}\!\!, N)}_{{n}_1^{\text{out}}\!,{n}_2^{\text{out}}-1,n_1^\text{in}\!,n_2^\text{in}\!,\vec{n}}-\bar{\gamma}_0\rho^{(N^{\text{out}}\!\!, N^{\text{in}}\!\!, N)}_{{n}_1^{\text{out}}\!,{n}_2^{\text{out}}\!,n_1^\text{in}\!,n_2^\text{in}\!,\vec{n}}\right]\\

+ n_1^{\text{in}}f_t s^X_t\rho^{(N^{\text{out}}\!\!, N^{\text{in}}-1, N)}_{{n}_1^{\text{out}}\!,{n}_2^{\text{out}}\!,n_1^\text{in}-1\!,n_2^\text{in}\!,\vec{n}}
\!\!
+ n_2^{\text{in}}\bar{f}_ts^X_t\rho^{(N^{\text{out}}\!\!, N^{\text{in}}-1, N)}_{{n}_1^{\text{out}}\!,{n}_2^{\text{out}}\!,n_1^\text{in}\!,n_2^\text{in}-1\!,\vec{n}}.

\end{array}}
\end{equation}
We can now decompose this equation into its input and output structure more explicitly. In fact, setting all the input indexes to zero, the equations above can be written as
\begin{equation}
   \label{eq:HEOMioquad_ver_new}
   \renewcommand{\arraystretch}{1.9}{
\begin{array}{lll}
\dot{\rho}^{(0, 0, N)}_{0,0,0,0,\vec{n}}&=&\displaystyle  \mathbb{I}_0\;\rho^{(0, 0, N)}_{0,0,0,0,\vec{n}} \\

\dot{\rho}^{(1, 0, N)}_{1,0,0,0,\vec{n}}&=&\displaystyle -{c}_0\rho^{(0, 0, N)}_{0,0,0,0,\vec{n}}s_t+(\mathbb{I}_0-{\gamma}_0)\rho^{(1, 0, N)}_{1,0,0,0,\vec{n}}\\

\dot{\rho}^{(1, 0, N)}_{0,1,0,0,\vec{n}}&=&\displaystyle\bar{c}_0s_t\rho^{(0, 0, N)}_{0,0,0,0,\vec{n}}+(\mathbb{I}_0-\bar{\gamma}_0)\rho^{(1, 0, N)}_{0,1,0,0,\vec{n}}\\

\dot{\rho}^{(2, 0, N)}_{1,1,0,0,\vec{n}}&=&\displaystyle -{c}_0\rho^{(1, 0, N)}_{0,1,0,0,\vec{n}}s_t+\bar{c}_0s_t\rho^{(1, 0, N)}_{1,0,0,0,\vec{n}}\\
&&+[\mathbb{I}_0-({\gamma}_0+\bar{\gamma}_0)]\rho^{(2, 0, N)}_{1,1,0,0,\vec{n}}\;,
\end{array}}
\end{equation}
 in which we introduced the symbol $ \mathbb{I}_0\equiv\text{HEOM}_{0}$ to improve clarity of presentation. For non-zero values of the input indexes, the last line in Eq.~(\ref{eq:HEOMioquad_new_in_full})  must be accounted for. However, this simply corresponds to follow Eq.~(\ref{eq:pure_input}) with a ``renormalized'' $\text{HEOM}_0$ matrix which includes the output indexes as described in Eq.~(\ref{eq:HEOMioquad_ver_new}). In other words, if we define
 \begin{equation} \mathbb{I}_{\text{out}}=
 \left(\begin{array}{cccc}
 \mathbb{I}_0&0&0&0\\
-\cdot c_0s_t &  \mathbb{I}_0-\gamma_0&0&0\\
\bar{c}_0 s_t \cdot&0&  \mathbb{I}_0-\bar{\gamma}_0&0\\
0&\bar{c}_0 s_t\cdot&-\cdot c_0s_t &  \mathbb{I}_0-{\gamma_0+\bar{\gamma}_0}\\
 \end{array}\right)\;,
 \end{equation}
 then Eq.~(\ref{eq:HEOMioquad_new_in_full}) can be written from Eq.~(\ref{eq:pure_input}) as
   \begin{equation}
  \label{eq:pure_input_simple}
     \renewcommand{\arraystretch}{1.9}{
\begin{array}{lll}
\dot{\tilde{\rho}}^{(0)}_{0,0}&=&\!\!\displaystyle \mathbb{I}_{\text{out}}\left[\tilde{\rho}^{(0)}_{0,0}\right]\\
\dot{\tilde{\rho}}^{(1)}_{1,0}&=&\!\!\displaystyle \mathbb{I}_{\text{out}}\left[\tilde{\rho}^{(1)}_{1,0}\right]\!+{f}_t[s_t,\rho^{(0)}_{0,0}]\\
\dot{\tilde{\rho}}^{(1)}_{0,1}&=&\!\!\displaystyle \mathbb{I}_{\text{out}}\left[\tilde{\rho}^{(1)}_{0,1}\right]\!+\bar{f}_t[s_t,\tilde{\rho}^{(0)}_{0,0}]\\
\dot{\tilde{\rho}}^{(2)}_{1,1}&=&\!\!\displaystyle \mathbb{I}_{\text{out}}\left[\tilde{\rho}^{(2)}_{1,1}\right]\!+{f}_t[s_t,\tilde{\rho}^{(1)}_{0,1}]+\bar{f}_t[s_t,\tilde{\rho}^{(1)}_{1,0}]\;,
\end{array}}
\end{equation}
 where the quantities on the left hand-side are defined as
 \begin{equation}
\tilde{\rho}^{(N^\text{in})}_{n^\text{in}_1,n^\text{in}_2}\equiv\left(\begin{array}{c}

{\rho}^{(0, N^\text{in}, N)}_{0,0,n^\text{in}_1,n^\text{in}_2,\vec{n}}\\

{\rho}^{(1, N^\text{in}, N)}_{1,0,n^\text{in}_1,n^\text{in}_2,\vec{n}}\\

{\rho}^{(1, N^\text{in}, N)}_{0,1,n^\text{in}_1,n^\text{in}_2,\vec{n}}\\

{\rho}^{(2, N^\text{in}, N)}_{1,1,n^\text{in}_1,n^\text{in}_2,\vec{n}}

\end{array}\right)\;.
 \end{equation}
The result in Eq.~(\ref{eq:pure_input_simple}) constitutes the promised specific version of Eq.~(\ref{eq:HEOM_extended_main_in_out_field_quadratic}). 

\section{Pure Dephasing}
\label{sec:PureDephasing}
In this section, we characterize the ``quadratic-input'' results in section \ref{sec:quadraticInput} for a specific pure dephasing limit in which the system coupling operators commute with the free system dynamics. In this case, the system interaction operators commute in the interaction picture and the evaluation of the influence superoperator can be done explicitly because of the trivial action of the time-ordering.

Specifically, here we define the pure dephasing limit by  the condition $[H_S,s]=0$, the reduced system dynamics in the absence of input, i.e., for a bath in equilibrium at inverse temperature $\beta$, can be written explicitly as
\begin{equation}
\rho^{\text{deph}}_S(t)=e^{\mathcal{F}_t^\text{deph}}\rho_S(0)\;,
\end{equation}
where
\begin{equation}
\label{eq:main_def_F_deph}
\mathcal{F}^\text{deph}_t=\int_0^t d t_2\;\left\{\Gamma_\text{deph}(t_2) D_s[\cdot]-i\Omega_\text{deph}(t_2) [H_\text{deph},\cdot]\right\}\;,
\end{equation}
where $D_s[\cdot]=2s\cdot s-s^2\cdot -\cdot s^2$, $H_\text{deph}=s^2$, and where
\begin{equation}
\begin{array}{lll}
\Gamma_\text{deph}(t_2)&=&\displaystyle\frac{1}{2}\int_0^{t_2} d t_1 [C(t_2-t_1)+C(t_1-t_2)]\\
&&\displaystyle\frac{1}{\pi}\int d\omega\; J(\omega) \coth(\beta\omega/2)\frac{\sin(\omega t)}{\omega}\\
\Omega_\text{deph}(t_2)&=&\displaystyle\frac{-i}{2}\int_0^{t_2} d t_1 [C(t_2-t_1)-C(t_1-t_2)]\\
&=&-\displaystyle\frac{1}{\pi}\int d\omega\; J(\omega) \frac{1-\cos(\omega t)}{\omega}\;,
\end{array}
\end{equation}
see, for example \cite{LuoSi}.
Here, the spectral density function $J(\omega)=\pi\sum_k g_k^2\delta(\omega-\omega_k)$, is written in terms of the coupling strengths $g_k$ characterizing the system-bath interaction, i.e., $X_t=\sum_k g_k (b_k e^{-i\omega_k t}+b^\dagger_k e^{i\omega_k t})$ in terms of the environmental bosonic operators $b_k$ with energy $\omega_k$. In this simplified case, the dynamics is determined by the correlation $C(t_2-t_1)=\langle X_{t_2} X_{t_1}\rangle$, which can also be written as
\begin{equation}
C(t)=\frac{1}{\pi}\int d\omega J(\omega)[\coth(\beta\omega/2)\cos{\omega t}-i\sin\omega t]\;,
\end{equation}
in terms of which, the pure dephasing influence superoperator can be written as
\begin{equation}
\label{eq:Fdeph_corr}
\begin{array}{lll}
\mathcal{F}^\text{deph}_t&=&\displaystyle\frac{1}{2}\int_0^t d t_2\int_0^{t_2}d t_1 C(t_2-t_1)\left\{D_s[\cdot]-[H_\text{deph},\cdot]\right\}\\
&+&\displaystyle\frac{1}{2}\int_0^t d t_2\int_0^{t_2}d t_1 C(t_1-t_2)\left\{D_s[\cdot]+[H_\text{deph},\cdot]\right\}.
\end{array}
\end{equation}
When the bath is prepared with an additional input, as in the case considered here where $\rho_B\rightarrow\varphi^\dagger\rho_B\varphi$, the reduced density matrix can be computed using Eq.~(\ref{eq:CO_1_in_field_quadr}), which, in the pure-dephasing limit, reads
       \begin{equation}
   \label{eq:deph_input}
   \begin{array}{lll}
\tilde{\rho}^{\text{deph}}_S(t)&=&\displaystyle\langle\varphi\varphi^\dagger\rangle\rho^\text{deph}_S(t)\\
&+&\displaystyle\int_0^td\tau\int_0^td\tau'\; C_{\varphi X}(\tau)C_{ X\varphi^\dagger}(\tau')D_s[\rho^\text{deph}_S(t)],
\end{array}
   \end{equation}
   in terms of the cross correlations
   \begin{equation}
   C_{\varphi X}(\tau)=\text{Tr}_B\left[\varphi X_\tau\right],\; \text{and}\;      C_{X\varphi^\dagger }(\tau)=\text{Tr}_B\left[ X_\tau\varphi^\dagger\right]\;.
   \end{equation}
   In order to use this setting to gain more intuition on the origin of the terms characterizing the input in the extended HEOM in Eq.~(\ref{eq:exHEOM_inp2}), we can consider a variation on the expression in Eq.~(\ref{eq:LrhoNnt_main_1_in_field_quadr}) defining the auxiliary density matrices as, omitting some time dependencies,
   \begin{equation}
   \label{eq:aux_deph}
   \begin{array}{lll}
   \rho^\text{deph}_{n^\text{in}_1,n^\text{in}_2,n}&=&\left[\mathcal{Y}_t^{\text{deph};\text{in};1}\right]^{n^\text{in}_1}\left[\mathcal{Y}_t^{\text{deph};\text{in};2}\right]^{n^\text{in}_2}\left[\mathcal{L}_t^\text{deph}\right]^n\rho^\text{deph}_S\\
   \end{array}
   \end{equation}
   with
      \begin{equation}
      \label{Y_deph}
   \begin{array}{lll}
\mathcal{Y}_t^{\text{deph};\text{in};1}&=&\displaystyle \int_0^td\tau\;C_{X\varphi^\dagger}(\tau) [s,\cdot]\\
\mathcal{Y}_t^{\text{deph};\text{in};2}&=&\displaystyle \int_0^td\tau\;C_{\varphi X}(\tau) [s,\cdot],
   \end{array}
   \end{equation}
Here, we modified the terms leading to the regular HEOM, which is not our main focus. In fact, the substitution $\Theta\rightarrow\mathcal{L}$ with respect to Eq.~(\ref{eq:LrhoNnt_main_1_in_field_quadr}) is motivated by the fact that, in the pure dephasing case, we have that $\dot{\mathcal{F}}^\text{deph}_t\equiv\mathcal{L}^\text{deph}_t$ with
   \begin{equation}
   \label{eq:L_deph}
   \mathcal{L}^\text{deph}_t=\Gamma_\text{deph}(t) D_s[\cdot]-i\Omega_\text{deph}(t) [H_\text{deph},\cdot]\;,
   \end{equation}
as it can be directly checked from Eq.~(\ref{eq:main_def_F_deph}).
Taking the derivative of  Eq.~(\ref{eq:aux_deph}) we obtain
   \begin{equation}
   \label{eq:HEOM_deph}
   \begin{array}{l}
\dot{ \rho}^\text{deph}_{n^\text{in}_1,n^\text{in}_2,n}=\displaystyle\rho^\text{deph}_{n^\text{in}_1,n^\text{in}_2,n+1}\\
+\displaystyle n^\text{in}_1 C_{X\varphi^\dagger}(t)\!\left[s,\rho^\text{deph}_{n^\text{in}_1-1,n^\text{in}_2\!,n}\right]+ n^\text{in}_2 C_{\varphi X}(t)\!\left[s,\rho^\text{deph}_{n^\text{in}_1,n^\text{in}_2-1\!,n}\right]
\end{array}
   \end{equation}
  which serves as an exemplification for the input terms in Eq.~(\ref{eq:exHEOM_inp2}) with respect to which the terms for the regular HEOM have, here, been rather simplified. Now, applying  Eq.~(\ref{eq:rho20}) to the present dephasing example, we can write
\begin{equation}
\tilde{\rho}^{\text{deph}}_S(t)=\langle\varphi\varphi^\dagger\rangle{\rho}^{\text{deph}}_{0,0,0}- {\rho}^{\text{deph}}_{1,1,0}\;.
\end{equation}
which, indeed, corresponds to Eq.~(\ref{eq:deph_input}) as it can be checked using the expressions for the auxiliary matrices in Eq.~(\ref{eq:aux_deph}) and the fact that $[s,[s,\cdot]]=-D_s[\cdot]$ so that, explicitly,
       \begin{equation}
   \begin{array}{lll}
{\rho}^{\text{deph}}_{0,0,0}&=&\displaystyle\rho^\text{deph}_S(t)\\
{\rho}^{\text{deph}}_{1,1,0}&=-&\displaystyle\int_0^td\tau\int_0^td\tau'\; C_{\varphi X}(\tau)C_{ X\varphi^\dagger}(\tau')D_s[\rho^\text{deph}_S(t)],
\end{array}
   \end{equation}
We can further check that the derivative of the second term in the expression above also gives
   \begin{equation}
   \begin{array}{lll}
\displaystyle\dot{\rho}^{\text{deph}}_{1,1,0}&=&\displaystyle\mathcal{L}^\text{deph}_t{\rho}^{\text{deph}}_{1,1,0}\\
&-&\displaystyle C_{ X\varphi^\dagger}(t) \int_0^td\tau'\;C_{\varphi X}(\tau')D_s[{\rho}^{\text{deph}}_{0,0,0}]\\
&-&\displaystyle C_{\varphi X}(t) \int_0^td\tau'\;C_{ X\varphi^\dagger}(\tau')D_s[{\rho}^{\text{deph}}_{0,0,0}]\\
&=&\displaystyle{\rho}^{\text{deph}}_{1,1,1}+C_{ X\varphi^\dagger}(t) [s,{\rho}^{\text{deph}}_{0,1,0}]+C_{\varphi X}(t)[s,{\rho}^{\text{deph}}_{1,0,0}]\;,
\end{array}
   \end{equation}
   which is, indeed, a specific instance of Eq.~(\ref{eq:HEOM_deph}).

As a further exemplification, when the bath is characterized by a single decaying exponential, i.e., when $C_0(t)=\lambda_0^2 e^{-i\Omega_0 t-\Gamma_0 |t|}$, in terms of frequency scales $\lambda_0$, $\Omega_0$, and $\Gamma_0$, Eq.~(\ref{eq:Fdeph_corr}) simplifies to
\begin{equation}
\begin{array}{lll}
\mathcal{F}^\text{deph}_t&\rightarrow&\mathcal{F}^{\text{deph}_0}_t=f_t (s[\cdot] s-s^2[\cdot])+\bar{f}_t (s[\cdot] s-[\cdot]s^2)\;,
\end{array}
\end{equation}
where 
\begin{equation}
f_t={[e^{-i\Omega_0 t- \Gamma_0 t} - 1 + (i\Omega_0+\Gamma_0) t]}/{(i\Omega_0 +\Gamma_0)^2}\;,
\end{equation}
is defined for $t\geq 0$.
In this case, we can interpret the correlation $C_0(t)$ as arising from a bath made by a single lossy harmonic mode $a_{k_0}$ having frequency $\Omega_0$, decay rate $\Gamma_0$, coupling rate   $\lambda_0$, and initially in equilibrium in its vacuum state. This allows us to further write  $\varphi\rightarrow a_{k_0}$, which, assuming $\rho_B=\ketbra{0}{0}$ in terms of the vacuum of the oscillator,  simplifies Eq.~(\ref{eq:deph_input}) to
          \begin{equation}
   \label{eq:deph_input_single_harmonic_osc}
   \begin{array}{lll}
\tilde{\rho}^{\text{deph}}_S(t)&=&\displaystyle\rho^\text{deph}_S(t)+ g_tD_s[\rho^\text{deph}_S(t)]\;,
\end{array}
   \end{equation}
   where 
   \begin{equation}
   g_t=\lambda_0^2\frac{(1-e^{-i\Omega t-\Gamma t})(1-e^{i\Omega t-\Gamma t})}{\Omega^2+\Gamma^2}\;.
   \end{equation}
In this setting, the auxiliary density matrices take a more explicit form when considering that Eq.~(\ref{eq:L_deph}) and Eq.~(\ref{Y_deph}) become
  \begin{equation}
  \begin{array}{lll}
   \mathcal{L}^\text{deph}_t&\rightarrow &   \mathcal{L}^{\text{deph}_0}_t=\dot{f}_t (s[\cdot] s-s^2[\cdot])+\dot{\bar{f}}_t (s[\cdot] s-[\cdot]s^2)\\
\mathcal{Y}_t^{\text{deph};\text{in};1}&\rightarrow&\displaystyle\mathcal{Y}_t^{\text{deph}_0;\text{in};1}= \int_0^td\tau\;C_{X\varphi^\dagger}(\tau) [s,\cdot]=\dot{f}_t[s,\cdot]\\
\mathcal{Y}_t^{\text{deph};\text{in};2}&\rightarrow&\displaystyle\mathcal{Y}_t^{\text{deph}_0;\text{in};2}=\displaystyle \int_0^td\tau\;C_{\varphi X}(\tau) [s,\cdot]=\dot{\bar{f}}_t[s,\cdot],
   \end{array}
   \end{equation}
   where $\dot{f}_t=\lambda_0{(1-e^{-i\Omega t-\Gamma t}})/{(\Gamma+i\Omega)}$ and where we used  $C_{\varphi X}(\tau)=\lambda e^{i\Omega t- \Gamma |t|}$ and  $C_{ X\varphi^\dagger}(\tau)=\lambda e^{-i\Omega t- \Gamma |t|}$ . We can use these expressions in Eq.~(\ref{eq:aux_deph}) to write  
            $\rho^\text{deph}_{n^\text{in}_1,n^\text{in}_2,n}\rightarrow\rho^{\text{deph}_0}_{n^\text{in}_1,n^\text{in}_2,n}$ and, more specifically,
             \begin{equation}
   \begin{array}{lll}
\rho^{\text{deph}_0}_{0,0,0}&=&\displaystyle\rho^\text{deph}_S(t)\\
\rho^{\text{deph}_0}_{1,1,0}&=&\mathcal{Y}_t^{\text{deph}_0;\text{in};1}\mathcal{Y}_t^{\text{deph}_0;\text{in};2}\rho^{\text{deph}_0}_S(t)\\
&=&-|\dot{f}_t|^2 D_s[\rho^{\text{deph}_0}_S(t)]\;,
\end{array}
   \end{equation}
which allows to check that Eq.~(\ref{eq:deph_input_single_harmonic_osc}) can be written as
\begin{equation}
\tilde{\rho}^{\text{deph}}_S(t)=\rho^{\text{deph}_0}_{0,0,0}-\rho^{\text{deph}_0}_{1,1,0}\;,
\end{equation}
thanks to the fact that $g_t=|\dot{f}_t|^2$, further confirming the consistency of the formalism regarding this example. This concludes the pure dephasing exemplification.

\section{Solution of the Markovian model in rotating wave form within the single-excitation sector}
\label{sec:analytical}
We now aim to solving the full Shr\"{o}dinger equation in the 0 and 1-excitation sector. To do this, we impose the ansatz
\begin{equation}
\ket{\Psi(t)}=c_0(t)\ket{0}+c_1(t)\sigma_+\ket{0}+\sum_\mathfrak{r}\int dp\; c_\mathfrak{r}(p;t) b_\mathfrak{r}^\dagger(k)\ket{0}\;,
\end{equation}
with $|c_0^2(t)|+|c_1^2(t)|+\sum_\mathfrak{r}\int dp\; |c_\mathfrak{r}^2(p;t)|=1$. Considering the action of the Hamiltonian in Eq.~(\ref{eq:1D}) on this state, the  Shr\"{o}dinger equation implies that
\begin{equation}
\begin{array}{lll}
\dot{c}_0(t)&=&\displaystyle 0\\
\dot{c}_1(t)&=&\displaystyle -i\omega_\text{S} c_1(t)-i\sum_\mathfrak{r}\int dp\; c_\mathfrak{r}(p;t) g_\mathfrak{r}(p)\\
\dot{c}_\mathfrak{r}(p;t)&=&\displaystyle -i\omega_\mathfrak{r}(p) c_\mathfrak{r}(p;t)-i c_1(t) \bar{g}_\mathfrak{r}(p)
\end{array}
\end{equation}
We then have $c_0(t)=c_0$, and, taking the Laplace transform for the remaining variables, we obtain
\begin{equation}
\label{eq:dyn_cont}
\begin{array}{lll}
s{c}_1[s]&=&\displaystyle c_1(0)-i\omega_\text{S} c_1[s]-i\sum_\mathfrak{r}\int dp\; c_\mathfrak{r}[p;s] g_\mathfrak{r}(p)\\
s {c}_\mathfrak{r}[p;s]&=&\displaystyle c_\mathfrak{r}(p,0)-i\omega_\mathfrak{r}(p) c_\mathfrak{r}[p;s]-i c_1[s] \bar{g}_\mathfrak{r}(p)
\end{array}
\end{equation}
We can now solve this equation for $c[p;s]$ to get
\begin{equation}
\label{eq:cps}
{c}_\mathfrak{r}[p;s]=\displaystyle \frac{c_\mathfrak{r}(p,0)}{s+i\omega_\mathfrak{r}(p)}-i \frac{\bar{g}_\mathfrak{r}(p)}{s+i\omega_\mathfrak{r}(p)} c_1[s]\;,
\end{equation}
which we can insert in the expression for $c_1[s]$ to obtain
\begin{equation}\label{eq:bath_an_sigma}
\left(s+i\omega_\text{S}+\Sigma[s]\right)c_1[s]=c_1(0)-i\sum_\mathfrak{r}\int dp\;\frac{g_\mathfrak{r}(p)c_\mathfrak{r}(p;0)}{s+i\omega_\mathfrak{r}(p)}
\end{equation}
in terms of the self-energy 
\begin{equation}
\Sigma[s]=\sum_\mathfrak{r}\int dp\;\frac{|g_\mathfrak{r}^2(p)|}{s+i\omega_\mathfrak{r}(p)}\;,
\end{equation}
We can now use the spectral flatness condition in Eq.~(\ref{eq:sp_flat_white_noise}) for both positive- and negative-energy modes to find that
\begin{equation}
\Sigma[s]=2\int_{-\infty}^{\infty} d\omega\;\frac{\Gamma}{2\pi}\frac{1}{s+i\omega}=\frac{\Gamma}{\pi}\int_{-\infty}^{\infty} d\omega\;\frac{s}{s^2+\omega^2}=\Gamma,
\end{equation}
so that Eq.~(\ref{eq:bath_an_sigma}) can be written as
\begin{equation}
\left(s+i\omega_\text{S}+\Gamma\right)c_1[s]=c_1(0)-i\sum_\mathfrak{r}\int dp\;\frac{g_\mathfrak{r}(p)c_\mathfrak{r}(p;0)}{s+i\omega_\mathfrak{r}(p)}
\end{equation}
which can be solved as
\begin{equation}
\begin{array}{lll}
c_1(t)&=&\displaystyle e^{-(i\omega_\text{S}+\Gamma)t} c_1(0)\\
&-i&\displaystyle\sum_\mathfrak{r}\int dp\; g_\mathfrak{r}(p) c_\mathfrak{r}(p;0) \;\frac{e^{-i\omega_\mathfrak{r}(p)t}-e^{-(i\omega_\text{S}+\Gamma) t}}{i[\omega_\text{S}-\omega_\mathfrak{r}(p)]+\Gamma}
\end{array}
\end{equation}
We can use it in Eq.~(\ref{eq:cps}) which in time-domain reads
\begin{equation}
\label{eq:cps_time}
{c}_\mathfrak{r}(p;t)=\displaystyle e^{-i\omega_\mathfrak{r}(p)t}c_\mathfrak{r}(p;0)-i\bar{g}_\mathfrak{r}(p)\int_0^td\tau\; e^{-i\omega_\mathfrak{r}(p)\tau} c_1(t-\tau),
\end{equation}
so that

\begin{equation}
\begin{array}{l}
{c}_\mathfrak{r}(p;t)
=\displaystyle e^{-i\omega_\mathfrak{r}(p)t}c_\mathfrak{r}(p;0)\\

\displaystyle-i\bar{g}_\mathfrak{r}(p)c_1(0)\frac{e^{-(i\omega_\text{S}+\Gamma)t}-e^{-i\omega_\mathfrak{r}(p)t}}{i\omega_\mathfrak{r}(p)-i\omega_\text{S}-\Gamma} \\

\displaystyle-\bar{g}_\mathfrak{r}(p)\sum_\mathfrak{\bar{r}} \int d\bar{p}\; \frac{g_\mathfrak{\bar{r}}(\bar{p}) c_\mathfrak{\bar{r}}(\bar{p};0)}{i[\omega_\text{S}-\omega_\mathfrak{\bar{r}}(\bar{p})]+\Gamma}\frac{e^{-i\omega_\mathfrak{\bar{r}}(\bar{p})t}- e^{-i\omega_\mathfrak{r}(p)t}}{i[\omega_\mathfrak{r}(p)-\omega_\mathfrak{\bar{r}}(\bar{p})]} \\

\displaystyle+\bar{g}_\mathfrak{r}(p)\sum_\mathfrak{\bar{r}} \int d\bar{p}\; \frac{g_\mathfrak{\bar{r}}(\bar{p}) c_\mathfrak{\bar{r}}(\bar{p};0)}{i[\omega_\text{S}-\omega_\mathfrak{\bar{r}}(\bar{p})]+\Gamma}\frac{e^{-(i\omega_\text{S}+\Gamma)t}-e^{-i\omega_\mathfrak{r}(p)t}}{i\omega_\mathfrak{r}(p)-i\omega_\text{S}-\Gamma}.

\end{array}
\end{equation}
In this formalism, the observable in Eq.~(\ref{eq:Obs_x}), i.e., 
\begin{equation}
O^x(t_\text{out})=\Delta x ~b_{+}^\dagger(x_\text{out}) b_{+}(x_\text{out})\;,
\end{equation}
can be written as
\begin{equation}
\label{eq:analytical_res}
O^x(t_\text{out})=\Delta x |c_{\mathfrak{p}}(p;t)+c_{\mathfrak{n}}(p;t)|^2\;.
\end{equation}

\nocite{apsrev41Control}
\bibliographystyle{apsrev4-2}
\bibliography{bib}
\end{document}